\documentclass[acmtog]{acmart}
\usepackage{booktabs} 
\usepackage[normalem]{ulem}
\usepackage{algorithm} 
\usepackage{algorithmicx}
\usepackage[noend]{algpseudocode}
\usepackage{wrapfig}
\usepackage{comment}
\usepackage[final]{pdfpages}

\acmSubmissionID{210}

\def\argmin{\mathop{\rm argmin}}

\setcitestyle{square}

\begin{document}

\newcommand{\ourmethod}{Constitutive Cloth }
\newcommand{\todo}[1]{\textcolor{blue}{\textbf{Todo: #1}}}
\newcommand{\towrite}[1]{\textcolor{purple}{\textbf{Towrite: #1}}}
\newcommand{\fanfu}[1]{\textcolor{green}{\textbf{ff: #1}}}
\newcommand{\danny}[1]{\textcolor{orange}{\textbf{d: #1}}}
\newcommand{\minchen}[1]{\textcolor{blue}{\textbf{mc: #1}}}

\newcommand{\del}[1]{}
\newcommand{\add}[1]{#1}

\title{Codimensional Incremental Potential Contact}

\author{Minchen Li}
\affiliation{%
  \institution{University of California, Los Angeles, University of Pennsylvania, \& Adobe Research}}
\email{minchernl@gmail.com}

\author{Danny M. Kaufman}
\affiliation{%
  \institution{Adobe Research}}
\email{danny.kaufman@gmail.com}
  
\author{Chenfanfu Jiang}
\affiliation{%
  \institution{University of California, Los Angeles \& University of Pennsylvania}}
\email{chenfanfu.jiang@gmail.com}

\renewcommand{\shortauthors}{Li et al.}

\begin{abstract}
We extend the incremental potential contact (IPC) model\ \cite{Li2020IPC} for contacting elastodynamics to resolve systems composed of codimensional degrees-of-freedoms in arbitrary combination. 
This enables a unified, interpenetration-free, robust, and stable simulation framework that couples codimension-0,1,2, and 3 geometries seamlessly with frictional contact. 
Extending the IPC model to thin structures poses new challenges in computing strain, modeling thickness and determining collisions. 
To address these challenges we propose three corresponding contributions. 
First, we introduce a $C^2$ constitutive barrier model that directly enforces strain limiting as an energy potential while preserving rest state. This provides energetically-consistent strain limiting models (both isotropic and anisotropic) for cloth that enable strict satisfaction of strain-limit inequalities with direct coupling to both elastodynamics and contact via minimization of the incremental potential. 
Second, to capture the geometric thickness of codimensional domains we extend the IPC model to directly enforce distance offsets. Our treatment imposes a strict guarantee that mid-surfaces (respectively mid-lines) of shells (respectively rods) will not move closer than applied thickness values, even as these thicknesses become characteristically small. This enables us to account for thickness in the contact behavior of codimensional structures and so robustly capture challenging contacting geometries; a number of which, to our knowledge, have not been simulated before.
Third, codimensional models, especially with modeled thickness, mandate strict accuracy requirements that pose a severe challenge to all existing continuous collision detection (CCD) methods. To address these limitations we develop a new, efficient, simple-to-implement  \emph{additive CCD} (ACCD) method that \add{applies conservative advancement \cite{mirtich1996impulse,zhang2006interactive} to iteratively refine a lower bound for deforming primitives,}  \del{that iteratively refines a 
lower bound} converging to time of impact. 
In combination these contributions enable \emph{codimensional} IPC (C-IPC).
We perform extensive benchmark experiments to validate the efficacy of our method in capturing intricate behaviors of thin-structure contact and resulting bulk effects.
In our experiments C-IPC obtains feasible, convergent, and so artifact-free solutions for all time steps, across all tested examples -- producing robust simulations. We test C-IPC across extreme deformations, large time steps, and exceedingly close contact over all possible pairings of codimensional domains. Finally, with our strain-limit model, we confirm C-IPC guarantees non-intersection and strain-limit satisfaction for all reasonable (and well below -- verified down to $0.1\%$) strain limits throughout all time steps.  

\end{abstract}

\ccsdesc[500]{Computing methodologies~Physical simulation}

\keywords{Strain Limiting, Contact Mechanics, Mixed-Dimensional Elastodynamics, Constrained Optimization}

\begin{teaserfigure}
    \centering
    \includegraphics[width=\textwidth]{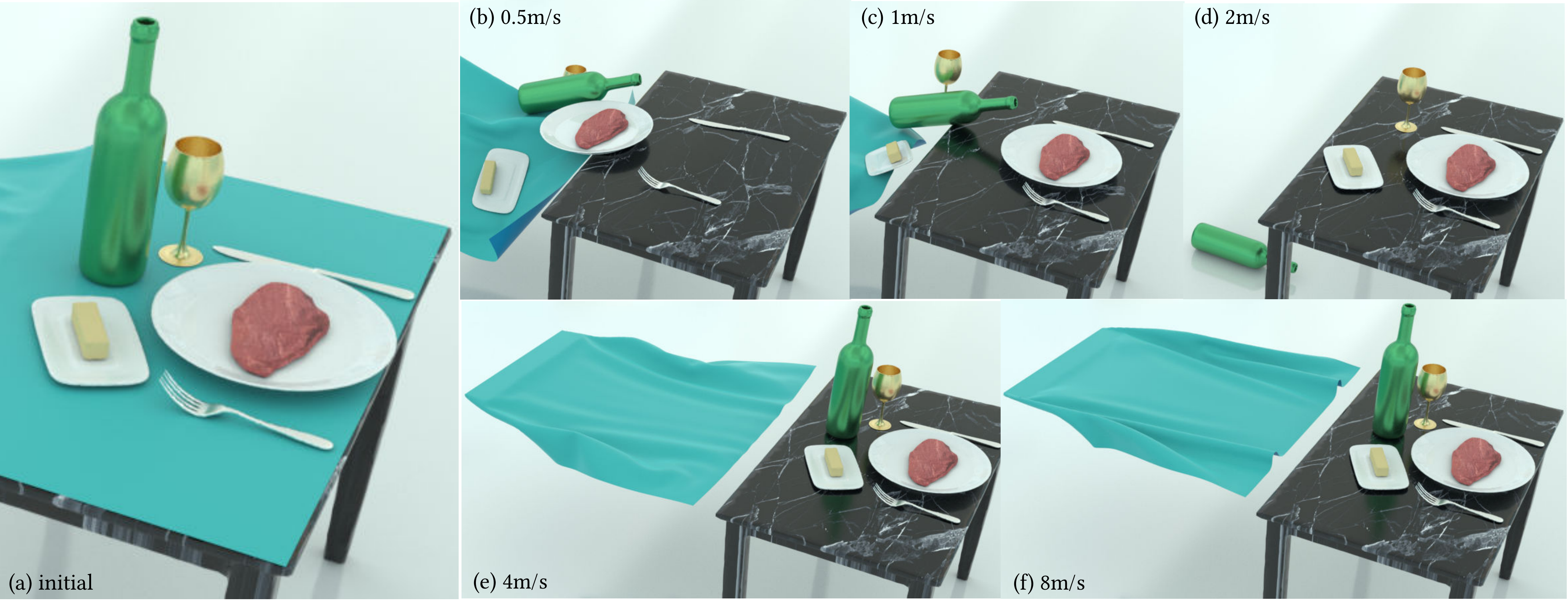}
    \caption{\textbf{Table cloth pull:} Codimensional Incremental Potential Contact (C-IPC) enables high-rate time stepping of codimensional models with intersection-free, accurate, controllable strain-limiting, thickness modeling, and frictional contact. Here, stepping shell and volumetric FE models at $h=0.01s$, a square cloth is laid on a table with heavy, stiff volumetric dinnerware (a). C-IPC stably responds to rapid pulling at varying speeds (b-f) without stretching beyond limits while accurately resolving tight contacts between table edges and dinnnerware.
With a slower pull in (b) of $0.5m/s$ nearly all objects pull off due to friction. As pull speed increase at $1m/s$ (c) and $2m/s$ (d), less dinnerware fall off with some just pulled a bit.
In (e) and (f) pulling at higher speeds $4m/s$ or $8m/s$ leaves the stetting on table with rapid acceleration overcoming friction. Here too, combined strain-limiting and resolution of elastodynamics generates detailed wrinkling behaviors in the cloth for these faster pulls as it flies off the table.
    \label{fig:teaser}}
\end{teaserfigure}

\maketitle
\section{Introduction}

\del{Cloth, hair and sand are everywhere. Simulating thin materials like these has long remained a critical task in computational modeling and animation. } \add{Thin materials like cloth, hair and sand are everywhere. Simulating them has long remained a critical task in computational modeling and animation.} Each such solid is generally best modeled by a corresponding reduced codimensional formulation; respectively by shells, rods and particles. Doing so enables improved efficiency with many less simulated degrees-of-freedom (DOF). Likewise it mitigates the numerical conditioning and severe convergence challenges posed by \emph{\add{shear }locking} that we would otherwise encounter if simulating thin materials volumetrically\ \cite{yang2000shells}. 
But while we can well-capture the \emph{elastic} behavior of individual thin materials with codimensional models, accurately and consistently simulating their collective and contacting behavior on practical size meshes, with real-world materials and conditions remains challenged. We enumerate these challenges here.

First and foremost simulations should remain \emph{intersection free} at all times. Steps that permit even slight intersections of a codimensional geometry can and will produce tangles that simulations can not recover from. 

Second, we require energetically consistent, controllable and accurate strain-limiting. While codimensional discretizations mitigate \add{shear} locking they \del{do not remove it}\add{can still suffer from severe \emph{membrane locking}}. For example, shell models continue to suffer severe numerical stiffening in bending modes for cloth materials unless midsurface meshes apply extremely high (and generally impractical size) resolutions\ \cite{quaglino2017numerical}. 
\add{Here higher-order elements can also help further reduce locking artifacts but they add prohibitive computational cost.}
Strain-limiting allows softer cloth materials coupled with tight limits to avoid these artifacts while still recovering proper cloth response. However, if strain-limiting is not accurately and consistently applied, and carefully coupled to elastodynamics, it often produces more numerical artifacts than it fixes.

\begin{figure}[t]
    \centering
    \includegraphics[width=\linewidth]{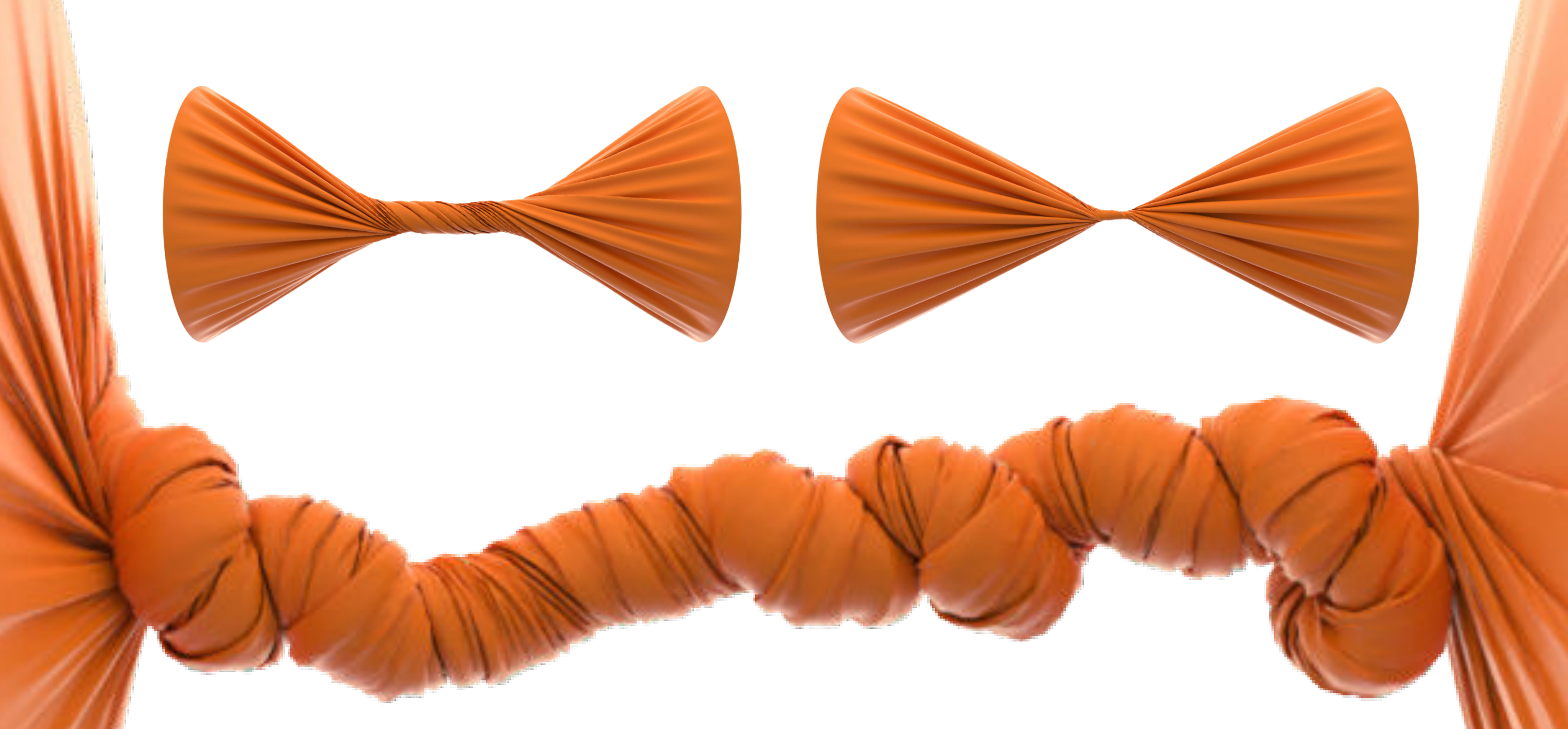}
    \caption{\textbf{Twisted cylinder.} Testing thickness modeling, contact resolution and buckling under extreme and increasing stress we start twisting a $1m$-wide cloth cylinder with 88K nodes, a thickness offset of $1.5mm$ and $\hat{d}$ of $1mm$. Time-stepped at $h=0.04s$, C-IPC  quickly obtains interesting folds and soon after forms a thick central cylinder of wound cloth supported by C-IPC's finite thickness offset (upper left\del{and bottom}). In contrast, (upper right) without C-IPC's thickness offset the correct geometry can not form (nor can it later capture the material's buckling behavior) further emphasizing the importance of consistent thickness modeling for shells. After $32.96s$ of further twisting, our offset-thickened cloth continues to support complex contact-driven behaviors including the final buckled geometry \add{(bottom)}; see our supplemental video \add{for the detailed trajectory}.}
    \label{fig:twistCylinder}
    \vspace{-0.5cm}
\end{figure}

Third, geometric thickness must be captured in contact. While thin structures typically have small relative thicknesses, whose \emph{elastic} behavior can be indirectly captured by codimensional DOFs, correctly and directly modeling their thickness in contact is crucial for 
\setlength{\columnsep}{5 pt}
\setlength{\intextsep}{5 pt}%
\begin{wrapfigure}[5]{r}{0.5\linewidth}
\vspace{-4 pt}
\centering
\includegraphics[width=0.9\linewidth]{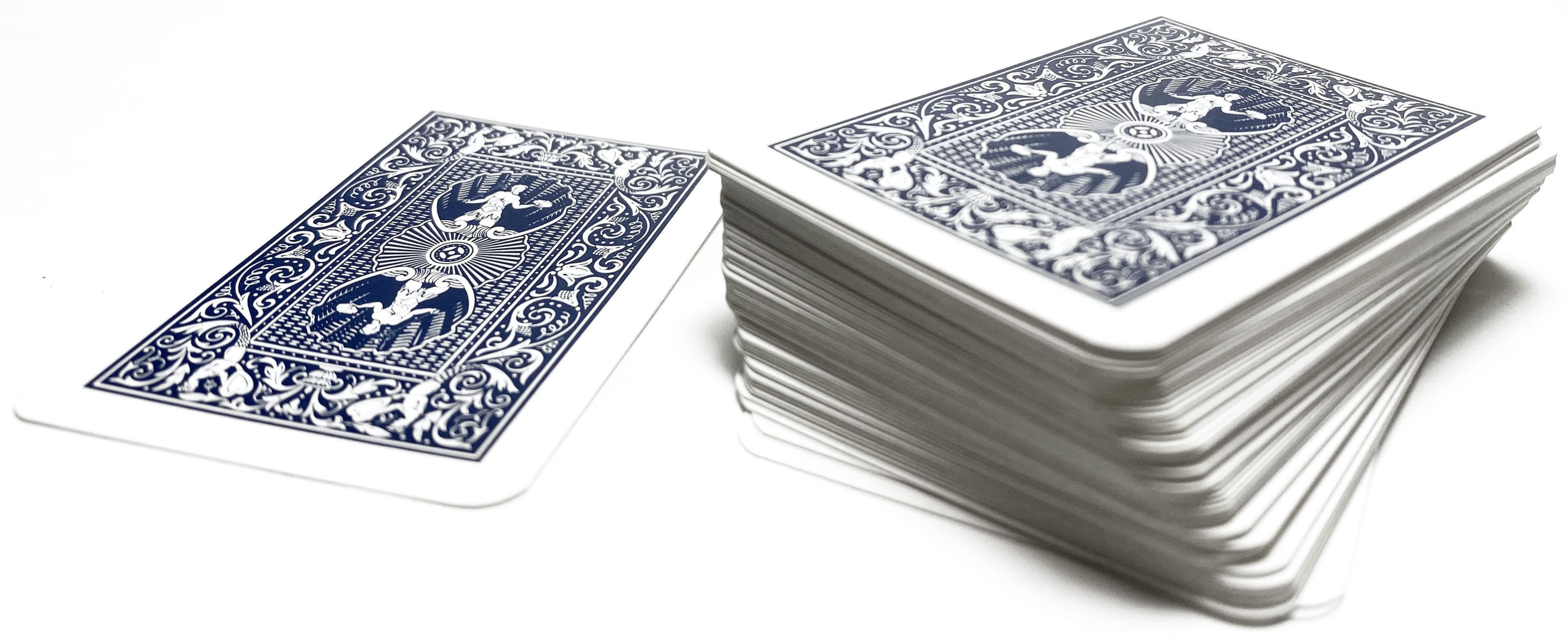}
\end{wrapfigure} 
capturing accurate geometries and bulk behaviors. 
Consider, for example, a deck of cards. Each card has a nearly imperceptible thickness when considered individually. However, when stacked, their combined thickness is large and clear, and can not be ignored. 
While contact-processing strategies often introduce thickness parameters, they are generally applied as a nonphysical failsafe to mitigate collision-processing inaccuracies. As analyzed in Li et al.\ \shortcite{Li2020IPC}, these thicknesses can not be consistently enforced and moreover must be changed heuristically per scene and example (e.g., based on collision speeds) to avoid simulation failures.

Fourth, all codimensions (including volumes) should be seamlessly simulated in a common framework without distinction nor special casing. Accurate frictional contact forces should directly couple interactions between all geometric types irrespective of how close the contact or how thin the modeled thicknesses are.

Fifth, contact modeling between codimensional models with small \emph{but finite} thicknesses challenges all collision-detection routines.
Most critically we see unacceptable failures and/or inefficiencies in all available \del{existing} CCD methods and codes. We see this both for standard floating point root-finding CCD methods as well as more recent developments in exact CCD methods. These issues are attributable to the well-known numerical sensitivity of the challenging underlying root-finding problems posed. Here, most specifically, we see that IPC is reliant on a bounded guarantee of non-intersection for all CCD evaluations. Large enough conservative bounds can help\ \cite{Li2020IPC} account for these inaccuracies, ensuring non-intersection, but they do so at the cost of progress and so can even stall convergence altogether when it comes to codimensional DOF.

Finally, simulations should not \add{artificially} snag in contact on sharp and codimensional geometries and  should be able to robustly solve time-step problems to convergence \del{irrespective of} \add{across variations in} step size, scene, materials, contact and boundary conditions.

Here we propose a method that, to our knowledge, is the first to directly resolve all six above challenges with a consistent and reliable solution for frictionally contacting dynamics. Our method guarantees strict satisfaction of fully coupled strain-limits, with intersection-free trajectories, and controllable geometric thickness resolution (at thin material scales) independent of time step size and severity of contact conditions.

\begin{figure*}[t]
    \centering
    \includegraphics[width=\linewidth]{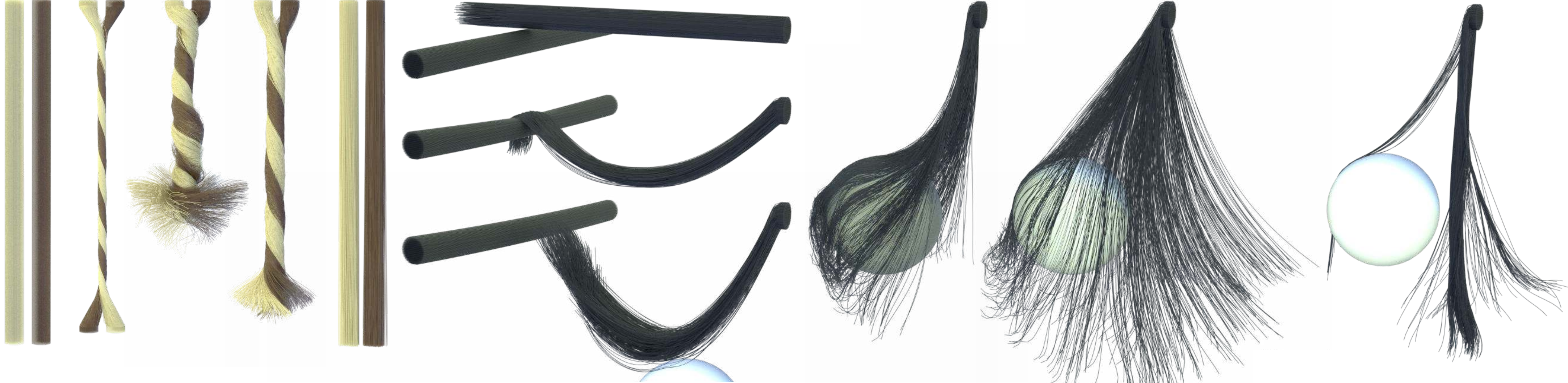}
    \caption{\textbf{Hairs.} Left: two hair clusters are twisted to form a tightly wound braid without intersection (processing up to 1.5M contacts per time step); the bottom end is then released and the braid unwinds.
    Right: a hair cluster simulation example inspired by McAdams\ \shortcite{mcadams2009detail} where one end of a hair cluster is fixed, and so falls under gravity upon another hair cluster with two ends fixed and then across a sphere.}
    \label{fig:hairs}
\end{figure*}

\subsection{Contributions}

Specifically, to address these challenges we extend the incremental potential contact model (IPC)\ \cite{Li2020IPC} for contacting elastodynamics to resolve systems composed of arbitrary combinations of codimensional DOF with three critical components.

\begin{description}
  \item[Constitutive strain limiting.] We introduce a $C^2$ constitutive barrier model that directly enforces strain limiting as an energy potential while preserving rest state. This provides energetically-consistent strain limiting models (both isotropic and anisotropic) for cloth that enable, for the first time, strict satisfaction of strain-limit inequalities for all iterations and so for all time steps (verified down to $0.1\%$), while fully and directly coupling to both elastodynamics and frictional contact via minimization of the incremental potential. Thus, as demonstrated in Section \ref{sec:SLSplittingExp}, we avoid artifacts  generated by force-splitting errors in traditional strain-limiting methods.

  \item[IPC thickness model.] To capture the geometric thickness of codimensional domains in contact we extend the IPC model to directly enforce distance offsets. Our treatment imposes a strict guarantee that mid-surfaces (respectively mid-lines and points) of shells (respectively rods and particles) will not move closer than applied thickness values, even as these thicknesses become characteristically small; see Sec.\ \ref{sec:thickness}. This enables us to account for thickness in the contacting behavior of codimensional structures and so capture challenging contacting effects; a number of which, to our knowledge, have not been simulated before. 
  
  \item[Additive CCD.] To provide the strict accuracy required of CCD to resolve codimensional models with thickness, we develop a new, efficient and simple-to-implement \emph{additive CCD} (ACCD) method \add{utilizing conservative advancement\ \cite{zhang2006interactive,zhang2007continuous,tang2009c}}. ACCD iteratively accumulates a lower bound converging to time of impact and is stable for the exceedingly challenging evaluations required. While we most immediately focus here on ACCD's application within our C-IPC framework, as we show in the following sections, ACCD is exceedingly simple and so easy to implement when compared to all prior CCD routines (exact and floating point) with both improved performance, robustness and guarantees, and so is suitable for easy replacement in all applications where CCD modules are employed.
\end{description}

Together these form the core of codimensional IPC (C-IPC). C-IPC enables unified simulation of all codimensions including elastic volumetric bodies, shells, rods and particles all coupled together via accurately solved contact and friction. Along with these core contributions we test, compare and analyze C-IPC across many new and pre-existing benchmarks. We confirm C-IPC provides tightly controllable geometric thickness behaviors and guarantees feasibility at all time steps. Finally, C-IPC remains intersection-free and strictly satisfies all set strain limits across all computed trajectories, likewise verified in our extensive benchmark testing.  

\section{Related Work}
\label{sec:rel}

\subsection{Shells and Rods}

Beginning with the pioneering work of Terzopoulos et al.\  \shortcite{terzopoulos1987elastically} the simulation of codimensional models, especially shells and rods, has remained a focus in computer graphics. Efficiently modeling the complex behaviors of cloth\ \cite{baraff1998large,volino2000implementing,bridson2002robust,grinspun2003discrete,harmon2009asynchronous,narain2012adaptive,li2020p,bender2013fast} and hair\ \cite{mcadams2009detail,selle2008mass,muller2012fast,choe2005simulating,ward2003adaptive,adn14,deul2018direct,kugelstadt2016position,daviet2011hybrid} are now particularly critical tasks across applications.

Pipelines for their simulation most commonly adopt implicit or else linearly implicit time-integration methods\ \cite{baraff1998large,bridson2002robust,harmon2008robust,otaduy2009implicit,narain2012adaptive,tang2016cama,tang2018cloth,li2018implicit,li2020p,kim2020baraff} with a diverse array of collision filters and penalties applied to help resolve contact processing.

To accelerate performance, extensions to the GPU\ \cite{tang2013gpu,schmitt2013multilevel,tang2016cama,tang2018cloth,li2020p}, fast projection\ \cite{goldenthal2007efficient,english2008animating} and multigrid \cite{tamstorf2015smoothed,xian2019multigrid,wangmg18} are all being actively explored. While, to improve the fidelity of models, data-driven material estimation\ \cite{bhat2003estimating,miguel2012data,clyde2017modeling}, alternate spatial discretizations\ \cite{jiang2017anisotropic,guo2018material,weidner2018eulerian} and even hybrid models combining yarn and shell simulation\ \cite{casafranca2020mixing,sperl2020homogenized} are being investigated. Likewise, the differentiability of cloth simulation is now also becoming critical\ \cite{liang2019differentiable} for neural training applications.

A key challenge then has been to reliably simulate shells with guaranteed results. Harmon et al.\ \shortcite{harmon2009asynchronous} introduce an explicit time-stepping method for simulating shells that offers a guarantee of non-intersection across all timesteps. This method then requires small timesteps and simplified friction. Here we provide a complementary, fully implicit, differentiable simulation method, unified for all codimensional models, with accurate frictional contact, guaranteeing non-intersection, as well non-inversion and strain-limit satisfaction (when desired), that can be stepped at all reasonable time step sizes.

\subsection{Strain Limiting}

\begin{figure}[t]
    \centering
    \includegraphics[width=\linewidth]{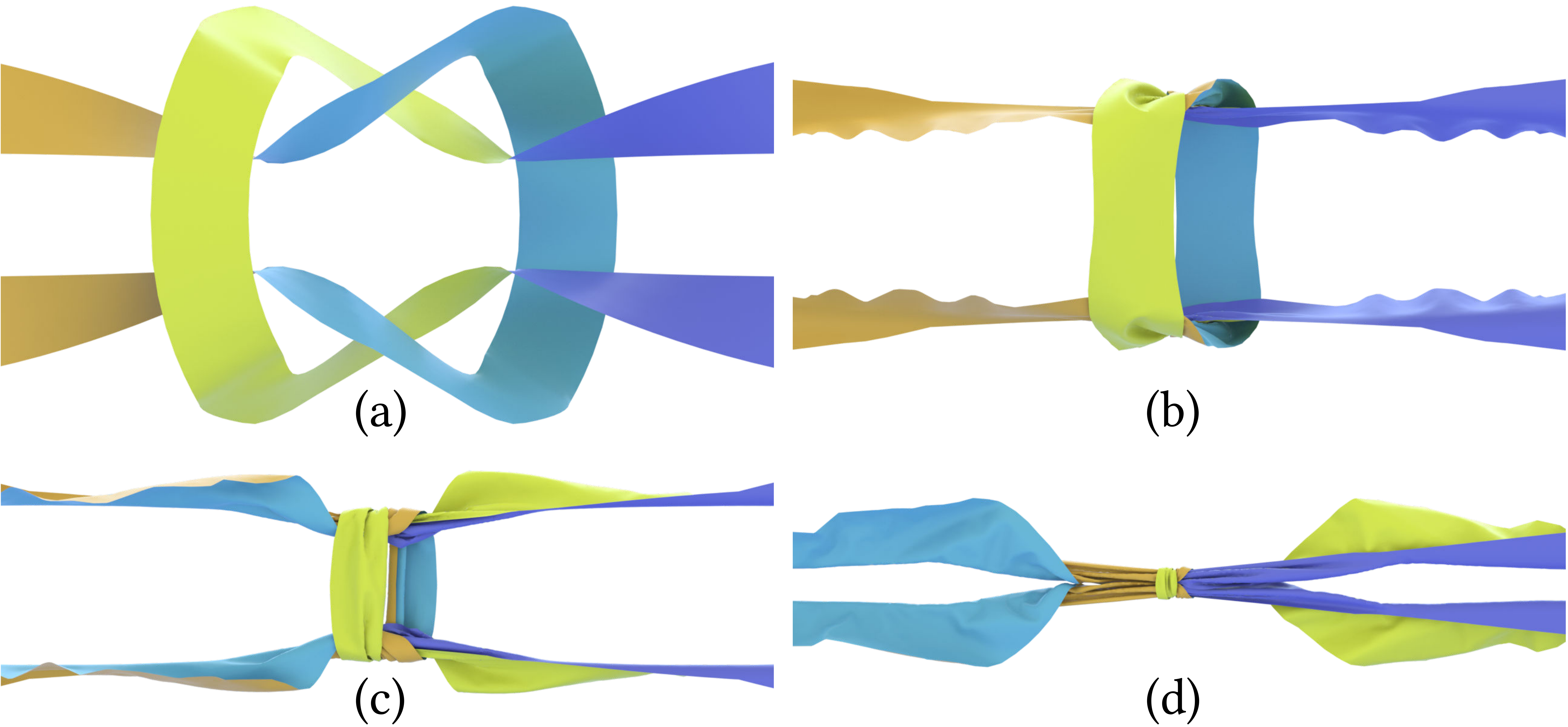}
    \caption{\textbf{Reef Knot.} 
    To exercise extreme compression, contact and friction with large stresses, we extend the the reef knot test from Harmon et al.\ \shortcite{harmon2009asynchronous} by simulating with \add{a larger system} 100K nodes ($10\times$ that of the original), \del{applying} \add{adding} a strain limit of $1.134$, friction with $\mu=0.02$, and \del{pulling even} tighter \add{pulling}. Two ribbons are initially intertwined in (a) and then stretched in (b)-(c) to form a final tight knot as the ribbons are pulled to nearly their strain limit in (d).}
    \label{fig:ribbonKnot}
\end{figure}

Strain-limiting methods seek to impose bounds on membrane deformation. A wide range of models and algorithms have long been proposed to provide strain limiting\ \cite{bridson2002robust,provot1995deformation}. To understand recent methods' behaviors and limitations we categorize them by: 1) constraint choice (limits imposed as either equality or inequality); 2) type of DOFs constrained (these are most commonly edge-based or singular values); 3) splitting model; and 4) solver applied to enforce these constraints. 

With this breakdown we see that equality constraints on length-based measures remain a consistent constraint choice. Goldenthal et al.\ \shortcite{goldenthal2007efficient} apply bilateral constraints on quadmesh edge lengths and propose a fast projection method to correct predicted displacements.
English and Bridson\ \shortcite{english2008animating} adopt a non-conforming strategy, applying equality constraints on triangle-edge midpoint distances, and extend fast projection to support a BDF-2 integrator. Thomaszewski et al.\ \shortcite{thomaszewski2009continuum} apply lagged, corotational small strain on triangles to define equality constraints and enforce limits by post-projection via Gauss-Seidel or Jacobi iterations. Chen and Tang\ \shortcite{chen2010fully} likewise define equality constraints on triangle edge lengths.

Improvement over edge-based constraints can often be obtained by constraining singular values of the deformation gradient tensor (generally per triangle) to a finite range.
However, irrespective of details, equality constraints always remain active and so simple DOF counting generally explains the membrane locking artifacts often encountered when these strain constraints are 
enforced. Here, alternate constraint DOF choices\ \cite{english2008animating, chen2019locking} can help alleviate this issue by reducing the ratio of constraints to DOFs but also can introduce new challenges, e.g. via non-conforming meshes. 
As an alternative to bilateral enforcement inequality constraints on strain, in the form of upper and lower bounds\ \cite{wang2010multi,narain2012adaptive,hernandez2013anisotropic,wang2016efficient} have long been applied; while, even more recently, applying just upper bounds\ \cite{jin2017inequality} has also been considered. Irrespective of bounds, unilateral constraints are only activated when their strain measures are at their limits and so the potential for an over-constrained system is reduced.

While approaches thus clearly vary we observe that all methods, with the exception of Chen and Tang's\ \shortcite{chen2010fully} least-squares approximation for frictionless elastostatics, currently employ time-step splitting. Here we see that two-step splits are often applied to first solve elastodynamics before applying a constraint projection to simultaneously resolve contact constraints and strain limits. Alternately, introducing even more errors, we also see three-step splits where sequential solves of elastodynamics, contact and strain limiting are each decoupled and so resolved independently per time step\ \add{\cite{narain2012adaptive}}. Thus, for the latter three-step approach strain limits and contact constraints cannot be simultaneously satisfied, while for the former two-step methods errors introduced by splitting elasticity and constraints also generate unacceptable artifacts; see Section\ \ref{sec:exp}. 

Finally, we note that, to our knowledge, no method (irrespective of constraint-type or splitting choice) employs a constraint solver that can guarantee enforcement of strain limits. As constraint enforcement errors then vary across simulation scenes, and even individual time steps, this leads to uncontrollable and inconsistent variations in material behavior per scene and step. To enable consistent, effective strain-limiting C-IPC introduces a fully coupled, inequality based model solved with a strict guarantee of strain-limit satisfaction.

  \begin{figure*}[t]
    \centering
    \includegraphics[width=\linewidth]{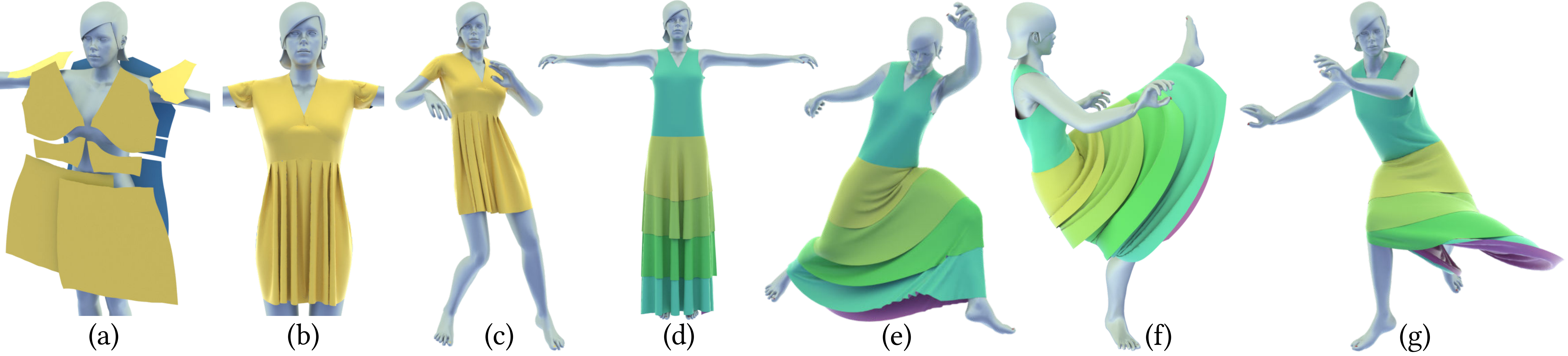}
    \caption{\textbf{Garments.} A yellow dress with challenging folded and seamed knife pleats (produced by FoldSketch\ \cite{Li2018FoldSketch}) is staged (a) and then draped on a mannequin to static equilibrium (b) with stitches seamed together. \del{and} \add{It is} then simulated on a Rhumba dancing sequence (c). A multilayer skirt is draped in the same way (d), and then used to simulate an input character animation sequence with large and fast motions (e,f,g). As the mannequin rapidly kicks, intricate garment details and high speed collisions are captured by C-IPC simulated with large time steps ($h=0.04s$).}
    \label{fig:garments}
    \vspace{-0.3cm}
\end{figure*}

\subsection{Modeling Thickness}

Thin bodies modeled by codimensional geometries generally have thicknesses orders-of-magnitude smaller than other dimensions. It is thus tempting (and common) to ignore thickness in contact. However, in order to correctly capture thin material interactions we must account for the geometric effects of thickness in contact.
We see this everywhere, for example when playing cards stack in a pile (Figure\ \ref{fig:cardShuffle}) or when noodles fill a bowl (Figure\ \ref{fig:noodles}).

A direct strategy for capturing thickness is then \add{the solid shell method\ \cite{hauptmann1998systematic} which models} \del{to model} thin materials volumetrically with full translational DOF. However, in doing so linear elements suffer from well-known \add{shear-}locking artifacts. And while higher-order elements can alleviate these convergence issues, shear-locking artifacts are still not easily avoided altogether, while computational costs increase significantly.
To address these challenges solid shell methods commonly employ reduced integration with hourglass mode stabilization\ \cite{reese2007large} or assumed natural strain methods\ \cite{cardoso2008enhanced} to mitigate \add{shear }locking with linear elements. However, this added complexity generally does not compete with codimensional modeling.

To resolve contact with codimensional models, contact-processing strategies often introduce thickness-like parameters to offset constraints and so reduce collision-processing inaccuracies; see e.g. Narain et al.\ \shortcite{narain2012adaptive} and Li et al. \shortcite{li2018implicit} for recent examples. However, as analyzed in Li et al.\ \shortcite{Li2020IPC}, these thicknesses can not be consistently enforced and moreover must be changed heuristically per scene and example (e.g., based on collision speeds) to avoid simulation failures from intersections and instabilities. In turn, while \add{they are} sometimes helpful to improve contact resolution, these tuned parameters can not reliably be applied to model consistent and changing thickness behaviors; see Section\ \ref{sec:thickness_exp}.

We extend the IPC model to capture the geometric thickness of codimensional domains with offset barriers that guarantee a requested small minimal separation from mid-surface as well mid-line and point geometries. In turn this enables reliable and consistent modeling of thickness behaviors in the interactions between thin contacting materials.

\subsection{Continuous Collision Detection}

Continuous Collision Detection (CCD) methods have long been employed as a check against intersecting mesh boundaries. Provot\ \shortcite{provot1997collision} formulated finding first times of impacts for linearly displacing point-triangle and edge-edge pairs by first testing for coplanarity via solution of a cubic equation and then performing overlap checks to detect collision.

This by now standard formulation of floating point CCD has been broadly applied and fine tuned over time\ \cite{harmon2009asynchronous,tang2011volccd}. However, while root-finding with floating-point arithmetic can be made efficient, significant numerical errors will still generate unacceptable, false results (both negative and positive). This is most commonly encountered when distances are small and/or configurations are unavoidably degenerate. Here, while switching from floating-point to rationals can help avoid round-off issues, cost can unacceptably increase if special care is not taken.

Recent methods address CCD accuracy\ \cite{brochu2012efficient,tang2014fast,wang2015tightccd} by application of exact arithmetic for improved robustness with efficiency. Alternately, others\ \cite{harmon2011interference,lu2019scalable} perform conservative CCD with a requested small conservative separation distance to better avoid interpenetration. This latter strategy is employed in volumetric IPC\ \cite{Li2020IPC} using a robust, floating-point CCD implementation\footnote{https://github.com/evouga/collisiondetection} as a base solver.

Here, for modeling thickness in C-IPC, we require CCD queries that can maintain finite separation distances between mid-surface, mid-line and point elements. In turn, we rely on accurately preserving orders-of-magnitude smaller distances between the offset surfaces for computing accurate contact forces. This challenges the accuracy and robustness of CCD queries and we see unacceptable errors, resulting in failures, in all available existing CCD methods and codes. We see this both for standard floating-point root-finding CCD methods as well as more recent exact CCD methods. In turn we see that these failures can generate intersections, slow convergence, and often stop simulation progress altogether; see Section\ \ref{sec:CCDComp}.

\add{As an alternative to root-finding, conservative advancement (CA) methods have been developed to instead iteratively advance rigid and/or articulated bodies until they are closer than some pre-defined small distance. Starting with Mirtich's work on convex rigid bodies\ \shortcite{mirtich1996impulse} this is performed by repeated calculation of a lower-bound on time of impact (TOI) followed by a conservative step taken with the bound\ \cite{zhang2006interactive,zhang2007continuous,tang2009c,tang2013hierarchical,lu2014collision}.}

For robust CCD evaluation \add{of deformable body trajectories} we derive \del{a useful} lower bounds on time of impact \add{for deforming mesh primitive pairs with arbitrary displacements} and apply them \add{in the CA framework} to develop a new, simple to compute, numerically robust, floating-point, additive CCD (ACCD) algorithm. 
\add{Under the CA framework, }ACCD monotonically approaches times of impact without error-prone, direct root-finding. In Section\ \ref{sec:exp} we confirm ACCD efficiently and accurately succeeds on a wide range of challenging cases where all other methods fail and find similar and often better performance in the cases where floating-point CCD methods can succeed. Finally we verify that ACCD is also suitable for replacement in CCD modules outside of the C-IPC framework with improved efficiency and robustness. 

\vspace{-0.2cm}
\subsection{Unified Codimensions}

\del{With the diverse range of geometries commonly found in environments the u} Unified simulation of all codimensions in \del{one} \add{a common} framework is critical for simulation efficiency and accuracy. 
Martin et al.\ \shortcite{martin2010unified} focus on a unified elasticity model. They derive elastons -- a higher-order integration rule to measure stretch, shear, bending and twisting along all axes without distinction between codimensions.
Elastons accurately capture a wide range of elastoplastic behaviors, while contact forces are determined by point-wise penalties, with objects represented by a set of spheres.
Chang et al.\ \shortcite{chang2019unified} address unification for mixed-dimensional elastic bodies 
by defining all connections between domains of varying codimension via equality constraints. Here, 
Bridson et al.'s\ \shortcite{bridson2002robust} collision processing algorithm is then applied to resolve contact.
The material point method (MPM)\ \cite{jiang2017anisotropic} also offers general-purpose modeling of all codomains and contact between them. MPM discretizes elasticity on Lagrangian particles while solving momentum balance on the DOFs of an Eulerian grid. Contact between codimensional objects is then directly resolved via the Eulerian grid as a velocity flow. However, sticking and gap errors in contact are well known artifacts in MPM and can be unacceptable if grid resolutions are too low.
\add{These artifacts are investigated and mitigated with Lagrangian DOF by Han et al.\ \shortcite{han2019hybrid}.}

Position-based dynamics (PBD)\ \add{\cite{stam2009nucleus,muller2007position,bender2015position,bender2014position}} also enables seamless, unified coupling of bodies with varying codimensions. Here both constitutive model and contacts are resolved as constraints that are iteratively processed for efficient time-integration at the cost of controllable accuracy as constraint complexity increases\ \cite{Li2020IPC}. 
Extensions of PBD with XPBD\ \cite{macklin2016xpbd}, Projective Dynamics\ \cite{Bouaziz2014Projective}, and its further generalization via ADMM\ \cite{overby2017admm} all similarly offer platforms for co-simulating a diversity of codomains. Recent enhancements now provide improved friction\ \cite{ly20pdf} and increased efficiency\ \cite{daviet2020simple}. 
\del{However, the fundamental trade-off of accuracy and robustness (in the resolution of elasticity and contact) for efficiency in computation remains\ \cite{ly20pdf,li2019decomposed}.}
\add{However, these methods both lack convergence guarantees and apply fixed upper iteration caps so that the fundamental trade-off of accuracy and robustness (in the resolution of elasticity and contact) for efficiency in computation remains\ \cite{ly20pdf,li2019decomposed}. In practice this means that numerical instabilities and explosions can and will be encountered, especially in challenging scenarios, e.g. with large time step size, stiffness, deformation, or velocities, as demonstrated in Li et al.\ \shortcite{li2019decomposed,Li2020IPC}.}
\add{Instead, targeting on gauaranteed convergence and stability, }C-IPC enables the direct simulation of all codimensions simultaneously. Coupling is provided by the interaction between any and all codimensional pairings via accurate, intersection-free contact. 
\section{Formulation} 

\begin{figure}[b]
    \centering
    \includegraphics[width=\linewidth]{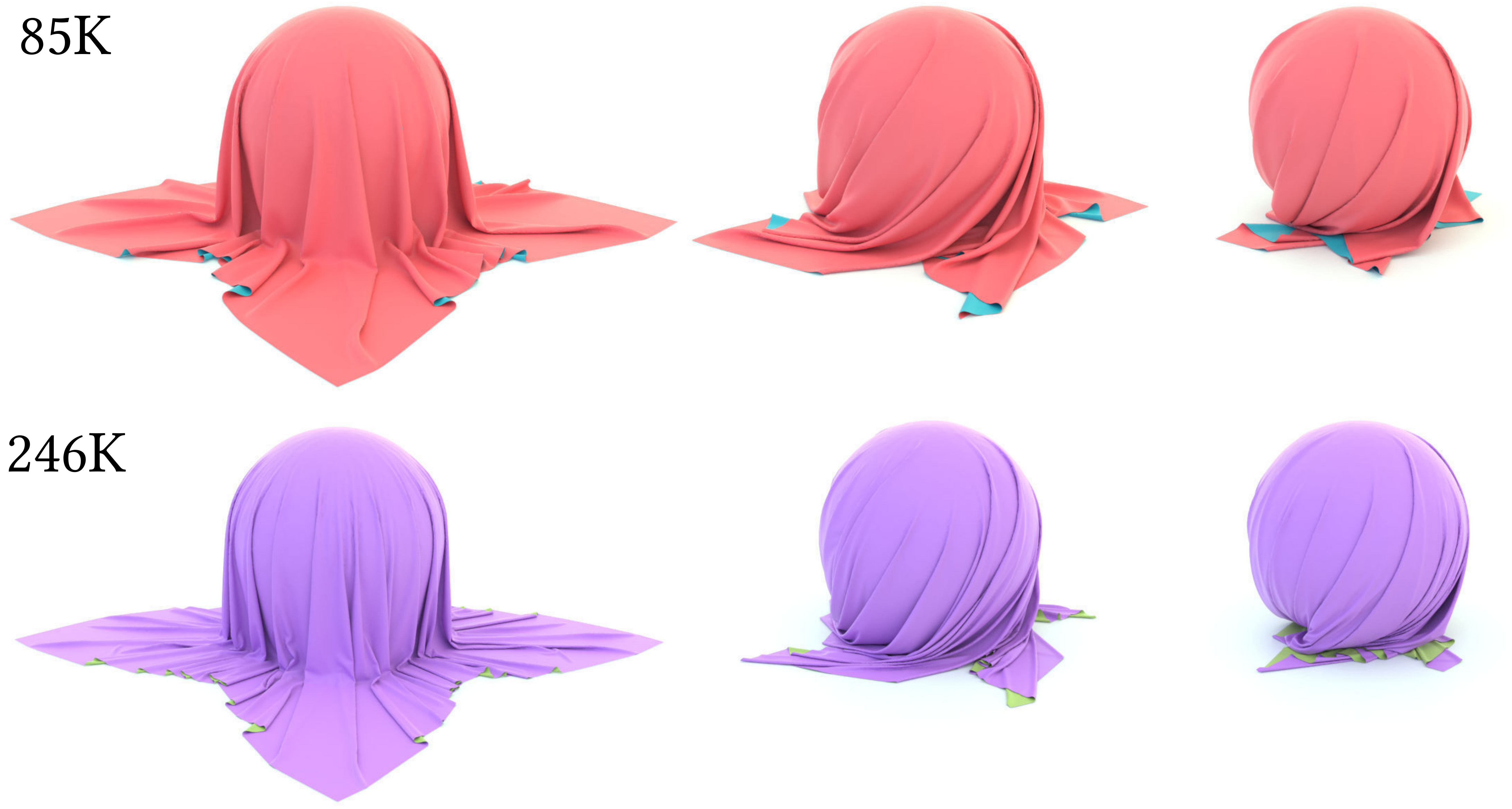}
    \caption{\textbf{Cloth on rotating sphere.} A square cloth (85K-node 1st row, 246K-node 2nd row) is dropped upon a sphere set on the ground with friction ($\mu=0.4$ for both). As the sphere starts to rotate the cloth follows and winds tightly while preserving strain limits throughout. 
    Differing moduli are used for the $85K$- and $246K$-node cloth respectively to produce intricate wrinkling behaviors with differing frequency. Here from left to right we show the 25th (right before rotation), 50th, and 75th frames.}
    \label{fig:clothOnRotSphere}
\end{figure}

We focus on the combined solution of meshed, codimensional models simulated jointly and so coupled arbitrarily via contact. To do so we generalize the IPC model to codimensional DOF and so must address the challenges introduced by thin models both coupled by, and stressed via contact and large, imposed boundary conditions. Here we first cover our generalization of IPC to mixed codimensional models and then, in the following sections, construct the key components of our method that enable their simulation.

\paragraph{Elastodynamics with contact.} For elastodynamics, we perform implicit time stepping with optimization time integration\ \add{\cite{brown2018accurate,Kaufman:2012ex,gast2015optimization,liu2016towards,overby2017admm,li2019decomposed,wang2020hierarchical}} to minimize an Incremental Potential (IP)\ \cite{kane2000variational} with line search, ensuring stability and global convergence\footnote{\add{Global convergence indicates convergence to a local optimum for arbitrary (feasible) initial configurations\ \cite{nocedal2006numerical}.}} .
For a wide range of time steppers and assuming  hyperelasticity, the IP to update from time step $n$ to $n+1$ is defined as
\begin{equation}
    E(x) = \frac{1}{2}||x-\hat{x}^n||^2_M + \alpha h^2 \Psi( \beta x + \gamma x^n)
    \label{eq:basicIP},
\end{equation}
with the timestep update given by $x^{n+1} = \argmin_x E(x).$
Here $h$ is time step size, $M$ is the mass matrix, and $\Psi$ is an elastic energy potential. Scaling factors $\alpha, \beta, \gamma \in [0,1]$ and explicit predictor $\hat{x}^n$ then determine the time step method applied. For example, here we focus primarily on the graphics-standard implicit Euler with $\alpha, \beta = 1$, $\gamma = 0$ and $\hat{x}^n = x^n + h v^n + h^2 M^{-1} f_\text{ext}$. 
Alternate implicit time stepping, e.g., with implicit midpoint or Newmark, are similarly matched by applying alternate scalings and predictors\ \cite{Kaufman:2012ex}.

Incremental Potential Contact (IPC) then augments the IP with contact and dissipative potentials\ \cite{Li2020IPC}:
\begin{equation}
    E(x) = \frac{1}{2}||x-\hat{x}^n||^2_M + \alpha h^2 \Psi(\beta x + \gamma x^n) + B(x) + D(x).
    \label{eq:fullIP}
\end{equation}
Here $B(x)$ and $D(x)$ are respectively the IPC contact barrier and friction potentials. The contact barrier, $B$, enforces strictly positive distances between all primitive pairs while $D$ provides corresponding friction forces. Following Li et al.~\shortcite{Li2020IPC} we apply a custom Newton-type solver to minimize each IP with Continuous Collision Detection (CCD) filtering executed in each line search of every Newton iteration to ensure intersection-free trajectories throughout. Please see Li et al.\ \shortcite{Li2020IPC} for details on the base IPC algorithm and solver implementation.

\paragraph{Mixed-dimensional hyperelasticity.}
To enable a unified simulation framework for coupled volumetric bodies, shells, rods and particles, we compute mass and volume for all codimensional elements by treating them as continuum regions with respect to standard discretizations 
and so construct the total elasticity energy $\Psi(x)$ for the IP as
\begin{equation}
    \Psi(x) = \Psi_\text{vol}(x) + \Psi_\text{shell}(x)
    + \Psi_\text{rod}(x).
    \label{eq:PsiSum}
\end{equation}
Here, without loss of generality, as representative examples, we apply fixed Corotated elasticity~\cite{stomakhin2012energetically} for volumes ($\Psi_\text{vol}$); the Discrete Shells hinge bending energy\ \ \cite{grinspun2003discrete,tamstorf2013discrete} 
combined with either isotropic or orthotropic StVK\ \cite{chen2018physical,clyde2017modeling}, or neo-Hookean membrane models for shells ($\Psi_\text{shell}$); and the Discrete Rods stretch and bending model\ \cite{bergou2008discrete} for rods ($\Psi_\text{rod}$).
We select these models as best for comparison and evaluation given models standard in existing \del{, available} codes. \del{While,} More generally, the C-IPC framework is agnostic to a broad range of elasticity and time stepper choices as demonstrated in Section \ref{sec:exp} and \cite{Li2020IPC}. 
Along with properly integrating all 
 energies with an accurate volume weighting per element, we further parameterize rod bending moduli via Kirchhoff rod theory \add{following Bergou et al.\ \shortcite{bergou2010discrete}} for direct material settings \add{(see our supplemental document for details)}. \del{As, to our knowledge, this is not previously covered in the literature\ \cite{bergou2008discrete}, we report the necessary details in our supplemental document.}
With per-domain elasticity summed in (\ref{eq:PsiSum}) our IPC model now couples objects of arbitrary codimensions directly (without splitting) via frictional contact. Specifically all codimensional DOF are now associated with their respective discrete inertial and potential energies and so are free to move by time stepping. In turn they are coupled by IPC-type barriers. C-IPC barriers now include all point-triangle and edge-edge pairings from all surfaces (both volumes' and shells'), rods and particles; point-edge pairs from all rod nodes and particle-to-rod segments; and finally point-point pairs between all particles.

\section{Constitutive Strain Limiting}
\label{sec:strain_limiting}

Here we begin by constructing a new constitutive barrier model that directly enforces strain limiting while maintaining rest-state consistency. We start with a formulation that augments existing membrane energies with an added strain-limiting potential for the general, isotropic case (Section\ \ref{sec:strain_limiting_iso}). We then demonstrate its extension to directly augment orthotropic STVK membranes with strain limits in Section\ \ref{sec:strain_limiting_aniso}.

\del{Because of its potential-based model, our isotropic (respectively anisotropic) constitutive strain limiting is then easily and directly applied in C-IPC, as just a simple addition to (respectively modification of) the elasticity potential, $\Psi$, in (\ref{eq:fullIP}).
This provides a mesh-based simulation framework that produces trajectories simultaneously ensuring intersection-free and strain-limit satisfying steps. In turn, the resulting dynamics satisfy accurate momentum balance, fully coupling frictional contact forces, strain-limiting and elasticity.}

\subsection{Isotropic Constitutive Strain Limiting}
\label{sec:strain_limiting_iso}

We define isotropic strain-limiting constraints per element $t$ as 
\begin{equation}
    \sigma^t_i < s, \> \> \forall i, 
\end{equation} 
with singular value decomposition of each triangle $t$'s deformation gradient, $F^t = U^t \Sigma^t {V^t}^T$, giving $\Sigma^t = \text{diag}(\sigma^t_1, \sigma^t_2)$. The imposed constraint bound, $s$, is then the requested strain limit with practical bounds generally selected for cloth with $s \in [1.01, 1.1]$\  \cite{provot1995deformation,bridson2002robust}.

\del{Inspired by IPC's contact barrier formulation, we then construct a local energy that enforces this unilateral strain-limit. 
We could begin with a typical log-barrier strategy. This would give energies
\begin{equation}
\kappa_s\sum_{i} -\ln(s-\sigma^t_i),
\label{eq:basicSLBarrier}
\end{equation}
with $\kappa_s$ defining the barrier stiffness for limit $s$.
However, while ensuring strain-limits this barrier is problematic. First, from the modeling perspective, it applies forcing everywhere. Specifically, constitutive behavior is unnecessarily changed far from strain limits where the underlying membrane model should be untouched and, worse yet, applies ghost forces at rest. Second, from the computational perspective, because this barrier is globally non-zero, evaluating it (and its derivatives) in optimization is prohibitively expensive.}

\del{From these issues we then observe that strain-limit forcing for triangles far from the strain limit is then both unacceptable and unnecessary. Instead, we want local support for nonzero forces just close to where these limits are achieved.
One simple and seemingly plausible remedy, to achieve this localized support region, is to clamp the barrier in (\ref{eq:basicSLBarrier}) to zero beyond a small strain threshold $\hat{s}$ (e.g. $\hat{s}=1$) giving barrier energies
\begin{equation}
    \kappa_s \sum_{i} b^{st}_i(x), \quad
   b^{st}_i(x) = \begin{cases}
        -\ln(\frac{s-\sigma^t_i}{s-\hat{s}}) &\text{$\sigma^t_i>\hat{s}$} \\
        $0$ &\text{$\sigma^t_i\leq\hat{s}$}
    \end{cases}.
\end{equation}
Unfortunately, while providing local support these energies are only $C^0$ continuous and so do not allow for practical gradient-based optimization necessary for the efficient solution of the IP.}

\begin{wrapfigure}[8]{l}{0.4\linewidth}
\includegraphics[width=\linewidth]{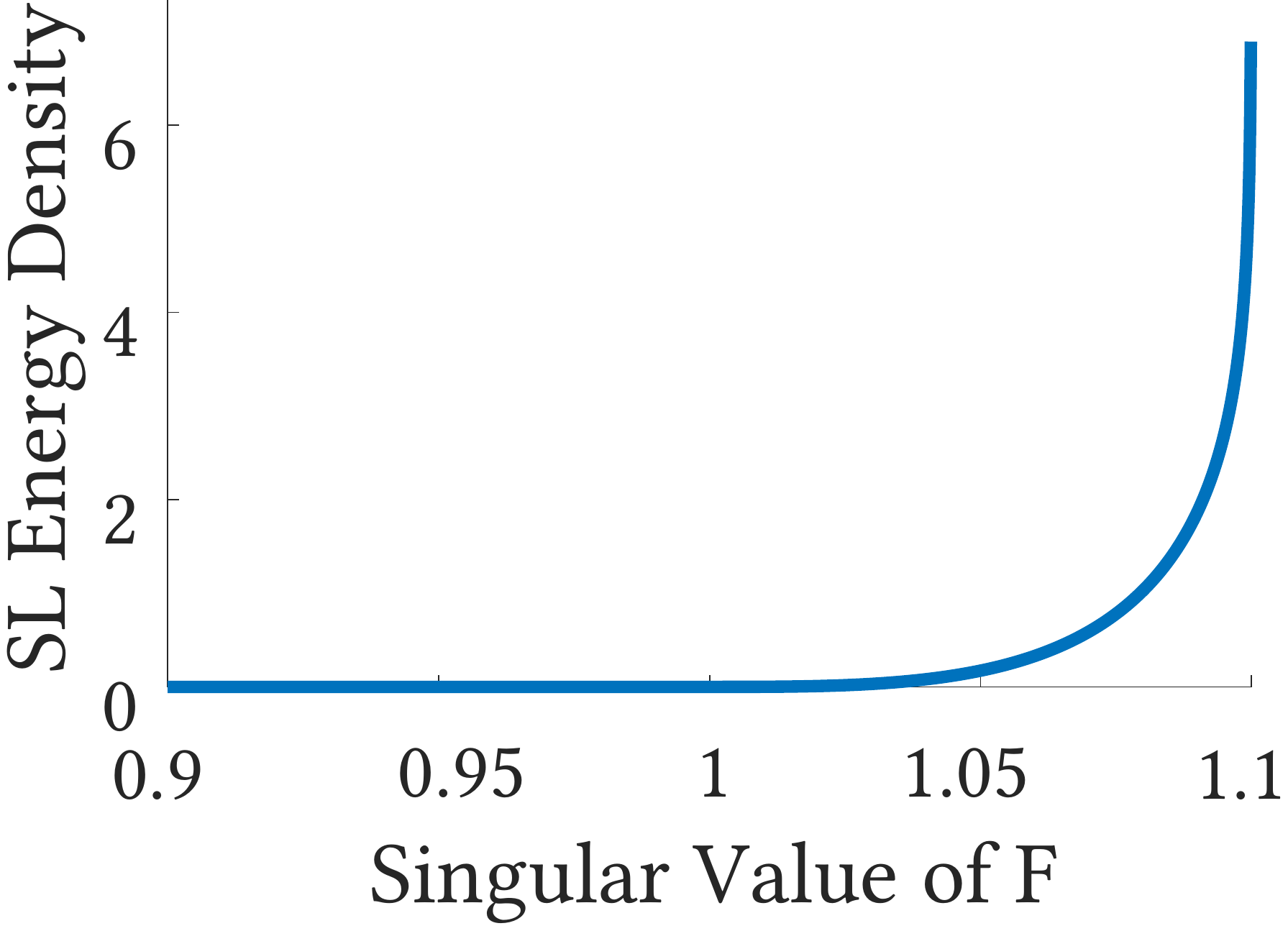}
\end{wrapfigure}
\del{To smooth our strain-limiting while providing local support we then propose a $C^2$ smoothly clamped barrier for strain-limiting}
\add{Strain-limiting requires local support. It should only exert constraint forcing during stretch when strain is close to an applied upper bound. Otherwise, it should leave an underlying membrane model unchanged. This is analogous to the application of contact forces which should only exert when boundaries are nearly touching. As such, we begin with IPC's contact barrier to construct a comparable $C^2$, smoothly clamped barrier for strain limiting. Each barrier per $\sigma^t_i$ is then}
\begin{equation}
    b^{st}_i(x) = \begin{cases}
        -(\frac{\hat{s}-\sigma^t_i}{s-\hat{s}})^2 \ln(\frac{s-\sigma^t_i}{s-\hat{s}}) &\text{$\sigma^t_i>\hat{s}$} \\
        $0$ &\text{$\sigma^t_i\leq\hat{s}$}
    \end{cases},
\end{equation}
\add{and so is only activated when strain exceeds an imposed, small-strain threshold $\hat{s}$. E.g., see inset for the barrier energy with strain limit $s=1.1$ and threshold $\hat{s}=1$.}

Now, with barriers in hand, we could potentially next consider treating them \del{these barriers} directly as constraints (as in primal barrier and interior point methods). However, doing so would then simply sum these barriers over elements and so would obtain inconsistent behavior as we change meshes. Instead, to provide consistent behavior we impose strain limiting constitutively as an energy density integrated over the volume of cloth to obtain the potential
\begin{equation}
    \begin{aligned}
        \Psi_\text{SL}(x) & = \sum_i \kappa_s \int_\Omega b^{st}_i(x) dV 
         \approx \kappa_s \sum_{t,i} V^t b^{st}_i(x),
    \end{aligned}
\end{equation}
\add{where $\kappa_s$ is the barrier stiffness in $Pa$.}
For our approximation, we use $V^t = A^t \xi^t$, with $A^t$ and $\xi^t$ respectively the area and thickness of triangle $t$. Our final isotropic strain-limiting potential, $\Psi_\text{SL}$, is then $C^2$ with local support and can simply be added to our total potential in Equation\ \ref{eq:PsiSum}. In turn, strain limiting is then directly handled at each time step by optimization of the IPC in Equation\ \ref{eq:fullIP}, while ensuring that strain-limit barriers are not violated in each line-search step. During steps when most triangles remain under the strain limit threshold, little additional computation is then required. While, when stretch increases, \del{we can adjust} the necessary nonzero terms from newly activated barriers \add{are added to the system} and so, as we show in Section \ref{sec:exp}, strictly enforce strain limits while balancing all applied forces.

\begin{figure}[t]
    \centering
    \includegraphics[width=\linewidth]{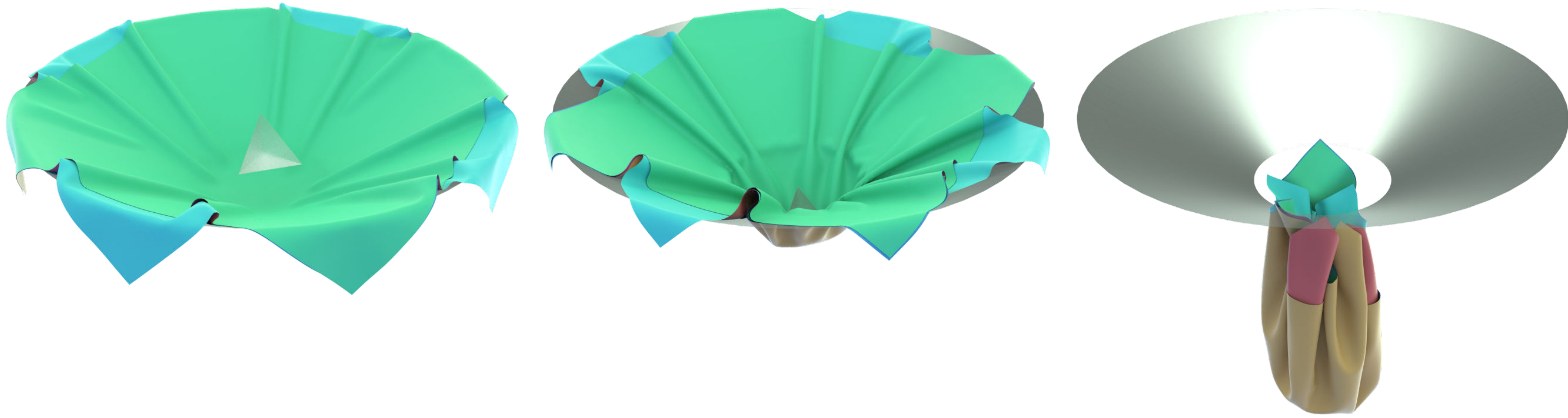}
    \caption{\textbf{Funnel.} Three cloth panels ($\mu=0.4$, 26K nodes each, strain upper bound at $1.0608$) are dropped on a funnel. Under friction they rest stably on top until a sharp, scripted collision object rapidly pulls the panels down.}
    \label{fig:funnel}
    \vspace{-0.4cm}
\end{figure}

\paragraph{Strain-Limit Stiffness}
With our barrier potential defined we now specify setting its stiffness.
The mollified \add{clamped barrier} \del{clamping strategy} we apply for our strain-limit model is inspired by IPC's smoothing of contact barrier energies\ \cite{Li2020IPC}. 
Here, however, notice an extra ingredient for strain-limiting is the applied $1/(s-\hat{s})^2$ factor. This scaling enables us to apply our barrier directly to a limit-normalized strain, measured by $y = (\hat{s}-\sigma^t_i)/(s-\hat{s})$ with 
\begin{equation}
    \hat{b}^{st}_i(y) = \begin{cases}
        -y^2 \ln(1+y) &\text{$y < 0$}, \\
        $0$ &\text{$ y \geq 0$}.
    \end{cases}
\end{equation}
In turn, as our applied strain-limit ($s$) and/or clamping threshold ($\hat{s}$) is varied with application, the barrier function w.r.t. $y$ remains unchanged. Only the gradient of the linear map from $\sigma^t_i$ to $y$ varies.  
This allows us to apply a single, consistent initial barrier stiffness $\kappa_s$ (we use $1KPa$ for all examples) across all choices of differing strain limits and clamping thresholds. Here the barrier potential curve is then simply linearly rescaled each time, to a different strain range, and so provides consistent conditioning to the system. 
Finally, to avoid numerical issues introduced by tiny gaps between a current strain and the imposed strain limit (e.g., when extreme boundary conditions are imposed as in Figures\ \ref{fig:needlebedpull} and \ref{fig:teaser}) we adapt barrier stiffness when needed \add{following IPC's barrier stiffness adjustment strategy}. Starting with our initial $\kappa_s$, we increase it by $2 \times$ (up to max bound of $\kappa_s = 0.1MPa$) whenever the strain gap $s - \sigma^t_i$ of a triangles $t$ is less than $10^{-4}(s-\hat{s})$ over two consecutive iterations.
\add{Notice that, when applied, this adjustment varies the strain-limit energy. This occurs, however, solely near the actual limit in extreme cases. Here strain-limit forces impose constraint while this adaptivity provides improved numerical conditioning. Once away from the limit the barrier returns to a consistent energy with no forcing at all below the threshold $\hat{s}$.}

\subsection{Anisotropic Constitutive Strain Limiting}
\label{sec:strain_limiting_aniso}

Above we have constructed strain-limiting as an isotropic constitutive model. We formed a barrier energy that can be integrated and so directly added to potential energy in order to augment existing membrane models with hard strain limits. The key to making this work is the application of our $C^2$ continuous clamping so that the application of strain-limits does not alter rest-shape gradients nor rest-shape Hessians. 

Alternately, we can apply a comparable constitutive strain-limiting strategy to directly modify membrane elasticity models to include strain limits. To do so we simply substitute an anisotropic membrane model with a barrier energy that prevents violation of strain limits, while matching the original membrane energy gradient and Hessian at rest. 

This second strategy is particularly motivated by the need to incorporate strain-limiting into available data-driven models that have been constructed to carefully fit measured cloth data. Here we demonstrate specific application to the anisotropic, data-driven model of Clyde and colleagues\cite{clyde2017modeling,clyde2017numerical}. 

Their constitutive model energy is
\begin{equation}
\begin{aligned}
\tilde{\psi}(\tilde{E}_{11}, \tilde{E}_{12}, \tilde{E}_{22})  = &
\frac{a_{11}}{2} \eta_1(\tilde{E}_{11}^2) +
\frac{a_{22}}{2} \eta_3(\tilde{E}_{22}^2) \\
&+  a_{12} \eta_2(\tilde{E}_{11} \tilde{E}_{22}) +
G_{12} \eta_4(\tilde{E}_{12}^2),
\end{aligned}
\end{equation}
where $\tilde{E} = D^TED$ is the reduced and aligned Green-Lagrangian strain, $\tilde{E}^{i3} = \tilde{E}^{3i} = 0$, and column vectors of $D$ are the tangent and normal bases of the shell in material space. Functions $\eta$ are then a sum of polynomial functions with real-valued exponents $\alpha_{ji}$ that satisfy $\eta(0)=0$ and $\eta'(0)=1$. Here the first constraint enforces zero-energy, zero-stress rest configurations, while the second constraint allows a natural correspondence between the parameters $a_{\alpha\beta}$ and $G_{12}$ and linear elasticity at infinitesimal strain. Clyde et al. use
\[ \eta_j(x) = \sum_{i=1}^{d_j} \frac{\mu_{ji}}{\alpha_{ji}} ((x+1)^{\alpha_{ji}} - 1). \]
Their measured data is then restricted to deformations within the fracture limit or elasticity range of the cloth. Beyond this range meaningful extrapolation is then unlikely, due to possible overfitting inside the range (exponents $\alpha_{ji}$ from the fitting can range up to $10^5$). 
To address this limitation Clyde et al. propose a quadratic extrapolation for simulation. However such extrapolation is not physically meaningful. To usefully apply data-driven modeling either fracture should be captured beyond this regime or else a stable and controllable strain limit imposed to stay in bounds.
Here, we focus on controllably imposing strain limit while respecting the underlying model fitting.

\paragraph{Barrier Formulation}

Starting from the same constitutive model we redefine the $\eta$ basis with barriers 
\[\eta(E)=-E^{max}\log((E^{max} - E)/E^{max}).\]
For consistency note this basis again satisfies both $\eta(0)=0$ and $\eta'(0)=1$ and so is valid for elasticity. This barrier also diverges at $E^{max}$ and so ensures that $E$ never exceeds measured strain limit bounds. It thus captures data-driven strain limits and at the same time avoids extrapolation. Here, as the underlying model is anisotropic, warp, weft, and shearing directions can all impose different, measured $E^{max}$ bounds, enabling preservation of measured anisotropy for all strain limits.

Free model parameters $a_{\alpha\beta}$ and $G_{12}$, are then matched at $\nabla^2 \psi(0,0,0)$ with the underlying model's fit; see our supplemental document for details. Note this formulation frees us from computing the sensitive (and potentially expensive) polynomials with large real-valued exponents. See Section\ \ref{sec:anisoSLParam} for experiments on the proposed model. Finally we note that, in contrast to the strategy we demonstrated in our isotropic case from the last section (where we worked directly on an upper bound w.r.t. the deformation gradient), here this second energy is symmetric for stretch and compression. 

\subsection{Line-Search Filtering for Strain Limits} As discussed above, CCD-based filtering is critical to ensure line-search is consistent with our contact barriers. Now that we add strain-limiting barriers we must additionally ensure that every line search will also not violate imposed strain limits. For isotropic limits, although $2\times 2$ SVDs have a closed form solution, it does not provide a polynomial \add{equation} for feasible step size computation. Thus, in C-IPC line search, after we compute an intersection-free \add{starting} step size, we simply half the step length if we detect a strain-limit violation during backtracking. We repeat until we obtain a step size satisfying energy decrease and strain limits. In practice, because our Newton step includes higher-order information of our strain-limits from Hessian and gradients, we observe IPC search directions are effective at naturally avoiding strain-limits. In all cases, when required, we so far observe at most three \del{bisections} \add{backtracking steps} for strain-limiting are required. For our anisotropic model we also currently apply the same \del{bisection}\add{backtracking} strategy. However, while currently not needed for efficiency, we do note that here 
we can formulate quadratic equations amenable to directly computing largest feasible strain-limit satisfying step sizes. 

\section{Modeling Thickness}
\label{sec:thickness}

In its original form IPC maintains intersection-free paths for volumetric models by enforcing positivity of unsigned distances $d_k$, as an invariant between all non-adjacent and non-incident surface primitive pairs $k$\ \cite{Li2020IPC}. This is suitable for volumetric contact where this constraint permits arbitrarily close, but never intersecting surfaces. For codimensional models, however, this constraint is no longer sufficient. 
When the 3D deformation of thin materials is reduced to the deformation of a 2D surface or 1D curve, elasticity can be well-resolved on surfaces and curves, but contact can not. Neglecting to account for finite thickness in codimensional contact generates unacceptable artifacts (see e.g. Figure\ \ref{fig:twistCylinder}) and clearly fails to capture geometries formed by thin-structure interactions (see e.g. Figures\ \ref{fig:ballOnClothStack} and \ref{fig:cardShuffle}).

We begin by observing that applying a larger $\hat{d}$ \add{(the threshold distance at which IPC contact force application begins)}, to codimensional geometries in IPC, models a thin responsive layer that resists compression in the normal direction and so forms an elastic thickness. In concert with this elastic layer we also generally require a core thickness beyond which further compression is not \del{possible} \add{allowed}. This is needed to guarantee a minimum finite (and so e.g. visible) thickness even when deformation is extreme (see e.g. Figure\ \ref{fig:twistCylinder} \add{bottom}).

To model thickness in contact for codimensional objects we build an inelastic thickness model that combines the purely elastic layer with a hard, and so inelastic offset to the contact barrier. 
For a nonzero offset, $\xi$, geometries are then guaranteed to be separated from each other by $\xi$. Elastic contact forces exert when distances are below $\xi+\hat{d}$ and diverge when distances are at $\xi$. Here larger $\hat{d}$ generates greater elastic response, while nonzero $\xi$ guarantees a minimum thickness. This holds even under extreme compression \del{(see e.g. Figure\ \ref{fig:twistCylinder}e)}.

We equip each surface element $i$ with a finite thickness $\xi_i$. The distance constraint for primitive pairs $k$ on the surface, formed between element primitives $i$ and $j$, are then  
\[d_k(x)>\xi_k=\frac{\xi_i + \xi_j}{2}.\]
\setlength{\columnsep}{5 pt}
\setlength{\intextsep}{5 pt}%
\begin{wrapfigure}[6]{r}{0.3\linewidth}
\vspace{-4 pt}
\includegraphics[width=\linewidth]{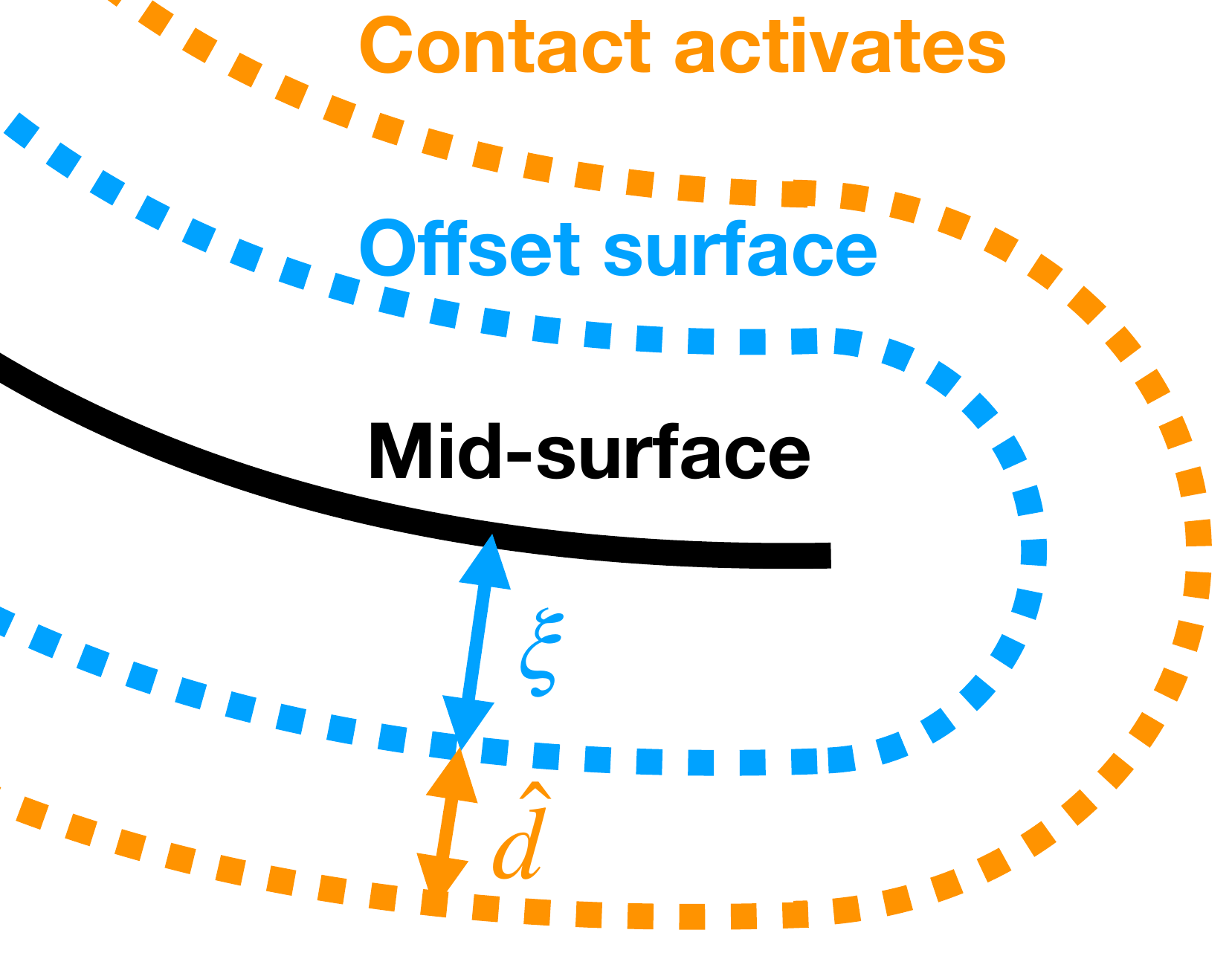}
\end{wrapfigure} 
Boundaries of codimensional materials are then approximated with a rounded cross-section, while for interaction between zero-thickness materials our distance constraints reduce to that of the original IPC constraint (e.g. for volume-to-volume contact). Finally for volume-to-codimensional contact, volumes can continue to maintain zero-thickness boundaries while interacting with finite thickness codimensional boundaries.

For scenarios where modeling thickness matters (see e.g. Figures\ \ref{fig:twistCylinder}, \ref{fig:hairs}-\ref{fig:allIn}), 
we then set $\xi_k$ to the true thicknesses of the material(s) or slightly smaller, and then set $\hat{d}$ near $\xi_k$ to compensate for the remaining thickness depending on how much compression, and so elastic response, is reasonable or required per application.

\begin{figure}[b]
    \centering
    \includegraphics[width=\linewidth]{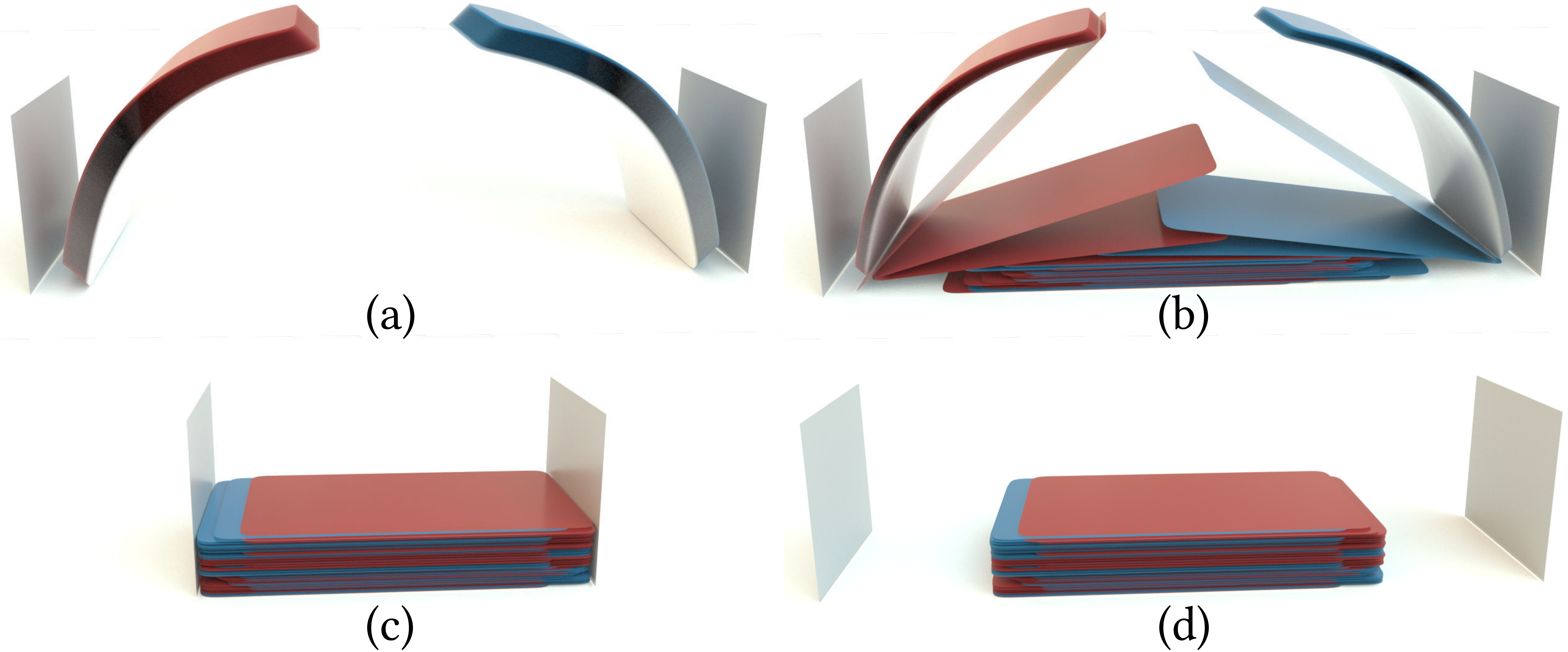}
    \caption{\textbf{``Precision'' card shuffle.} We divide fifty-four playing cards into two separate piles and and bend them in preparation for a ``precision'' bridge shuffle. Unlike a human-performed bridge finish, where cards begin interleaved, here we increase challenge by keeping the two bridged piles apart and then (a) precisely, one-by-one, bottom up they are rapidly released in turn to shuffle the deck (b). We partially square the deck in (c), obtaining a fully shuffled stack with non-intersecting, interleaved cards that is 15.4 mm heigh and so well-matching the height of a standard deck (d).}
    \label{fig:cardShuffle}
\end{figure}

\subsection{Solving IPC with Thickened Boundaries}
For a numerically robust and efficient implementation, IPC applies squared distances (with appropriate rescalings) to compute with equivalent barrier functions.
This change swaps the model's derived contact barrier $b(d_k(x), \hat{d})$ to an equivalent rescaled impl\add{e}mentation using $b(d^2_k(x), \hat{d}^2)$\ \cite{Li2020IPC}.
Here we directly apply our thickened contact barrier via squared distances 
\[b_\xi(d_k(x), \hat{d}) = b(d^2_k(x)-\xi_k^2, 2\xi_k\hat{d}+\hat{d}^2),\]
so that contact forces correctly diverge at $d_k(x)=\xi_k$, and nonzero contact forces are only applied between mid-surface pairs closer than $\hat{d}+\xi_k$.
\add{Recall that $b(d^2, \hat{d}^2) = -(d^2 - \hat{d}^2)^2 \ln(d^2/\hat{d}^2)$ if $d^2 < \hat{d}^2$, and is $0$ otherwise\ \cite{Li2020IPC}.}

Here then the barrier function, gradient, and Hessian only need to be evaluated with the modified input distance, while distance gradient and Hessian remain unchanged as $$\frac{\mathbf{d} (d^2_k(x)-\xi_k^2)}{\mathbf{d}x}=\frac{\mathbf{d} d^2_k(x)}{\mathbf{d}x}.$$ 
Evaluation of contact force magnitudes for computing friction forces as well as adaptive stiffness ($\kappa$) updates for contact barriers follow similarly.

Most constraint set computations then require simple and comparable modifications. For spatial hashing, bounding boxes of all mid-surface primitive $i$ are extended by $\xi_i/2$ in all dimensions on left-bottom and top-right corners before locating voxels. This enables hash queries with $\hat{d}$ to remain unchanged. Next, for broad-phase contact pair detection, query distances $\hat{d}$ are now updated to $\hat{d}+\xi_k$ to check for bounding box overlap. Or else, equivalently, bounding box primitives are extended as in the spatial hash. Finally, to accelerate continuous collision detection (CCD) queries, spatial hash construction also requires similar bounding box extension, while for its broad phase, applied gaps are $\xi_k$ (rather than $0$).

\subsection{Challenges for CCD}

While the above initial modifications for thickness are straightforward, finite thickness and codimensional DOF introduce new computational challenges for CCD queries. Here we analyze these challenges and develop a new CCD method to address them.

\begin{figure}[t]
    \centering
    \includegraphics[width=\linewidth]{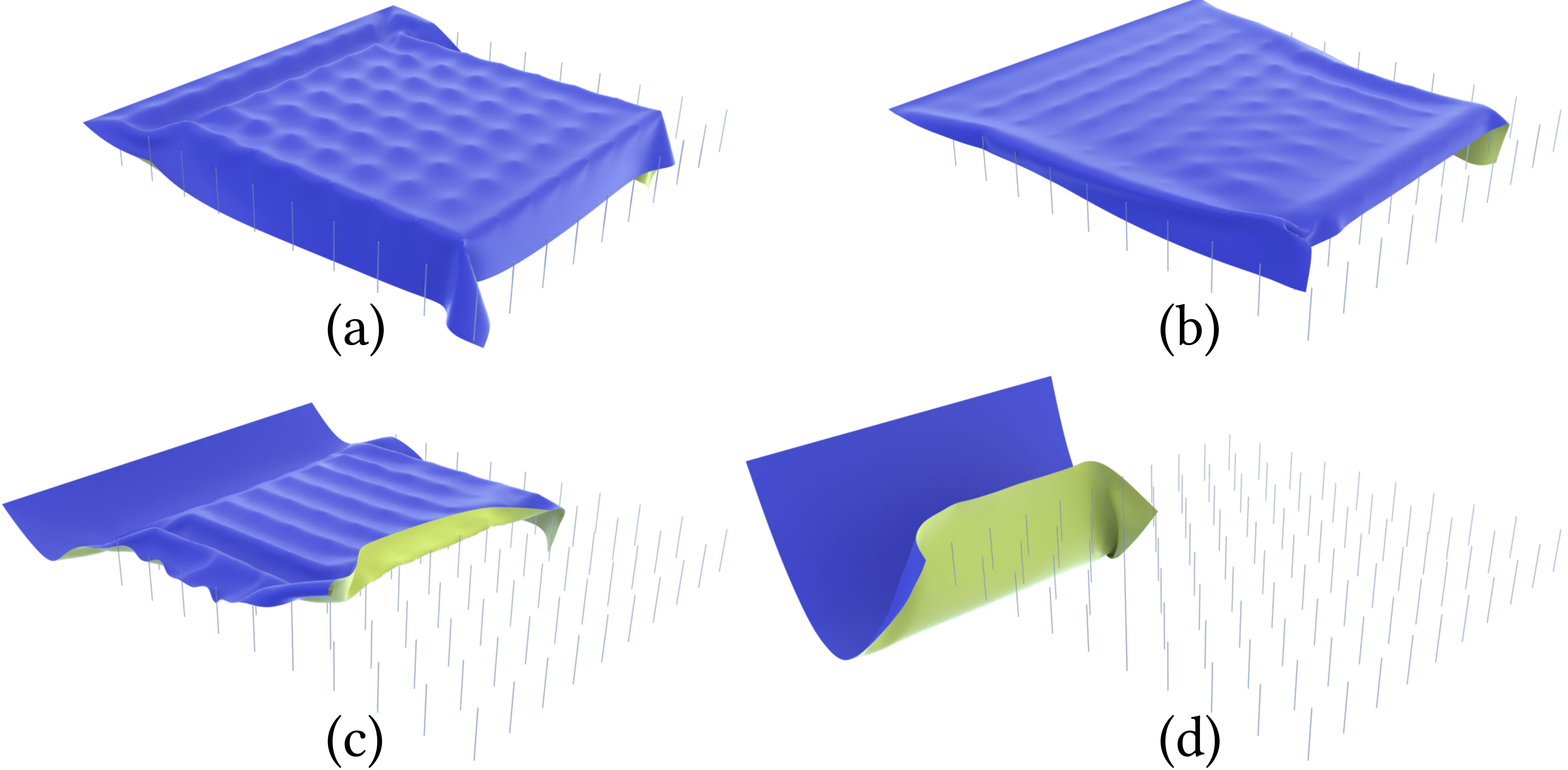}
    \caption{\textbf{Cloth over codimensional needles.} 
    To test contact across sharp boundaries we drop and then drag a cloth panel across a needle bed formed by line segments. With a large time step size of $h=0.02s$ and strain limit of $1.0608$, the cloth comes safely to a static rest on the segments without jittering (a). We then pull the left side of the cloth at $1m/s$ across the segment tips (b) and observe rapid sliding along (c), and then off (d) the top of the bed without snagging nor stretching artifacts (strain limits are satisfied throughout).}
    \label{fig:needlebedpull}
\end{figure}

Standard-form IPC applies position updates in each iteration to obtain minimal separation distances of $s \times$ the current separation distances, $d_k^{cur}$, at time of CCD evaluations. Here $s \in (0,1)$ is a conservative rescaling factor (generally set to $s=0.2$ or $0.1$) that allows CCD queries to avoid intersection, even when surfaces are very close\ \cite{Li2020IPC}. To similarly resolve finite thicknesses, our conservative distance bound is now $s (d^{cur}_k-\xi_k)+\xi_k$.

Concretely, the barrier evaluated parameters, $d^2_k(x)-\xi_k^2$, must always remain positive. In turn we require all CCD evaluations, for each displacement $p$, to provide us with sufficiently accurate times $t \in (0,1]$ to satisfy positivity. If impact would occur for a pair along $p$, we require a time $t$ so that the scaled displacement, $t p$, ensures distance remains greater than $\xi_k$ and is as close as possible to the target separation distance $\xi_k + s(d_k^{cur}-\xi_k)$.

Doing this with finite thickness is then much more challenging than without thickness. In contact $d_{cur}-\xi_k$ can be as small as $10^{-8}m$ (e.g. under large compression) so that for $s \geq 0.1$ absolute errors in the order of $10^{-9}m$ are acceptable. While, in practice, $\xi_k$ is then regularly at the scale of $10^{-4}m$ (e.g. thickness values of $\sim 3\times10^{-4}m$ for cloth). Without thickness ($\xi_k=0$) updated distances are targeting $s d^{cur}_k$ (and are only required to be strictly greater than $0$). In other words any step size that prevents primitive-pair intersection is valid and so relative errors less than $100\% \approx 10^{-9}m/(s 10^{-8} m)$ are acceptable. On the other hand, with thickness, updated distances are targeting $\xi_k + s(d_k^{cur}-\xi_k)$ and so, for standard values of $\xi_k$, would require relative errors from the CCD evaluations near $0.01\% \approx 10^{-9}m/(10^{-4}m + s(10^{-8}m))$ in order to avoid interpenetration.

\begin{figure*}[t]
    \centering
    \includegraphics[width=\linewidth]{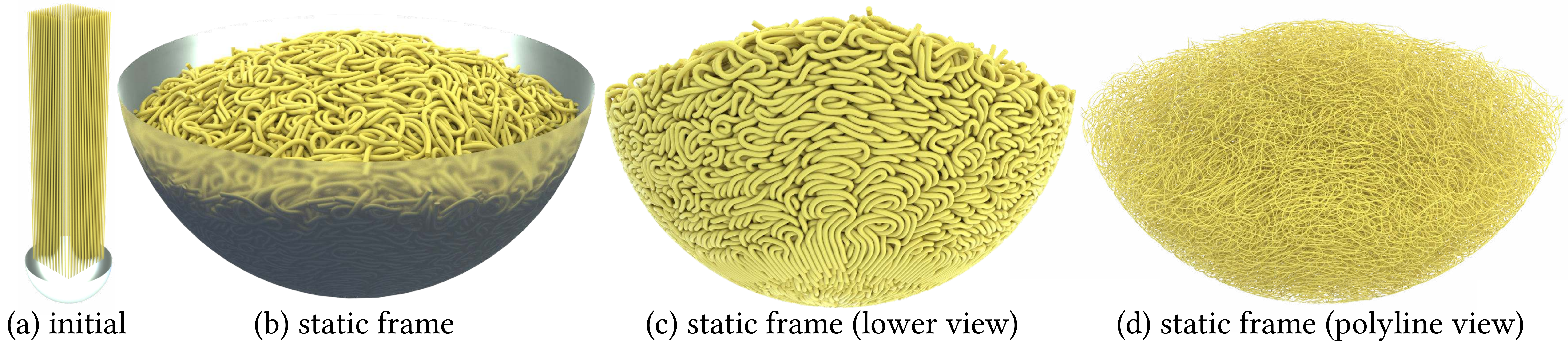}
    \caption{\textbf{Noodles.} (a) We drop 625 $40cm$-long noodles, modeled by discrete rods with C-IPC offset, into a bowl. 
    (b) We set offset and $\hat{d}$ to model noodle thickness enabling the bowl to fill when they reach  equilibrium. (c) We then remove the bowl to show the final intertwined, intersection-free configuration, noting that thickness offsets are satisfied throughout. In (d) we then remove the volume render to highlight the geometry is simulated solely with discrete rods polyline geometry.}
    \label{fig:noodles}
    \vspace{-0.3cm}
\end{figure*}

Obtaining CCD evaluations to this accuracy is then extremely challenging for available methods. As a starting example we tested the floating point CCD solver from Li et al.\ \shortcite{Li2020IPC}, requesting a $s(d_{cur}-\xi_k)+\xi_k$ minimal separation for two challenging codimensional examples with thickness (see Figures\ \ref{fig:noodles} and \ref{fig:sprinkles}) with $\xi_k>0$. Here the CCD solver returns $t=0$ time of impact (TOI) erroneously even if we remove the conservative scaling factor (i.e. set $s=0$) altogether (See Figure\ \ref{fig:CCDComp}). Note that in doing so this error stops simulation progress altogether.

\subsection{CCD Lower Bounding}
\label{sec:lower-bound}
Next we derive a useful lower bound value for CCD queries that is numerically robust and can be efficiently evaluated in floating point.
This lower bound provides a conservative, guaranteed safe step size and also a clear measure to test a CCD query's validity: any CCD-type evaluation with smaller TOI has clearly failed.
In Section\ \ref{sec:accd} we then apply this bound to derive a simple, effective and numerically accurate explicit CCD solver that is robust and efficient for progress even in the challenging CCD evaluations we require for thin material simulations.

Here, without loss of generality, we will focus on the edge-edge case between edge pairs ($x_0,x_1$) and ($x_2,x_3$) with corresponding displacements $p_0,p_1,p_2$ and $p_3$. 
The distance function between arbitrary points parameterized respectively by $\gamma$ and $\beta$ on each edge at any time $t$ is then
\begin{align}
\begin{split}
	f(t, \gamma, \beta) &=||d(t, \gamma, \beta)||, \text{with} \\
	d(t, \gamma, \beta)&=(1-\gamma)x_0(t) + \gamma x_1(t) - ((1-\beta)x_2(t) + \beta x_3(t)),\\
	x_i(t) &= x_i(0) + t p_i, \text{and} \ t,\gamma,\beta \in [0, 1].
	\end{split}
\end{align}
A CCD evaluation then seeks the smallest positive real $t$ satisfying
\[ f_1(t) = \min_{\gamma,\beta}f(t,\gamma,\beta) = 0. \]
If such $t$ exists we call it $t_a$. Parameters $(\gamma_a,\beta_a) = \argmin_{\gamma,\beta}f(t_a,\gamma,\beta)$ in turn give the corresponding points colliding at time $t_a$ on each both edges.

We can then express $t_a$ as
\[ t_a =\frac{f(0,\gamma_a,\beta_a)}{||(1-\gamma_a)p_0 + \gamma_a p_1 - ((1-\beta_a)p_2 + \beta_a p_3)||}. \]
Challenges to CCD evaluations then occur in determining $\gamma_a$ and $\beta_a$, when degeneracies and numerical error make floating-point methods both prone to false positives and false negatives.

We can however, directly and accurately perform a distance query w.r.t. the primitives at start (t=0) of evaluation, to find the distance, $f_1(0)$, that certainly satisfies $f_1(0) \leq f(0,\gamma_a,\beta_a)$. 
Then triangle inequality gives
$ ||(1-\gamma_a)p_0 + \gamma_a p_1 - ((1-\beta_a)p_2 + \beta_a p_3)|| \leq ||(1-\gamma_a)p_0 + \gamma_a p_1|| + ||(1-\beta_a)p_2 +\beta_a p_3|| $,
and since $\gamma_a,\beta_a \in [0, 1]$ we get
$ ||(1-\gamma_a)p_0 + \gamma_a p_1|| \leq \max(||p_0||, ||p_1||) $
and
$ ||(1-\beta_a)p_2 +\beta_a p_3|| \leq \max(||p_2||, ||p_3||) $.
Put together we then have the practical and directly computable bound on TOI
\begin{equation}
    t_a \geq \frac{f_1(0)}{\max(||p_0||, ||p_1||) + \max(||p_2||, ||p_3||)}.
    \label{eq:CCD_LB}
\end{equation}
More generally, for any query between any two primitives types, our bound on $t_a$ is correspondingly given by the appropriate distance function in the numerator (e.g., $f_1(0) = min_{\gamma_a,\beta_a, \alpha_a} f(0,\gamma_a,\beta_a, \alpha_a)$ for point-triangle and $f_1(0) = min_{\gamma_a} f(0,\gamma_a)$ for point-edge) and the sum of the max displacement from each of the two primitives in the denominator. 
We then note that even when there is no smallest positive time satisfying $f_1(t) = 0$ (and so no impact on the interval), our bound remains well-defined as a conservative step size.

We next observe that, perhaps surprisingly, state-of-the-art floating-point CCD solvers can and will return TOI results smaller than our lower bound, and so are clearly in error (See Figure\ \ref{fig:CCDComp}).

Finally, to improve our bound, we observe that it holds independently of the choice of the reference frame. Thus we can further tighten the bound on each CCD query independently by picking frames that reduce the norm of displacement vectors $p_i$. For example by subtracting each $p_i$ by the average $\frac{1}{4}\sum_i p_i$.

\subsection{Additive CCD}
\label{sec:accd}
\del{We now apply our lower bound to build a new CCD algorithm}
\add{With our computed lower bound we now apply a CA strategy\ \cite{mirtich1996impulse,zhang2006interactive} to build a new CCD algorithm with finite offsets for deformable bodies}
that iteratively updates and adds our lower bound over successive conservative steps towards TOI. The resulting \emph{additive} CCD (ACCD) method then robustly solves for a bounded TOI, monotonically approaching each CCD solution, while only requiring explicit calls to evaluate distances between updated primitive positions.

\begin{algorithm}[ht]
\caption{Additive CCD}
\label{alg:LBCCD}
\begin{algorithmic}[1]
    \Procedure{AdditiveCCD}{$x_i$, $p_i$, $\xi$, $s$, $t_c$, $t$}

    \State $\bar{p} \leftarrow \sum_i p_i / 4$
    \For{$i$ in $\{0,1,2,3\}$}
        \State $p_i \leftarrow p_i - \bar{p}$
    \EndFor
    \vspace{0.15cm}
    \State $l_p \leftarrow \max_{i \in \text{1st primitive}} (||p_i||) + \max_{i \in \text{2nd primitive}} (||p_i||)$
    \If{$l_p = 0$}
        \State \textbf{return} false
    \EndIf
    \vspace{0.15cm}

    \State $d_{sqr} \leftarrow \text{computeSquaredDistance}(x_0, x_1, x_2, x_3)$
    \State $g \leftarrow s (d_{sqr} - \xi^2) / (\sqrt{d_{sqr}} + \xi) $
    \vspace{0.15cm}

    \State $t \leftarrow 0$
    \State $t_l \leftarrow (1 - s) (d_{sqr} - \xi^2) / ((\sqrt{d_{sqr}} + \xi) l_p)$
    \While{true}
        \For{$i$ in $\{0,1,2,3\}$}
            \State $x_i \leftarrow x_i + t_l p_i$
        \EndFor
        \State $d_{sqr} \leftarrow \text{computeSquaredDistance}(x_0, x_1, x_2, x_3)$
        \If{$t > 0$ and $(d_{sqr} - \xi^2) / (\sqrt{d_{sqr}} + \xi) < g$}
            \State \textbf{break}
        \EndIf
        \vspace{0.15cm}

        \State $t \leftarrow t + t_l$
        \If{$t > t_c$}
            \State \textbf{return} false
        \EndIf
        \vspace{0.15cm}

        \State $t_l \leftarrow 0.9 (d_{sqr} - \xi^2) / ((\sqrt{d_{sqr}} + \xi) l_p)$
    \EndWhile
    \vspace{0.15cm}

    \State \textbf{return} true

    \EndProcedure
\end{algorithmic}
\end{algorithm}

At the start of each CCD query, to initialize the ACCD algorithm (see Algorithm\ \ref{alg:LBCCD}), we first center the collision stencil's displacement at origin to reduce our bound's denominator, e.g.,  $l_p = \max(||p_0||, ||p_1||) + \max(||p_2||, ||p_3||)$ for edge-edge pairs, and so increase the step size we can safely take. 
If there is no relative motion ($l_p=0$), we of course simply return no collision and so a full unit step is valid.
We then compute the requested minimal separation $g = s (\sqrt{d_{sqr}} - \xi)$ to the offset surface
based on the current squared distance $d_{sqr}$ and the scaling factor $s$. For this we use a formula that is more robust to cancellation error; see lines 8-9 in Algorithm\ \ref{alg:LBCCD}.

Starting with a most conservative time of impact $t=0$ (line 10) we create a \emph{local} scratch-pad copy of nodal positions, $x_i$, and initialize the current lower bound step, $t_l$, with Equation\ \ref{eq:CCD_LB} (line 11).

 We then enter our iterative refinement loop (lines 12-21) to monotonically improve our TOI estimate $t$. At each iteration, we update our local copy of nodal positions $x_i$ with the current step $t_l$ (lines 13-14). If this new position achieves our target distance to the offset surface (becomes smaller than $g$) we have converged and the previous $t$ is the time of impact that brings distance just up to $g$ (line 17). If not, we update our TOI estimate by adding the current $t_l$ to $t$ (line 18). Note that we always add our first lower bound step to $t$ (line 16) as it is guaranteed to not bring distance closer than $g$. If our TOI is now larger than $t_c$, the current minimum first time of impact (or can be simply set to $1$), we can return no collision (lines 19-20). Otherwise we compute a new local, lower bound, $t_l$, \add{(with $0.9$ scaling for improved convergence)} from the updated configuration (line 21) and begin the next iteration.

 ACCD thus provides an exceedingly simple-to-implement, numerically robust CCD evaluation. It requires only explicit calls to distance evaluations and so no numerically challenging root-finding operations. In turn, ACCD is able to support thickness offsetting and also controllable accuracy and so flexible tuning for performance vs. accuracy trade-offs in CCD applications. In Section\ \ref{sec:CCDComp} we compare ACCD with state-of-the-art CCD solvers, there we see that they all fail severely, resulting in intersection or tiny TOI that stalls the optimization in challenging examples. In turn, in easier cases where alternate CCD methods do succeed, we then see that ACCD achieves similar and often improved timing performance.
 Finally, we point out that ACCD's stopping criteria requires $s>0$ (we apply $s=0.1$ in all examples) to provide finite termination. In turn, ACCD does not target computing an exact TOI but rather (as actually required in contact-processing applications) obtains  reliable, intersection-free steps towards TOI. 
 
\paragraph{Worst-case performance} 
As discussed above, we find in practice and particularly throughout all our challenging testing that ACCD converges with solid performance (comparable or faster times when other methods succeed, efficient times when all other methods fail). It is worth analyzing, however, that it should be possible to have unbounded, worst-case performance. Recall we pick our reference frame so that displacements sum to $0$. However, if a primitive has a diverging displacement field which cannot be cancelled out, our denominator $l_p$ can remain large.  
Then, if a primitive also has a small starting distance $d$ and so correspondingly small numerator, iteration count for ACCD on this case can be large.
While we see that this scenario does not occur in practice for elastodynamics, extensions of ACCD to accelerate convergence for these possible cases, and so obtain bounded performance guarantees is an interesting future step.

\section{Evaluation}\label{sec:exp}

We implement our methods in C++, applying CHOLMOD\ \cite{chen2008algorithm}, compiled with Intel MKL LAPACK and BLAS for linear solves and Eigen for remaining linear algebra routines\ \cite{eigenweb}.
Necessary derivatives for C-IPC and our algebraic simplifications applied for efficiency and numerical robustness are detailed in our supplemental document. 
To enable future applications, development and testing we will release our implementation of C-IPC as an open source project.
All our experiments and evaluations are executed on either an Intel 16-Core i9-9900K CPU  3.6GHz $\times$ 8 (32GB memory), an Intel Core i7-8700K CPU @ 3.7GHz $\times$ 6 (64GB memory), or an Intel Core i7-9700K CPU @ 3.6GHz $\times$ 4 (32GB memory) as detailed per experiment below.

\begin{figure}[t]
    \centering
    \includegraphics[width=\linewidth]{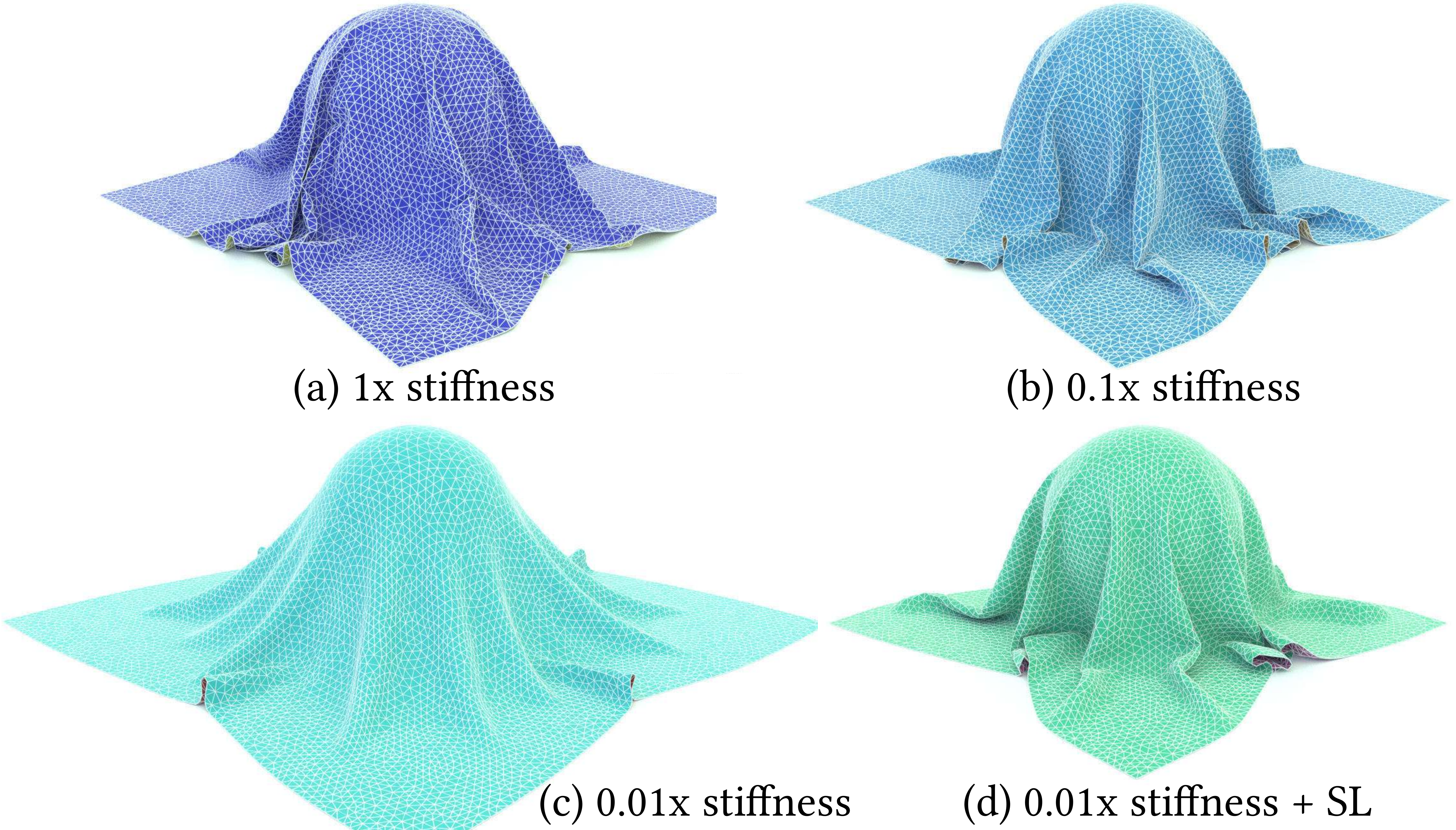}
    \caption{\textbf{Membrane Locking.} A cloth on sphere simulation (8K-node) with (a) $1\times$ stiff membrane of cotton material ($0.8MPa$) suffers from membrane locking where bending behavior is artificially stiffened and creased. Reducing membrane stiffness to (b) $0.1\times$ that of  measured cotton values removes artificial bending stiffness but we still observe overly sharp edge creasing. Next an even softer membrane at (c), $0.01\times$ cotton's, then avoids observable  membrane locking artifacts altogether, but results in nonphysical stretching inappropriate for cotton. However, (d) applying a softer $0.01\times$  membrane together with our constitutive strain limiting model to enforce measured cotton strain limits prevents membrane locking while obtaining natural stretch limits for the material.}
    \label{fig:membraneLocking}
    \vspace{-0.3cm}
\end{figure}

\paragraph{Experiments} Below we begin with a study illustrating and analyzing membrane locking behaviors for standard cloth materials and meshes (Section\ \ref{sec:membLockingExp}). We then demonstrate C-IPC's ability to strictly satisfy strain-limits while fully coupling all physical forces in Section\ \ref{sec:exactSLExp}. To our knowledge C-IPC is the first method to both enable strict satisfaction of strain-limits and to support full coupling of strain-limiting, elasticity and contact forces. To analyze the impact of strain-limit satisfaction and full coupling we next consider comparison against prior methods. As discussed in Section\ \ref{sec:rel}, existing methods in strain-limiting generate artifacts including locking, jittering and interpenetration. These issues result from two sources: 1) inaccuracies from applied splitting models and 2) inability to satisfy strain-limits in computation. Here, for the first time, we first separately analyze the problems that are created by splitting models (Sec. \ref{sec:SLSplittingExp}) independent of solver accuracy errors. Then in Section \ref{sec:exp_SLComp} we consider the artifacts and inaccuracies (generated by state-of-the art cloth code) that jointly result from both splitting errors \emph{and} inability of the method to enforce requested strain-limits. In Section\ \ref{sec:thickness_exp} and \ref{sec:CCDComp} we analyze C-IPC's thickness modeling, comparing with both existing cloth simulation codes and prior methods for CCD evaluation. Finally, with all components of C-IPC evaluated, we assess C-IPC's application to garment simulation (Section \ref{sec:garmentExp}) and its resolution of previous stress-test cloth benchmarks (Section\ \ref{sec:clothBenchmarkExp}). We then demonstrate C-IPC on simulations of fully coupled systems of arbitrary codimension with consistent thickness modeling\ (Section\ \ref{sec:generalCoDim}) and finally consider its performance on a new set of cloth simulation stress tests designed to exercise robustness and accuracy (Section\ \ref{sec:clothStressTestExp}).
\add{All experimental setup and statistics are listed in Figure\ \ref{fig:timingTB}. We report the total number of nodes involved in the system, including those that are kinematic (equality constraints). All examples are directly applying IPC's smoothed, semi-implicit friction (1 lagged iteration) and the default Newton tolerance if not otherwise mentioned.}

\subsection{Cloth Material and Membrane Locking}\label{sec:membLockingExp}

Here we first illustrate and examine the impact of membrane locking on real-world materials, and strain-limiting's ability to mitigate them. In Penava et al.\ \shortcite{penava2014determination}, density, thickness, and directionally dependent membrane stress-strain curves are measured and validated for a range of cloth materials. However, directly applying these real-world cloth parameters in simulation with standard-resolution meshes can produce severe membrane locking. This is unavoidable unless exceedingly high and so often impractically large mesh sizes are used. Here we examine this locking behavior, first using our IPC model \emph{without} strain-limiting as a demonstration. We also note that such locking behaviors are independent of algorithm and so, for example, are also easy to demonstrate in other codes like ARCSim\ \cite{narain2012adaptive} with their real-world captured material parameters\ \cite{wang2011data}; see Figure\ \ref{fig:SLComp_arcsim}d.

We start by considering the behavior of a simple $1m \times 1m$ unstructured mesh dropped on a fixed sphere and ground plane (with friction of $\mu=0.4$ for all). We apply the measured cotton cloth density ($472.6kg/m^3$) and thickness ($0.318mm$) from Penava et al.\ \shortcite{penava2014determination} and then consider varying membrane stiffness while keeping bending Young's modulus at $0.8MPa$ and Poisson ratios for both membrane and bending to the measured value for warp direction as $0.243$. Specifically Penava et al.\ \shortcite{penava2014determination}, find membrane Young's moduli ranging from $0.8MPa$ to $32.6MPa$ for varying in-plane directions.

For our example, we then find that even when applying an isotropic membrane model, with the smallest determined membrane stiffness ($0.8MPa$), we observe severe locking artifacts (see e.g. Figure\ \ref{fig:membraneLocking}a) where bending is artificially stiffened. 
If we next lower this smallest measured membrane stiffness by $0.1\times$, we then see that artificial bending artifacts are largely eliminated but we still obtain sharp creasing artifacts forced by the dominating membrane energy (see e.g. Figure\ \ref{fig:membraneLocking}b). 
Next, if we try an even smaller $0.01\times$ scaling of  membrane stiffness, observable membrane locking effects are now gone, but as expected the resulting material is much too stretched and  so not even close to the desired material behavior (Figure\ \ref{fig:membraneLocking}c). This simple example nicely demonstrates the challenges of membrane locking when simulating stiff real-world cloth materials. We then note that these artifacts are only exacerbated in more challenging simulations with, for example, moving boundary conditions and tight contact.

Next we apply a strict strain-limiting with C-IPC's isotropic model to constrain strain within the elasticity range measured by Penava et al.\ \shortcite{penava2014determination}. The broadest range in all directions for cotton allows a stretch factor up to $6.08$\%. Here, applying this bound, even with $0.01\times$ scaling of membrane stiffness, we now regain a simulation free from membrane locking and, since restricted to measured strain limits, also free of unnatural stretching artifacts (Figure\ \ref{fig:membraneLocking}d).

\add{Finally, it is important to reiterate that membrane locking is indeed a resolution-dependent artifact whose impact can decrease as mesh size increases. For example, locking artifacts for the cloth draped on sphere at $1\times$ bending stiffness significantly decrease with a mesh of 85K nodes and are imperceptible at 246K nodes. However, this improvement is highly dependent on scene parameters. For example, the same 246K mesh, in the exact same scene, exhibits significant locking artifacts if we just reduce bending stiffness by $0.01\times$ to match the material in the strain-limited simulation in Figure\ \ref{fig:clothOnRotSphere}, bottom row.
Please see our supplemental document for details and illustration of these cases.}

\subsection{Exact Strain Limits}\label{sec:exactSLExp}

\begin{figure}[t]
    \centering
    \includegraphics[width=\linewidth]{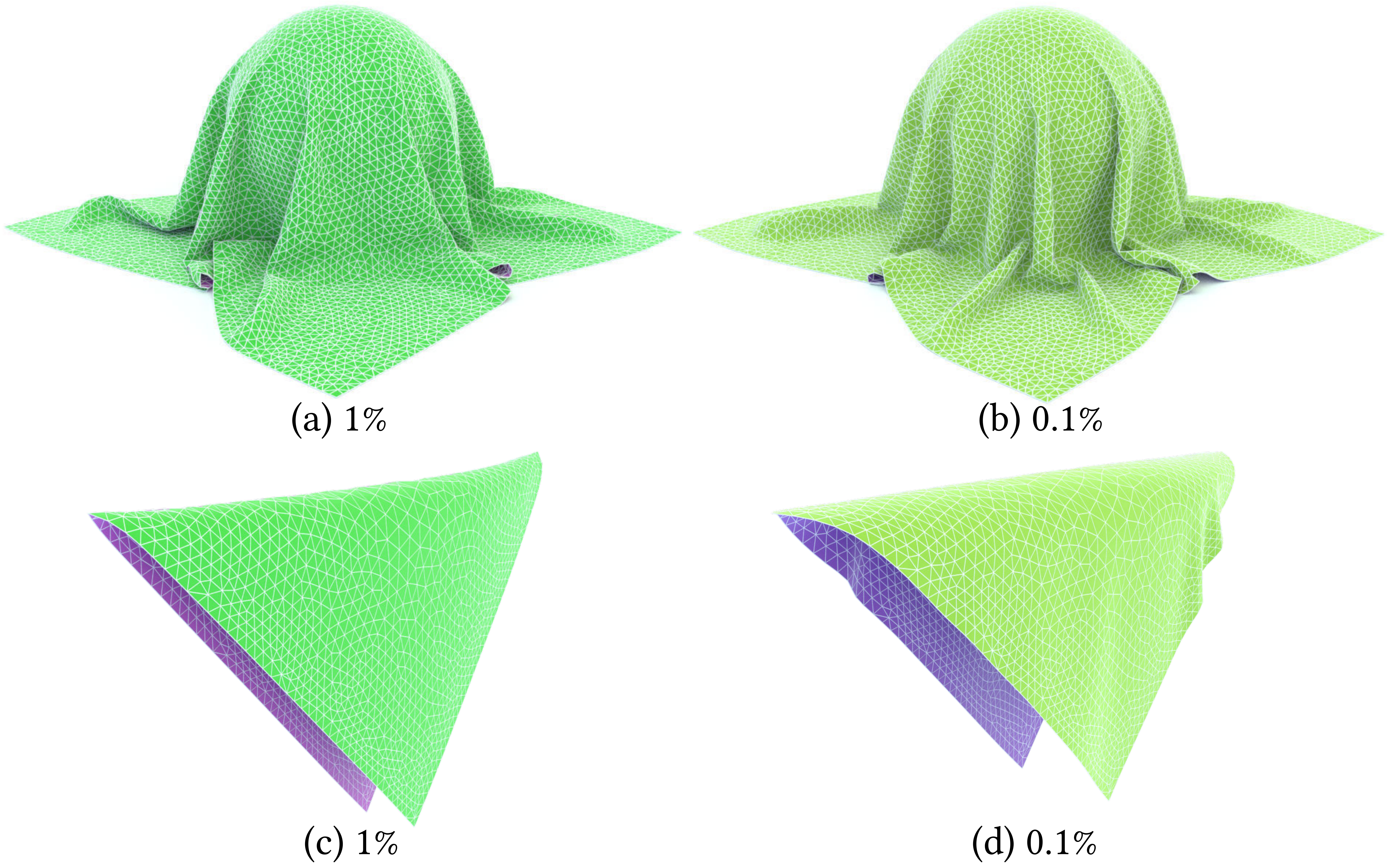}
    \caption{\textbf{Extreme Strain Limits.} We confirm C-IPC's constitutive strain limiting model provides robust and guaranteed satisfaction down to tightest measured and practically applied limits for cloth (and well below). Here we test C-IPC with strain limits of $1$\% (already well below most measured limits for cloth) in (a) and (c). To push the challenge further in (b) and (d) we then test C-IPC with an excessively small, largely nonphysical strain limit of $0.1$\%. Here the applied simulation meshes are low resolution, containing respectively only 8K (for a and b) and 2K (c and d) nodes, and so are challenging to avoid locking artifacts with real-world materials. We confirm that C-IPC preserves requested limits across all elements in all simulation time steps. We also note that, as expected, with the excessively small, nonphysical strain-limits 
    applied in (d) more constraints become active w.r.t. to the DOF count resulting in a small amount of visible membrane locking artifacts; see our discussion below.}
    \label{fig:SL_stressTest}
    \vspace{-0.3cm}
\end{figure}

Above we have demonstrated the well-known importance and impact of applying strain-limiting for simulating cloth materials. Here we investigate C-IPC's ability to accurately enforce tight strain limits 
across both our isotropic and anisotropic models. 

\paragraph{Accuracy across decreasing strain limits}

To confirm accuracy, controllability and robustness of C-IPC's constitutive strain limiting we test increasingly severe strain limits of $1$\% and $0.1$\% (well below measured limits in standard cloth materials\ \cite{clyde2017modeling,penava2014determination}). We apply them to both the sphere drape example considered in the last section as well as a simple membrane locking test with a two-corner pinned cloth \cite{jin2017inequality, chen2019locking}. Here, applying C-IPC, see Figure\ \ref{fig:SL_stressTest}, we observe stable simulation output with stretches on all triangles satisfying their prescribed limit constraints.
Closer examination of the draped sphere tests in Figure\ \ref{fig:SL_stressTest} also confirms that both are free from artificial bending stiffness. For contrast, compare these results with the artificial stiffening in Figure\ \ref{fig:membraneLocking}a, where the maximum stretch exceeds $4.5$\% despite applying the measured membrane stiffness directly. 

Here, as we are applying our strain-limits \emph{unilaterally}, this agrees with Jin et al.'s\ \shortcite{jin2017inequality} argument that unilateral strain-limits can 
avoid membrane locking better than bilateral enforcement -- at least to a certain degree. 
Even unilateral enforcement is not a perfect panacea if limits are too tight. For an extreme example we can push strain-limits to a very small and largely unrealistic ($0.1$\%) strain limit. This provides an extreme stress test and we confirm C-IPC does preserve these challenging limits across all time steps. As expected, however, here we can finally observe some slight but distinct sharp edge-creasing artifacts in Figure\ \ref{fig:SL_stressTest}b and similarly visible locking issues in Figure\ \ref{fig:SL_stressTest}d. Although such an extreme limit is unlikely in practice and generally not encountered for cloth, this experiment does highlight an important point: if most unilateral st\add{r}ain-limit constraints are active then they behave similarly to the bilateral case. Thus, when most strain limit constraints are at the bound active constraint numbers can again approach the number of DOFs, and so, as discussed in Chen et al.\ \shortcite{chen2019locking}, locking can once again be encountered.  
Here we focus on formulating a controllable and robust constitutive model for strain limiting.  We hope this will then enable further increasing the range of locking-free configurations via explorations of alternate discretization for the strain constraints as in Chen et al.\ \shortcite{chen2019locking}. However, we note that this can best be enabled when the underlying method, as proposed here, can accurately guarantee coupled, constraint satisfaction.


\paragraph{Anisotropic strain limiting}
\label{sec:anisoSLParam}

\begin{figure}[b]
    \centering
    \includegraphics[width=0.95\linewidth]{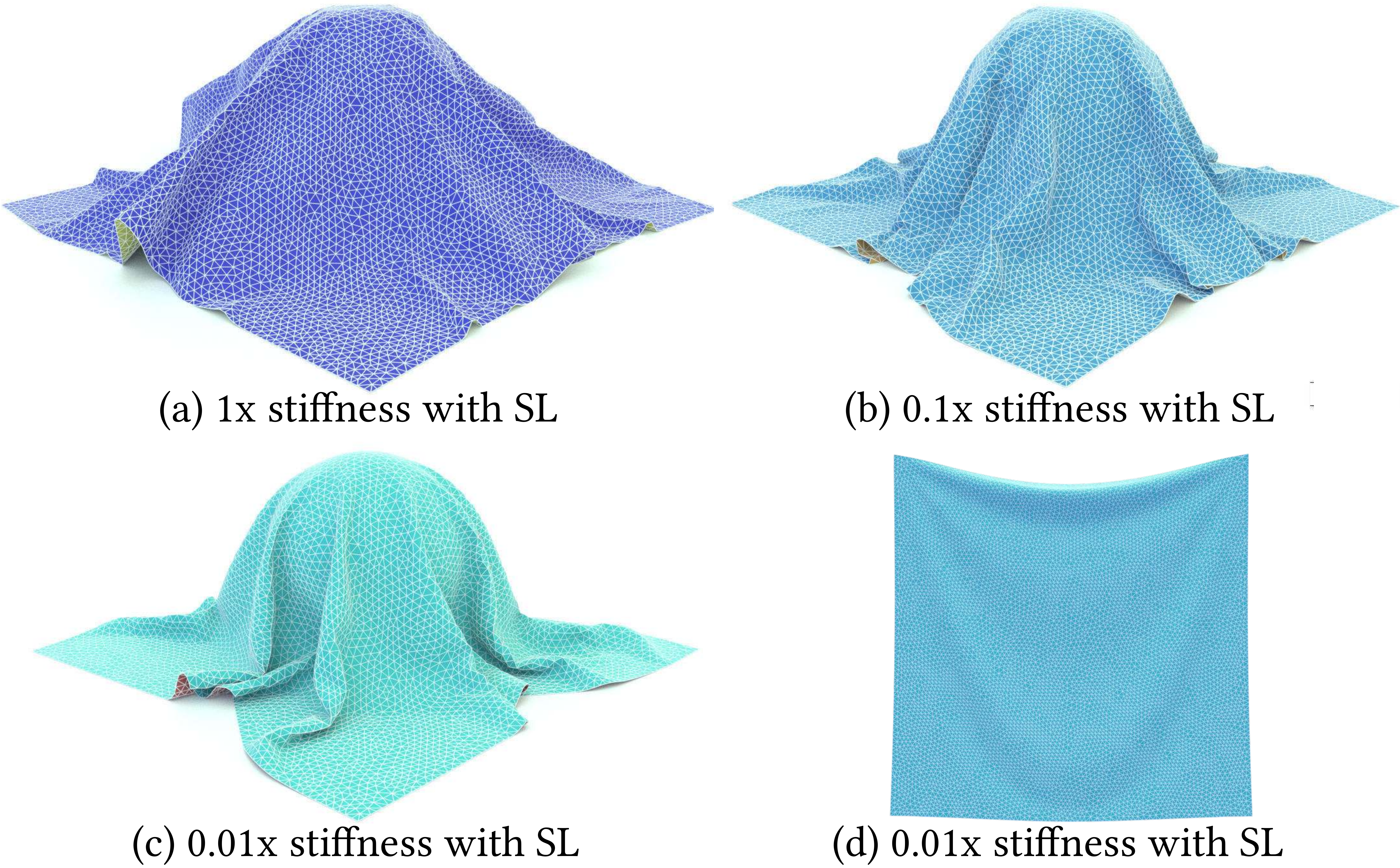}
    \caption{\textbf{Anisotropic strain limiting.} We apply the C-IPC anisotropic strain-limiting model to the sphere drape test in (a) through (c), varying only scaling of the measured membrane stiffnesses, and to the two-corner cloth hang test in (d). All are time stepped at $h=0.04s$ with their final equilibria shown here. In (c) and (d), with a $0.01\times$ scaling of the measured stiffness, we see that C-IPC's tight enforcement of the measured strain limits\ \cite{clyde2017numerical} obtains artifact-free drapes with differing wrinkle patterns resulting from the anisotropic model.}
    \label{fig:anisoSL}
\end{figure}

For anisotropy, the story is similar. In Figure\ \ref{fig:anisoSL} we demonstrate C-IPC's anisotropic model with both the sphere drape and two-corner hang tests. Again we observe that applying measured stiffness values, here from Clyde\ \shortcite{clyde2017numerical}, generates clear membrane locking artifacts (see Figure\ \ref{fig:anisoSL}a). Next in Figures\ \ref{fig:anisoSL}b and c we scale down membrane stiffness by  $0.1\times$ and $0.01\times$ respectively. Throughout we confirm C-IPC enforces Clyde's measured strain limits and again obtaining the expected improved results, both without locking artifacts and without nonphysical stretching, at $0.01\times$ the measured stiffness. Similarly, in Figure\ \ref{fig:anisoSL}d (again at $0.01\times$ reported stiffness) we apply C-IPC to preserve measured strain limits, obtaining artifact-free wrinkling with the anisotropic model.

\subsection{Comparison with Splitting Models}\label{sec:SLSplittingExp}

\begin{figure*}[t]
    \centering
    \includegraphics[width=0.9\linewidth]{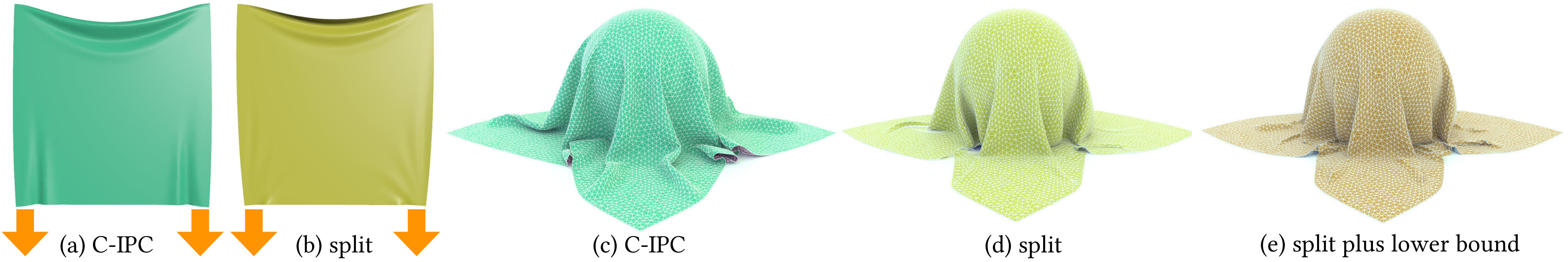}
    \caption{\textbf{Comparison of C-IPC's fully coupled strain limiting model with standard splitting models} highlights the differences between the fully coupled solutions in (a) and (c) and the artifacts generated by splitting strain-limiting solves from elastodynamics in (b), (d) and (e). Here we see that with splitting resulting large errors in elasticity (membrane and bending) can generate incorrect wrinkling directions (b), severe compression artifacts (d), or even numerical softening as in (d) and (e).}
    \label{fig:splitComp}
    \vspace{-0.2cm}
\end{figure*}

Existing strain-limiting methods introduce errors from two primary sources. First, at the modeling level, they split strain limiting from the resolution of elasticity. Second, irrespective of the splitting model employed, algorithms applied  to solve them are inaccurate and often unable, as we will see in the following section, to satisfy even moderate requested strain-limits. All prior methods then introduce errors from both of these sources and so it has remained unclear what problems originate from each source. Here, we first separately analyze the problems that are created by splitting models, by solving each step of the split to tight accuracy. This allows us to show that the splitting itself introduces errors that are unavoidable irrespective of how accurately limit constraints could be enforced. Then, next in Section \ref{sec:exp_SLComp}, we will examine solution accuracy and see that errors in the strain-limit solve itself then produce inconsistent and so often uncontrollable results for simulation.

To examine splitting errors we begin by applying the standard strain-limiting, splitting strategy and so divide each time step solve into two sequential steps. The first solves a predictor step that includes all forces for the whole system, \emph{with the exception of the strain-limit}, to obtain an intermediate configuration $\hat{x}$. Next, the second step \emph{projects} the predictor to satisfy the strain-limit. To do so we minimize our constitutive strain limiting energy summed with the mass-weighted $L^2$ distance from the final timestep solution to $\hat{x}$. 

In the simplest case we can consider the effect of this splitting when there are no contact forces. We start with a pinned and pulled square cloth. We fix its two top corners and apply heavy weights to pull its two bottom corners downwards; see Figure\ \ref{fig:splitComp}a and b. For comparison, without strain limiting the cloth is stretched well over $10\times$ (not shown) while with our constitutive strain limits, strains are restricted to the measured elasticity range and, in this fully coupled solution, we see the expected vertically aligned wrinkles from stretching at the two bottom pulled corners; see Figure\ \ref{fig:splitComp}a. On the other hand, in Figure\ \ref{fig:splitComp}b we see that splitting strain limiting from the elasticity solve produces obvious biasing artifacts at the two bottom corners where the wrinkles are aligned in non-physical directions. These errors, resulting from the decoupling of strain-limiting forces, are then increasingly severe as time step sizes increase (here we show with $h=0.04$s).

Next, to consider splitting errors with contact, we again consider the cloth on sphere example. Here we apply a neo-Hookean membrane elasticity to help the splitting method avoid possible triangle degeneracies. 
Note that even here our fully coupled C-IPC model does not actually require the neo-Hookean elasticity (as triangle degeneracies are prevented by point-triangle constraints between neighbors). However, for consistent comparison, we apply neo-Hookean models for both. Here in Figure\ \ref{fig:splitComp}d we then see that the two-step splitting with contact now suffers from severe compression artifacts, which again come from resolving elastodynamics in the first step and then applying strain limiting and contact forces in the second. For comparison, consider C-IPC's corresponding, fully coupled strain-limited solution in Figure\ \ref{fig:splitComp}c. In turn the error in the split solution suggests that an additional lower bound on strain (e.g. as applied in ARCSim) might be helpful to avoid these compression errors. If we then additionally add a strain-limit lower bound (here at $0.7$) then we indeed find the resulting splitting solution is now free from severe compression artifacts. Now however, due to the errors in membrane and bending, the cloth in the split solution still remains unnaturally flat against the floor; see Figure\ \ref{fig:splitComp}e.

Of course as with all time-step splitting methods, splitting errors can be reduced by applying increasingly smaller time step sizes. Here, for example, we find that visually noticeable errors between the split model and our fully coupled solve disappear at $h=0.004s$. However, as expected, the resultingly large and often unacceptable increase in compute time wipes out any expected gains from splitting (we will see this theme again in the next sections' comparisons with existing cloth codes). Likewise, the necessary decrease in time step size then varies with example and scene so that robust, controllable time-stepping with splitting, does not appear possible.

\subsection{Comparison with ARCSim's Strain-Limit Solver}
\label{sec:exp_SLComp}

\begin{figure*}[t]
    \centering
    \includegraphics[width=\linewidth]{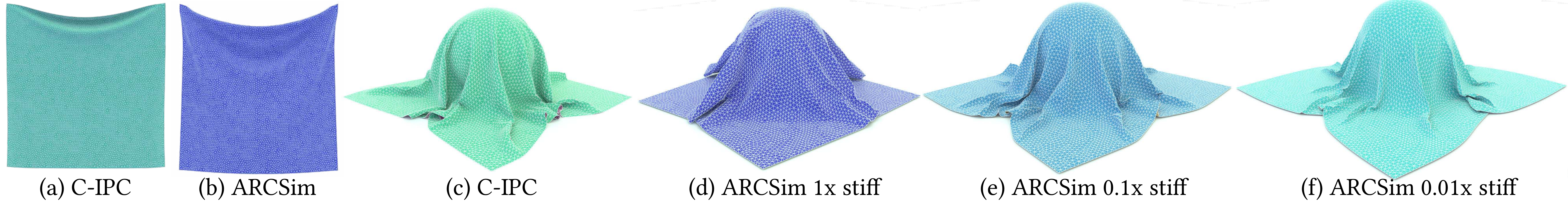}
    \caption{\textbf{Comparison of C-IPC and ARCSim results} illustrate the different behaviors of C-IPC's fully coupled constitutive strain limiting solutions in (a) and (c) and the output of ARCSim's split model and augmented Lagrangian strain-limiting solver in (b) and (d-f). Here frames from ARCSim near equilibrium exhibit excessive stretch due to lack of limit enforcement in (b), (e) and (f). At default material settings in (d) ARCSim exhibits the expected locking artifacts. Then, as we decrease ARCSim's membrane stiffness to avoid membrane locking in (e) and (f) its strain-limiting can not preserve bounds, while the split model introduces additional artifacts so that the over-compression and stretching artifacts are now most obvious. For reference compare with C-IPC's strain-limited, fully coupled solutions in (a) and (c) at comparable stiffness ($0.01\times$) and same times and see Figure \ref{tb:SLComp_arcsim} for summary of timings and limits achieved.}
    \label{fig:SLComp_arcsim}
\end{figure*}

\begin{figure}[t]
    \centering
    \includegraphics[width=\linewidth]{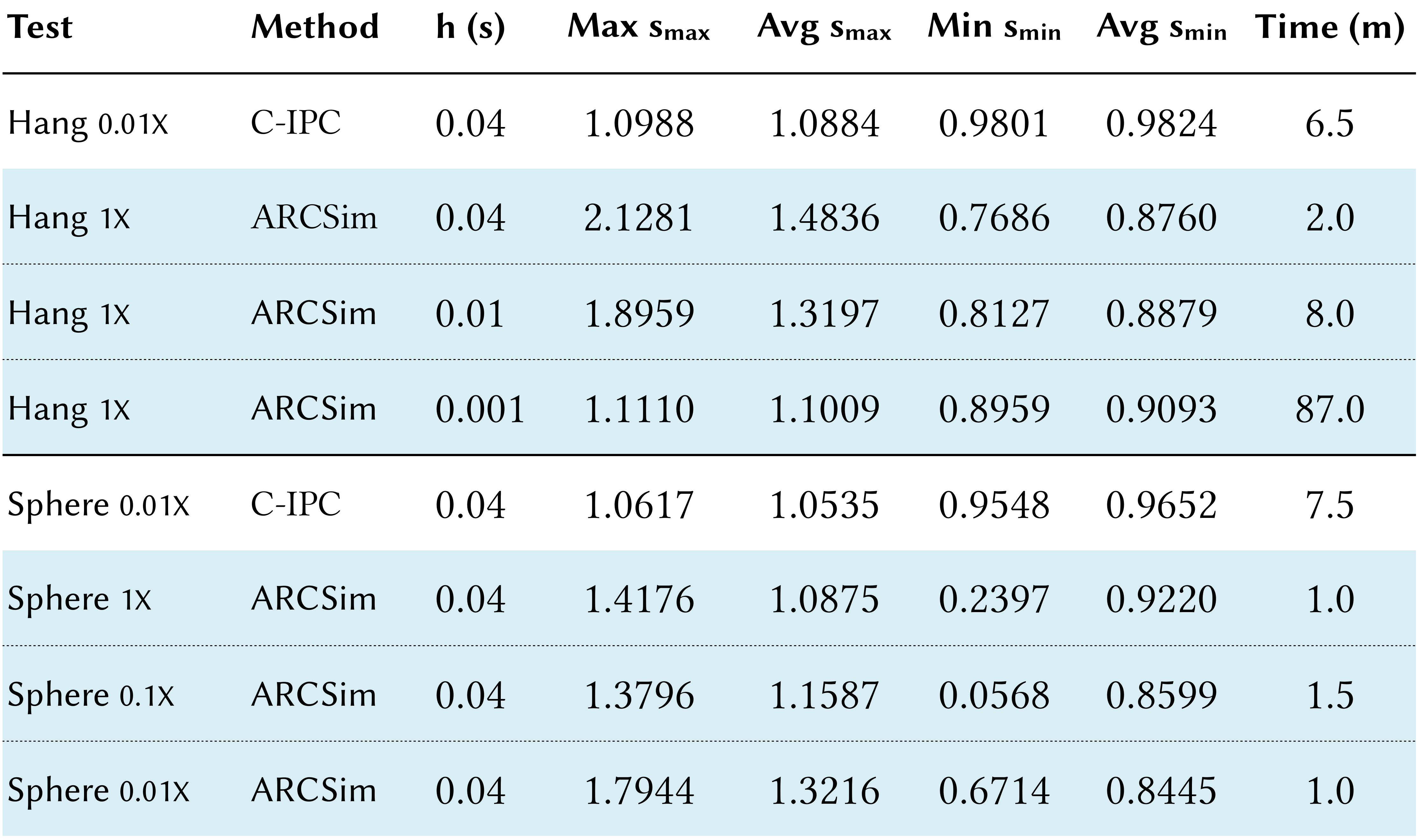}
    \caption{\textbf{Strain limiting comparison with ARCSim.}  We compare ARCSim and C-IPC strain limiting on two cloth simulation test cases: a pinned hanging cloth (without contact) and fixed sphere drape (with contact). For both methods we apply ARCSim's default strain limit settings targeting strains to $[0.9,1.1]$. Above we summarize statistics of the achieved max, $s_\text{max}$, (respectively, min $s_\text{min}$) strain-limit over all elements in a timestep as the averaged and max (respectively min) over all time steps. We apply C-IPC stepped at $h=0.04$ with stiffness rescaled by $0.01\times$ (necessary to avoid membrane locking artifacts and so requiring accurate strain-limiting to compensate) for both examples. We correspondingly consider ARCSim across a range of time-step sizes and stiffness re-scalings ($1 \times$ ,$0.1 \times$ and $0.01 \times$). See Figure \ref{fig:SLComp_arcsim} for visualization of results. In the hang test, even for the full membrane stiffness, we see ARCSim's strain limits are far from satisfied. Then, as step size decreases we observe ARCSim results better satisfy strain limits -- although never fully. Correspondingly, in the sphere drape, as we reduce membrane stiffness towards a value required to mitigate locking, we ARCSim results increasingly violate strain limits more significantly. In comparison we observe C-IPC satisfies strain limit satisfaction at the necessary soft membrane energy, across all elements and time steps for both tests.}
    \label{tb:SLComp_arcsim}
\end{figure}

Above we analyze the \emph{modeling} errors generated by time-step splitting with strain limiting. To do so we compare models for which each stage is consistently solved to tight accuracy in our common benchmark code. Here we next consider the joint impact that splitting errors combined with inaccurate constraint solves then have in existing cloth simulation codes. To do so we compare with ARCSim\ \cite{narain2012adaptive} which is, to our knowledge, the most robust shell simulator currently available to support strain limits with real-world cloth material parameters. See Figure\ \ref{tb:SLComp_arcsim} for statistics on the strain limits and timings achieved by ARCSim and C-IPC here. 

We again begin by considering the simpler, contact-free case. To do so we consider an even simpler cloth hang example: by dropping the pinned cloth \emph{without} weighted bottom corners. We apply ARCSim's default cotton material and default strain limit settings which restrict to the range $[0.9, 1.1]$. Here, even with this wide range of permissible strains and at full membrane stiffness, ARCSim results violate the limit bounds significantly at each time step. For visual reference in Figure\ \ref{fig:SLComp_arcsim}a we illustrate the fully coupled C-IPC solution at equilibrium satisfying the strain-limit bound. Compare with the ARCSim output at equilibrium in Figure\ \ref{fig:SLComp_arcsim}b where we easily see the artifacts resulting from over-stretching near the fixed corners forming a larger arc. Here both frames are taken at $4s$ and are time-stepped with $h=0.04s$. 

As we then decrease ARCSim's time-step size to $h=0.01s$ these errors still remain significant, and it is not until we reach at step size of $h=0.001s$ that strain limits are mostly (although still not entirely) satisfied. However, to do so ARCSim then requires $87$ minutes to simulate this simple $4s$ simulation sequence, resulting in a well-over $10\times$ slowdown compared to C-IPC's fully coupled, strain-limit-satisfying solution stepped at $h=0.04s$.  However, beyond the requisite slowdown, the required decrease in step size changes per scene and material and so it is always unclear, without many time-consuming simulation tests per scene, how to determine the necessary time step decrease to ensure ARCSim results satisfy a prescribed strain limit. In contrast C-IPC again maintains strain-limit constraint satisfaction across time step, material and scene settings. 
To investigate ARSim's constraint errors further we also observe that the augmented Lagrangian strain-limiting solver\ \cite{narain2012adaptive} employed in ARCSim applies a hard-coded max-iteration cap set to 100. Experimentally increasing this limit (and otherwise leaving the ARCSim code unchanged) we confirm that even increasing this cap by an order of magnitude, strain limit satisfaction is still not achieved.

Next we add contact for our ARCSim comparison and consider the 8K-node sphere drape test. Here we again observe that applying strain-limiting to reduce membrane locking does not work and instead see that strain-limit satisfaction worsens and artifacts actually increase as membrane stiffness decreases. 

Specifically we start by applying ARCSim's default cotton material and the same default strain-limit settings as above. For this stiff material we observe the expected membrane locking issues; see Figure\ \ref{fig:SLComp_arcsim}d. However, we do note that with this stiff membrane, strain limit constraints are certainly reasonably well satisfied except for a few time steps. 

Next, as standard, we reduce membrane stiffness in ARCSim with hope of mitigating the locking with the expectation that strain-limiting will compensate. If we \emph{reduce} stretching stiffness by $10\times$, the locking artifacts are indeed less; however, ARCSim's strain-limiting does not provide the intended compensation for the reduced stiffness. Instead, there are significantly more triangles violating the strain limit throughout the simulation (see Figure\ \ref{fig:SLComp_arcsim}e) resulting in stretching artifacts. If we then further reduce stiffness to $0.01\times$ the default cotton value (generally necessary in this example to remove all visible locking artifacts; e.g. compare with our comparable stiffness, C-IPC result in Figure\ \ref{fig:membraneLocking}d and \ref{fig:SLComp_arcsim}c) ARCSim's results in Figure\ \ref{fig:SLComp_arcsim}f, show even more significantly violated strain limits and so suffer from the same extreme compression issues exhibited by the split model analysis in the last section; see Figure\ \ref{fig:splitComp}e for comparison. In addition, we now also observe jittering resulting from ARCSim's further three-way split that further separates the strain limiting solve and the contact forces into two separate, sequential projection steps. We note that this last issue is not new to our analysis here and is discussed in Narain et al.\ \shortcite{narain2012adaptive}.

\subsection{Thickness Modeling}\label{sec:thickness_exp}

\begin{figure}[t]
    \centering
    \includegraphics[width=\linewidth]{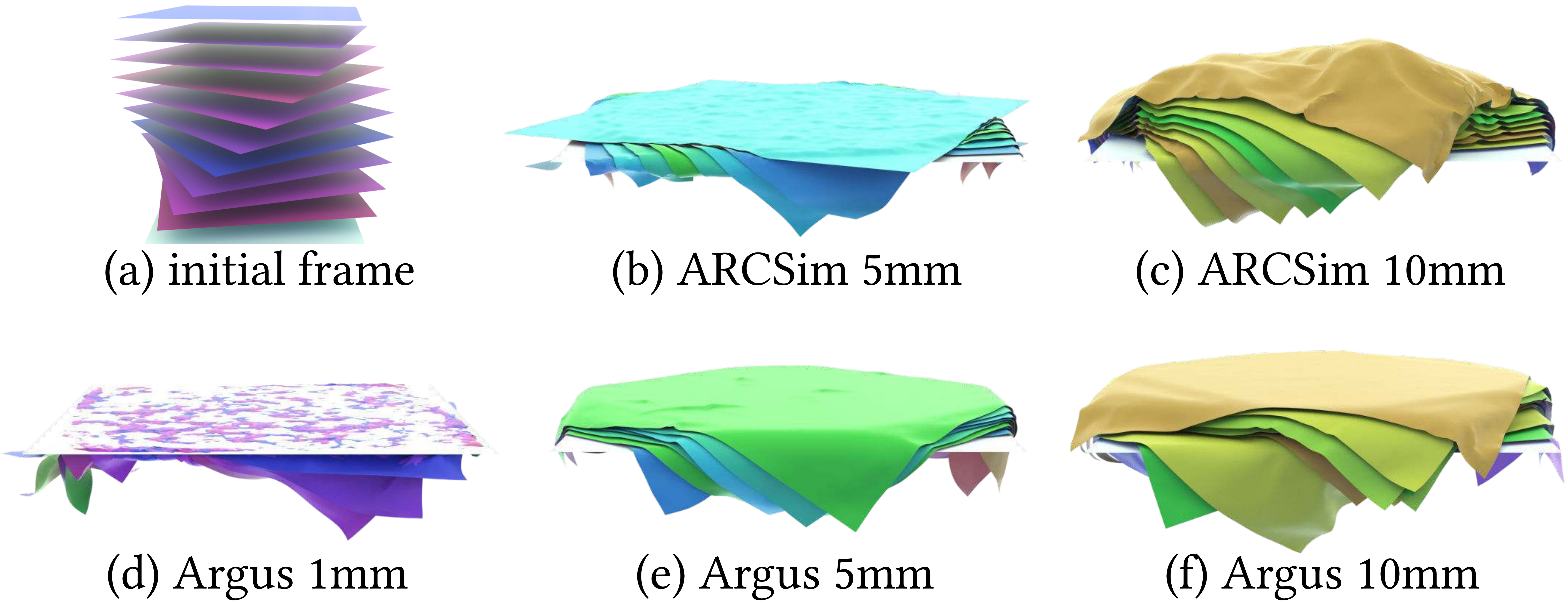}
    \caption{\textbf{Cloth stack comparisons.} Ten 8K-node square cloth panels are dropped in stack upon onto a square board with friction $\mu=0.1$ (a). 
    Here ARCSim\ \cite{narain2012adaptive} and ARGUS\ \cite{li2018implicit} are tested for varying cloth thicknesses. Time stepping at $h=0.001s$ ARCSim fails to resolve contact for a thickness of $5mm$ (b), and at $10mm$ there is no longer interpenetration but the dynamics and geometry have noticeable artifacts (c). 
    ARGUS can successfully simulate cloth with $5mm$ and $10mm$ thicknesses (e,f) at  $h=0.001s$ with some smaller artifacts; but at $1mm$ thickness it fails to resolve contact altogether.
    See the top row Figure\ \ref{fig:ballOnClothStack} for our corresponding IPC results for this example.}
    \label{fig:clothStackComp}
    \vspace{-0.3cm}
\end{figure}

Next we examine modeling small but finite thicknesses in contact. To illustrate differences between the C-IPC model and existing state-of-the-art shell simulators  
we examine a simple, yet challenging cloth stacking benchmark. 
We construct a scene with a stack of ten 8K-node square cloth panels aligned with vertical spacing and rotated orientation, dropped simultaneously onto a fixed square board. See Figure\ \ref{fig:clothStackComp}a for initial set up. This should form a cloth pile where as we vary material thickness the overall pile height and bulk properties should also correspondingly change. 

{\bf ARCSim}\ \cite{narain2012adaptive} computes\footnote{We use the most recent ARCSim v0.3.1 in our testing.} collision response applying Harmon et al's\ \shortcite{harmon2008robust} non-rigid impact zones and a repulsion thickness applied to both model cloth thickness and to stabilize collision-processing constraints. ARCSim's thickness parameter default is set to $5$mm and is exposed as a configuration that can be changed by users. This parameter is then associated in code with additional thickness parameters directly applied for collision projection, analysis and contact force computation. 
Starting with the default $5mm$ thickness setting, and testing with increasingly smaller time steps down to $h=0.001s$, ARCSim consistently reports collision handling failures and demonstrates severe artifacts in the simulation results; see Figure\ \ref{fig:clothStackComp}b. Next applying a $10mm$ thickness ARCSim reports success in resolving contacts at $h=0.001s$, but still generates large artifacts in both geometry and dynamics; see Figure\ \ref{fig:clothStackComp}c. We did not push ARCSim further to time steps smaller than $h=0.001s$, as this already makes ARCSim excessively slow for computation; e.g. as compared to C-IPC.

{\bf ARGUS}\ \cite{li2018implicit} 
updates ARCSim with improved contact and friction processing for shell simulation. As in ARCSim it also applies and exposes to users the same thickness parameters. \add{With default solver parameters, a}\del{A}t $h=0.001s$ ARGUS is able to simulate the cloth stack with the default $5mm$ material thickness (Figure\ \ref{fig:clothStackComp}e) and also with a thicker $10mm$ material (Figure\ \ref{fig:clothStackComp}f). Here we also observe that some height differences in the piles are captured. However, as we decrease thickness towards thinner, more standard garment material thicknesses, e.g. setting thickness to $1mm$,
ARGUS produces severe artifacts as shown in Figure\ \ref{fig:clothStackComp}d. As with ARCSim, for all three thickness settings ARGUS' timing remains much slower than C-IPC.

In summary we observe that state-of-the-art methods require small and often impractical time step sizes to avoid failures as modeled geometric thickness decreases. With smaller thickness and/or increasing collision speeds, the required time step size to avoid failure decreases, correspondingly increasing simulation cost and making it unclear what time step size is required for success per scene and thickness. 

\begin{figure}[b]
    \centering
    \includegraphics[width=\linewidth]{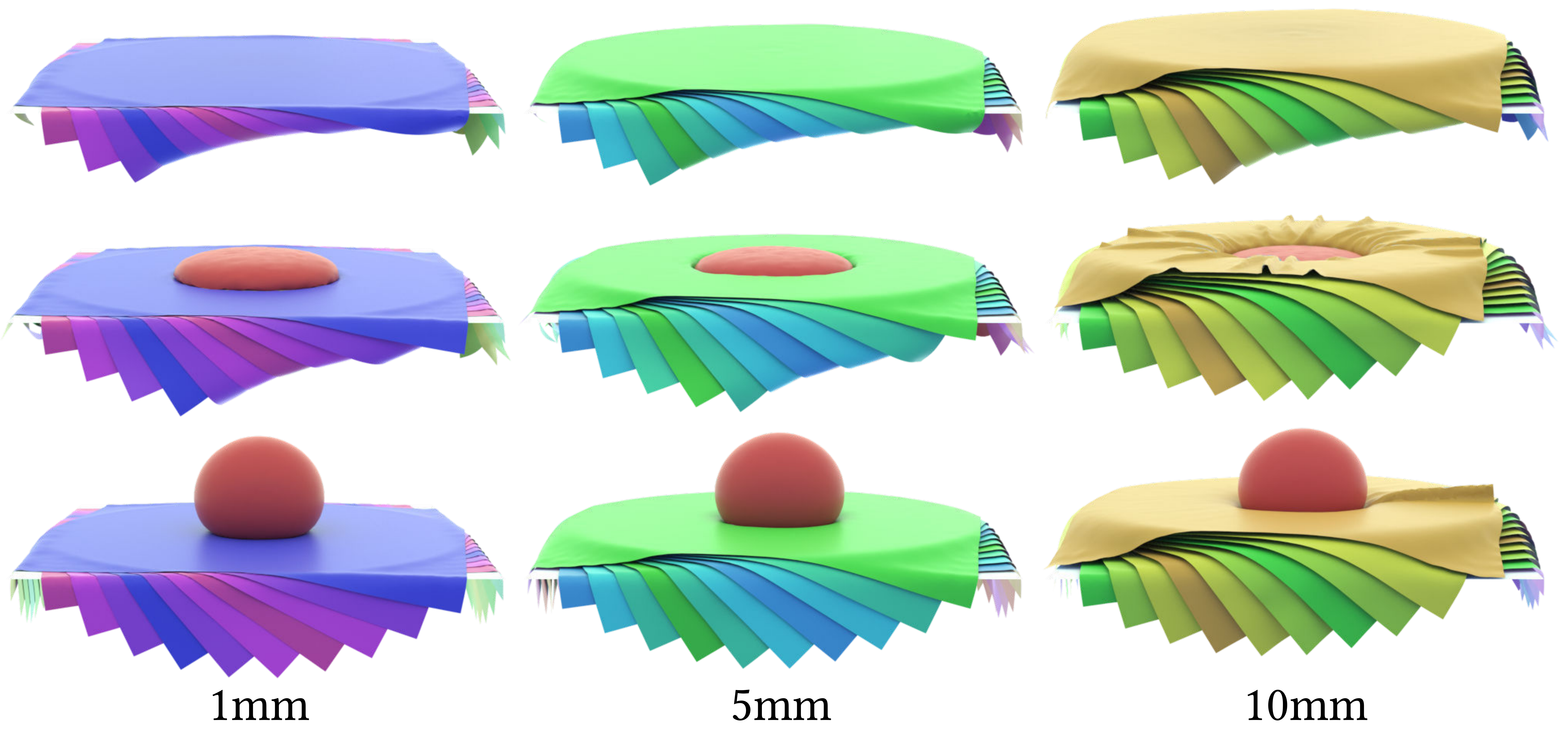}
    \caption{\textbf{Sphere on cloth stack.} Ten 8K-node square cloth are dropped onto a square board with friction ($\mu=0.1$). 
   Top: Varying IPC $\hat{d}$ per column, C-IPC is able to provide controllable thickness modeling for cloth. 
   Middle: A soft elastic volumetric sphere ($E=10KPa$) is dropped onto to the cloth stacks and is illustrated here at maximum compression; here with $10mm$ thickness (far right) we observe the resulting wrinkling enabled by the thicker depression. 
    Bottom: at rest equilibrium heights continue to vary with thickness under the weight of the ball.}
    \label{fig:ballOnClothStack}
\end{figure}

{\bf C-IPC} in contrast, for the same thickness benchmark consistently models the varying thickness by simply setting $\hat{d}$ (recall the distance we start exerting contact forces at) to effective thickness values $1mm$, $5mm$, and $10mm$. In the first row of Figure\ \ref{fig:ballOnClothStack} C-IPC successively models all varying material thicknesses and corresponding heights of the cloth stacks. 
C-IPC's model then further provides constitutive behavior for thickness in the normal direction. This is demonstrated in middle and bottom rows of Figure\ \ref{fig:ballOnClothStack} where we then drop an elastic ball on the stack. In the middle row we show maximum compression for each stack's collision, observing the increasing indentation with thickness. Also note the wrinkles forming solely in the thickest, $10mm$ case. In the bottom row we show the final frame at equilibrium for each simulation after collision, illustrating the different rest heights of the shell piles weighted by the volumetric FE ball model. This example also illustrates C-IPC's natural coupling of differing elastic models which we explore in detail in Section\ \ref{sec:generalCoDim}.
\add{These examples also help illustrate how computational cost for C-IPC can vary with the threshold $\hat{d}$; see  Figure\ \ref{fig:timingTB}. Generally we see that at larger threshold values, e.g. $\hat{d}=10mm$, C-IPC processes more contact pairs per step, making simulation more expensive than at smaller values, with less contacts per solve, e.g $\hat{d}$ at $1mm$ and $5mm$. At the same time setting $\hat{d}$ to values orders of magnitude smaller, e.g. $1\mu m$, leads to slower convergence, and so longer compute times, due to increased sharpness in the barrier functions.}

\subsection{CCD Benchmarking}\label{sec:CCDComp}

\begin{figure}[t]
    \centering
    \includegraphics[width=\linewidth]{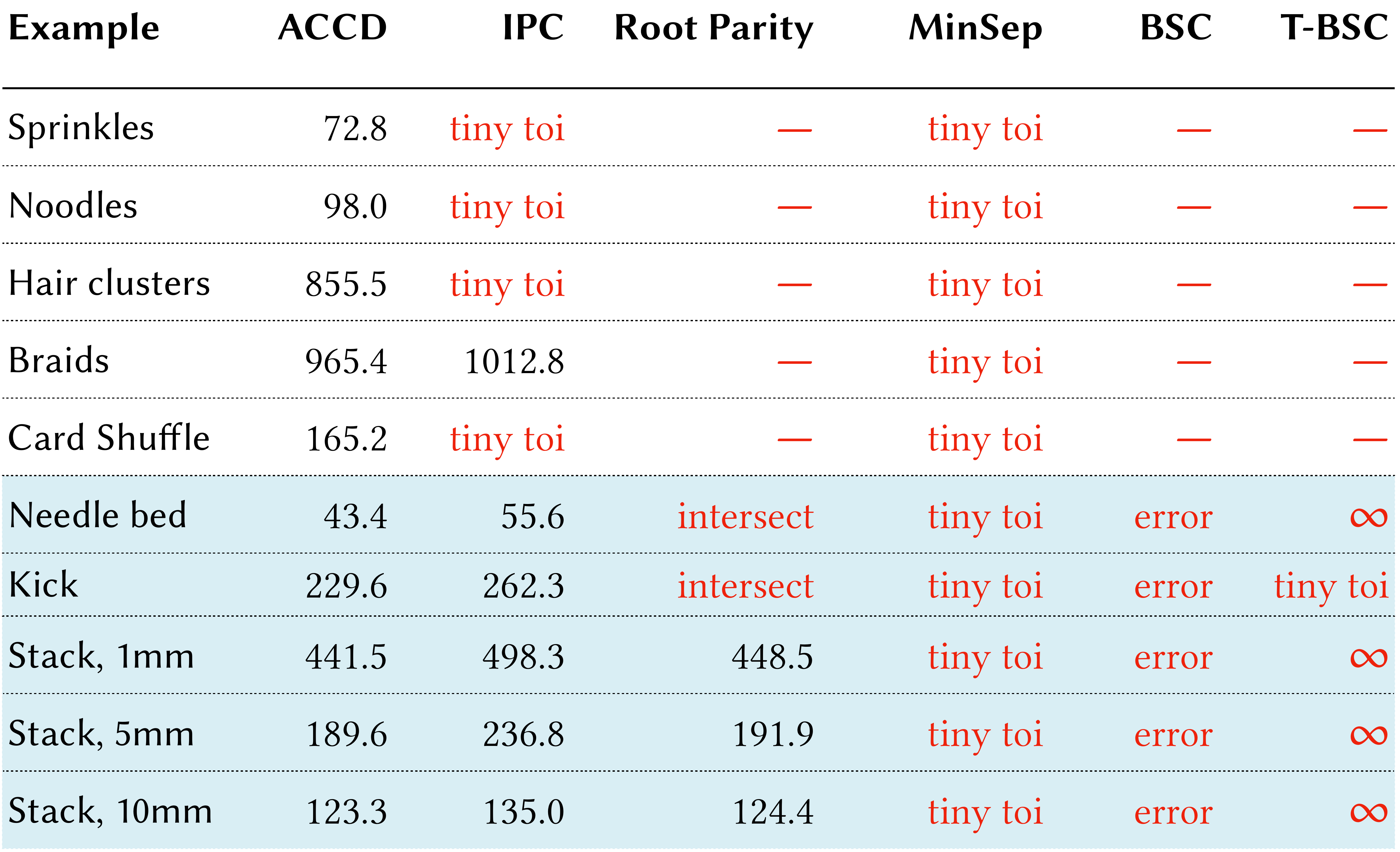}
    \caption{\textbf{CCD comparison statistics.} Here we summarize the per-Newton iteration, total CCD query time (excluding spatial hash) in milliseconds for ten simulation benchmark examples  across CCD methods. We compare our ACCD method, IPC's conservative floating point CCD\ \cite{Li2020IPC}, MinSep\ \cite{harmon2011interference,lu2019scalable}, Root Parity\ \cite{brochu2012efficient}, BSC\ \cite{tang2014fast}, and T-BSC\ \cite{wang2015tightccd}. First five examples in the benchmark include finite thickness offset, which can not be supported by Root Parity, BSC,  and T-BSC methods. When simulation cannot proceed from CCD errors we report type. Here ``$\infty$'' indicates unable to find non-intersecting step (endless line-search) and ``tiny toi''' too small a step size (no progress).}
    \label{fig:CCDComp}
    \vspace{-0.3cm}
\end{figure}

For the last section's comparisons it was sufficient to model thickness effects in C-IPC solely via $\hat{d}$ and so purely as an elastic behavior. More generally, as discussed in Section \ref{sec:thickness}, as thicknesses become smaller and contacts tighter, C-IPC's inelastic thickness model utilizing both $\hat{d}$ and our offset $\xi$ is required. 

For simulating codimensional objects these two mechanisms are combined together. Here larger $\hat{d}$ introduces increased elastic response while a nonzero $\xi$ provides guarantees for minimum thickness even under extreme compression; see e.g. Figure\ \ref{fig:twistCylinder}. For scenes where consistently modeling thickness matters; see e.g. Figures\ \ref{fig:twistCylinder}, \ref{fig:hairs}, \ref{fig:noodles}, \ref{fig:sprinkles}, \ref{fig:sandoncloth}, and \ref{fig:allIn}, we then set $\xi$ to the reported material thickness value of the object or slightly smaller and then set $\hat{d}$ near $\xi$ to control the degree of elastic response required.

As discussed in Section\ \ref{sec:lower-bound} simulating these examples produces much more degenerate contact pairs and requires higher precision to capture thickness offsets and so makes great demands on CCD accuracy. Here the conservative CCD strategy from IPC\ \cite{Li2020IPC} is no longer sufficient. Employing standard floating point CCD routines can and will simply return $0$ TOI for many queries, and so incorrectly stop the IPC optimization progress altogether. Likewise alternative CCD methods with higher reported accuracies similarly fail in most such cases.

This motivates our new ACCD method which, as demonstrated below, remains robust and accurate -- obtaining stable, forward progress in IPC-optimization for even our most complex examples where all available CCD alternatives fail. At the same time ACCD continues to provide comparable and generally faster timings on easier examples where alternate CCD methods are able to succeed. 

For comparison with ACCD we test a range of CCD methods: Minimal Separation (MinSep) CCD\ \cite{harmon2011interference,lu2019scalable}, Root Parity\ \cite{brochu2012efficient}, BSC\ \cite{tang2014fast}, T-BSC\ \cite{wang2015tightccd} and IPC's conservative floating point CCD\cite{Li2020IPC} on a set of ten benchmark examples with challenging codimensional collisions. The first five examples in the benchmark have thickness offsets and the remaining five do not; see Figure\ \ref{fig:CCDComp}.

We are able to directly test MinSep by applying it as a base solver within the same conservative CCD strategy as IPC\ \cite{Li2020IPC}. Here minimal separation is set to $s \times d_\text{cur}$ with the current scaling factor $s$ and distance $d_\text{cur}$. If the returned TOI is less than $10^{-6}$ the query is performed again without scaling ($s=0$) in order to attempt to make the query easier.
To test the Root Parity, BSC and T-BSC methods, which only decide whether an interval "has collision" or "has no collision" without computing TOI, we apply them with\del{in a monotonic bisection}\add{backtracking} for each query until finding a step size without collision. When found, we then return a conservative scaling of the step by $1-s$. As \del{bisection}\add{backtracking} cannot directly extend these methods to compute times or step sizes that \textit{first} bring a distance of a contact pair to $\xi$, 
we can not test these three methods on our examples with thickness offsets; and so restrict our testing of them to just the last five test examples in Figure\ \ref{fig:CCDComp}).

We summarize our comparisons in Figure\ \ref{fig:CCDComp}.
IPC's conservative strategy with floating-point CCD works well on simpler scenes, albeit with slightly slower timings than our ACCD method, but then fails on the even more complex ones.
We also note that while conservative floating point CCD handles degenerate cases like parallel edge-edge by querying additional point-edge and point-point pairs, ACCD only needs to query general pairs (e.g. point-triangle and edge-edge pairs for surface-only scenes).
On the other hand, MinSep always returns tiny values, sometimes even $0$, for all ten benchmark examples and so makes no progress.
Root Parity performs well with nice efficiency on most benchmark examples it can be applied to (those without thickness offsets) but unfortunately misses some collisions altogether leading to unacceptable interpenetrations in examples like Needle Bed and Garment.
For BSC we find it reports runtime errors for all benchmark examples it can be applied to with an internal message "Inflection points are not handled exactly in BSC", while T-BSC returns tiny values for the Garment example, and is unable to determine step sizes without collision (we confirm both BSC variants work well on simple cases) and so becomes trapped in line search in the other five benchmark examples it can be applied to. In contrast we see that ACCD enables all examples to complete to convergence while, for examples where alternate CCD routines are able to complete, it offers fastest times. Finally, as a preliminary investigation of ACCD application outside of the C-IPC framework we apply ACCD in place of ARGUS's default CCD routines and test with both 1mm and 5mm cloth stack examples. While, as expected, the simulation results are nearly identical (with no visual difference) CCD costs decrease by $\sim50\%$.

\subsection{Garment Simulation} 
\label{sec:garmentExp}

We next apply C-IPC to simulate garment drapes and dynamics. 
\begin{figure}[t]
    \centering
    \includegraphics[width=\linewidth]{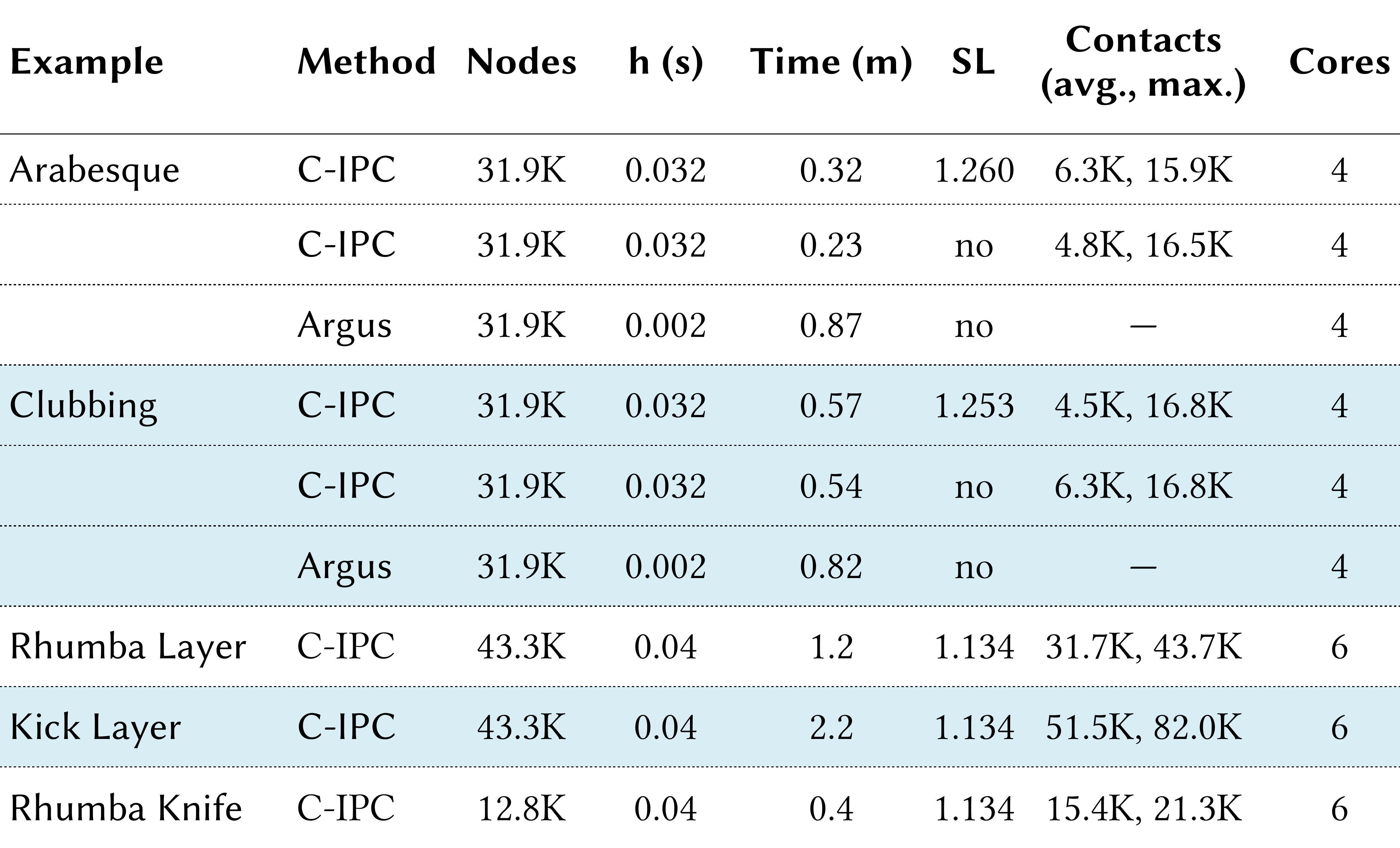}
    \caption{\textbf{Garment simulation \add{statistics} \del{tests}} \add{for}  \del{summary of garment simulation statistics} a pair of test dance sequences, Arabesque and Clubbing, from ARGUS and three additional tests with garment patterns not handled by ARGUS. Here {\bf SL} denotes the strain limit applied.}
    \label{fig:garmentTimingTB}
    \vspace{-0.2cm}
\end{figure}

\paragraph{ARGUS Comparisons} 
We start by considering C-IPC on a pair of test dance sequences, Arabesque and Clubbing, proposed for ARGUS\ \cite{li2018implicit}. Both examples use the highest resolution input meshes tested in Li et al.\ \shortcite{li2018implicit}. Here Clubbing exercises contact processing more extensively, with faster motions and transitions than Arabesque. For comparable results we run C-IPC and re-run ARGUS on the same machine, with fixed mesh resolution for ARGUS. Here we observe robust, smooth simulation results across the entire sequences for C-IPC both with and without strain limits. Notably, even with the addition of strict non-intersection enforcement, convergent, fully implicit time steps and resolving all contact stencils, not solely node-node as in ARGUS's treatment, we still observe 3.7X and 1.5X speedup for C-IPC on Arabesque and Clubbing over ARGUS results. Likewise, if we also enable C-IPC's constitutive strain-limiting, speedups remain similarly at 2.7X and 1.4X. See Figure\ \ref{fig:garmentTimingTB} for comparative timings and our supplemental video for results. 

\paragraph{Garment Simulation Challenges}

\begin{figure*}[t]
    \centering
    \includegraphics[width=\linewidth]{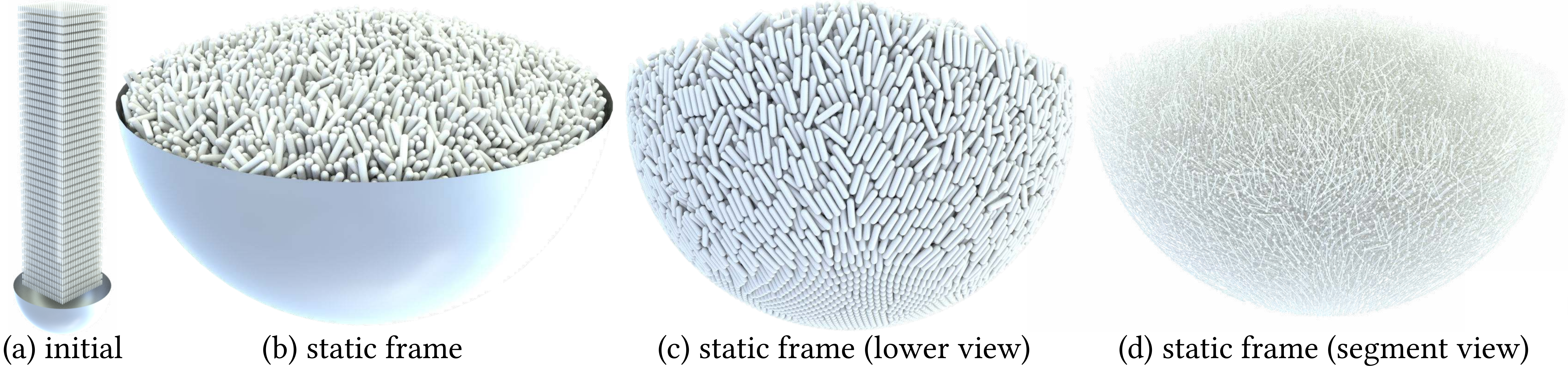}
    \caption{\textbf{Sprinkles.} We drop 25K $6mm$-long and $2mm$-thick ``sprinkles'' into a bowl (a). Setting IPC's offset and $\hat{d}$ to 1.5mm and 0.5mm respectively to model each sprinkle's thickness C-IPC fills the bowl's volume at rest (b). In (c) we remove the bowl to highlight the non-interpenetrating rest configuration while in (d) we remove volume rendering as well to show the underlying edge segments simulated.}
    \label{fig:sprinkles}
    \vspace{-0.2cm}
\end{figure*}

Two primary challenges in garment simulation are resolving complex, multilayered designs and simulating clothing worn by rapidly moving characters.
  To test the robustness of C-IPC for these cases we construct two new examples. Due to their complex stitching patterns these garments can not be simulated in the ARGUS and ARCSim codes.

  We simulate the yellow dress (15.7K nodes) produced by FoldSketch\ \cite{Li2018FoldSketch} with challenging-to-resolve knife pleats created where the cloth is folded upon itself and then stitched tight in contact. We start C-IPC with the flat pattern staged around the 12.8K-node mannequin (Figure\ \ref{fig:garments}a). We then drape by  stepping C-IPC forward with the stitching forces (springs) seaming the design together at large time step. C-IPC obtains the static equilibrium of the draped garment after just a few large time steps obtaining sharp knife pleats while preserving the correct layering order, free of intersection (Figure\ \ref{fig:garments}b). We then move the knife pleat through the steps of the Rhumba, preserving the flowing motion of the dress with knife pleats remain crisp and properly colliding throughout (Figure\ \ref{fig:garments}c). 

We then apply C-IPC to similarly initialize a multilayer skirt pattern (30.5K nodes) designed using Sensitive Couture\ \cite{umetani2011sensitive}. This obtains the initial pose static drape in Figure\ \ref{fig:garments}d. Next we script the mannequin mesh dressed in the layered garment through a fast-moving martial arts sequence\footnote{ We use a motion sequence from \del{a} Mixamo \url{https://www.mixamo.com/}, and apply Fang et al.\ \shortcite{fang2021gipc} to obtain an intersection-free mannequin sequence.} with large motions. 
In the bottom row of Figure\ \ref{fig:garments} we show three frames of the resulting C-IPC simulation featuring a portion of a rapid kicking sequence. Please also see our supplemental video. C-IPC stably resolves these rapid movements, continuing at a large time step of $h=0.04s$, while capturing intricate garment details, enforcing strain limits and non-intersection (and so preserving the layering order) throughout. Finally we take the same garment through the slower Rhumba sequence and observe a correspondingly reduced cost as less iterations are required to resolve the time steps with the gentler motion and correspondingly lower contact count.

\subsection{Benchmarks for Cloth Simulation}\label{sec:clothBenchmarkExp}
In this section we confirm C-IPC's performance on challenging benchmark tests from prior work.

\paragraph{Cloth on rotating sphere} 
In Figure \ref{fig:clothOnRotSphere} we test spinning sphere-type examples from  Bridson et al.\ \shortcite{bridson2002robust} designed to exercise friction, contact processing and tight wrinkling. We start by dropping an $86K$ node square cloth (unstructured mesh with $\mu=0.01$ and $1.0608$ strain upper bound) onto a sphere and ground (both $\mu=0.4$). As the sphere rotates, friction catches the cloth and fine wrinkling is captured as the cloth is pulled inward without locking and stretching artifacts. We confirm that as we increase resolution (to 246K nodes) and reduce bending stiffness (by $10\times$) higher frequency wrinkles resolve in the folds.

\paragraph{Funnel} 

To test layering, friction and tight contact with strain limiting we extend the funnel examples from Tang et al.\ \shortcite{tang2018cloth} and Harmon et al.\ \shortcite{harmon2008robust}. In Figure\ \ref{fig:funnel} we drop three cloth panels ($\mu=0.4$, 26K nodes each, strain upper bound at $1.0608$) upon a funnel. Under friction they rest stably on top until a sharp, scripted collision object (four-node tetrahedra) rapidly thrusts the panels down and through the funnel. C-IPC solve all time steps in the sequence to convergence while preserving an intersection free, strain-limit satisfying trajectory throughout.

\begin{figure*}[t]
    \centering
    \includegraphics[width=\linewidth]{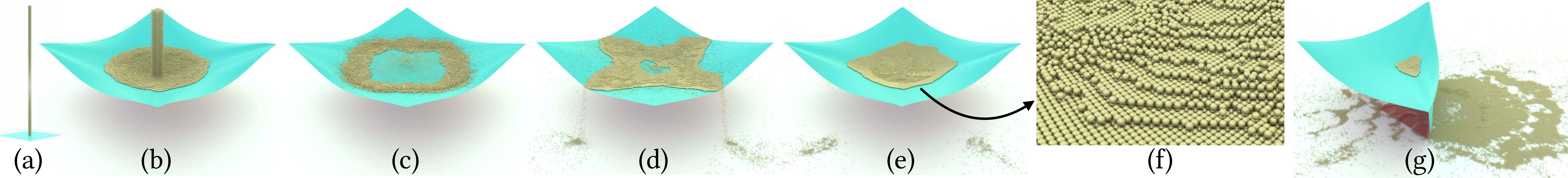}
    \caption{\textbf{Granules on cloth.}
    We drop a column of 50K particles onto a cloth with fixed corners. Each particle is represented by a single node, with C-IPC offset and $\hat{d}$ set to $2mm$ and $1mm$ respectively in order to capture the granule's volume at rest while the C-IPC barrier ensures particles do not tunnel through each other nor the cloth shell. Particles at rest capture tight spherical packings (see zoom in) confirming offsets are preserved throughout. Finally we release one corner of the cloth to let the granules fall onto the ground.}
    \label{fig:sandoncloth}
    \vspace{-0.2cm}
\end{figure*}

\paragraph{Reef knot}

  To combine extreme compression, contact and friction with large stresses we match the initial set up of the reef knot example from Harmon et al.\ \shortcite{harmon2009asynchronous}. Two ribbons are initially intertwined and then stretched to form a knot in Figure\ \ref{fig:ribbonKnot}. Here we extend the challenge of the original test by simulating with 100K nodes ($10\times$ that of the original), applying a strain limit of $1.134$, friction with $\mu=0.02$, and pulling much tighter -- stretching the ribbons to nearly their strain limit. This captures the frictional dynamics of the trajectory (without jitter) and results in extreme stress at the center, forming a much smaller knot than originally demonstrated.

\subsection{Shells, Rods, Particles and Coupling}
\label{sec:generalCoDim}
In this section we demonstrate C-IPC simulation across all codomains. 
This covers
volumetric FE materials, shells, rods and even, when desired, particles. 
Because these domains interact via IPC barriers we are able to seamlessly and directly couple all such mesh-based models in the same simulation. 
These tests also
highlight the advantages and suitability of our consistent and controllable thickness modeling for these thin material interactions.

\paragraph{Card shuffle}
We test accuracy and precision in very thin shell frictional contact during a complex, interleaving shuffle. We begin with fifty-four stiff shell playing cards ($63.5mm \times 88.8mm$) with IPC offset and $\hat{d}$ set to $0.1mm$ and $0.2mm$ respectively to resolve card thickness. In Figure\ \ref{fig:cardShuffle}a we break the cards into two piles and apply moving boundary conditions on their sides to bend them in preparation for a 
``precision'' bridge shuffle. Unlike a human-performed bridge finish, where cards begin interleaved, here we increase challenge by keeping the two bridged piles apart. We then release the held boundary conditions card-by-card from the bottom up, alternating between left and right side piles (Figure\ \ref{fig:cardShuffle}b). Released cards quickly regain flat rest shapes and fall on the ground to form a new, shuffled, intersection-free stacked pile. We then partially square the deck with two scripted collision boundaries (Figure\ \ref{fig:cardShuffle}c) obtaining a final pile 15.4 mm high and so well matching that of a real deck (Figure\ \ref{fig:cardShuffle}d).
\add{Here we applied $0.5\times$ the default tolerance to capture the intricate dynamics.}

\paragraph{Hairs}
Modeling hair with self-collision is a severe challenge for rod contact processing -- in part due to their extremely small cross-sections. We examine C-IPC's behavior on two challenging hair simulation tests. Hair thickness is small, but its bulk volumetric effects are important and have remained difficult to capture. Here C-IPC captures hair thickness by  enforcing an offset of $0.08mm$ for all rods with $\hat{d}=0.1mm$.
In Figure\ \ref{fig:hairs}(a)-(e) we first test C-IPC with braiding. We form a tight braid by twisting the bottom DOFs of two clusters of hair about each other. Despite undergoing large stress the resulting braid remains intersection free and preserve the thickness offset throughout. We also confirm this in dynamics by releasing the bottom held boundary conditions and letting the braid unwind without snagging artifacts nor instabilities. 
Next, we test hair self-collision and friction further by dropping a cluster of hairs with one free end across a similar perpendicularly hanging cluster\ \cite{mcadams2009detail} and over a fixed sphere  in Figure\ \ref{fig:hairs}(f)-(k).
C-IPC again simulates this scene with convergent, large time step solves throughout.

\paragraph{Noodles}
A large bowl of noodles is delicious but also challenging to simulate due to the large numbers of DOF and high contact counts. 
In Figure\ \ref{fig:noodles}a we drop a stack of 625 
$40cm$-long cooked noodles into a $13.85cm$-diameter bowl (fixed geometry). Each noodle is a 200-segment discrete rod with $\mu = 0.4$ and offset and $\hat{d}$ set to $1mm$ and $0.5mm$ respectively to model the noodles' $1.5mm$ thickness.
With C-IPC's thickness handling the dropped rods come to a rest (their total volume should be $4.42\times10^{-4}m^3$) largely filling the bowl (Figure\ \ref{fig:noodles}b) whose volume is $6.96\times10^{-4}m^3$. At equilibrium  In Figure\ \ref{fig:noodles}c we remove the bowl to show the final intertwined, interpenetration-free configuration. Finally in Figure\ \ref{fig:noodles}d  we remove the volume render to highlight that this complex non-intersecting volume is simulated solely with discrete rod polylines. 

\paragraph{Sprinkles}
Similarly we can do the same for just edge segments. We drop a grid of 25K ``sprinkles'' each formed by $6mm$-long, single edge segment into a bowl (Figure\ \ref{fig:sprinkles}a). The sprinkles' $2mm$ thickness are modeled by a C-IPC offset of $1.5mm$ and $\hat{d}$ of $0.5mm$. 
Making the offset $3\times$ larger than $\hat{d}$ models a strongly inelastic response for their collisions. As with the noodles we can estimate the total expected volume of the sprinkles as $5.76\times 10^{-4}m^3$ which at equilibrium fill the bowl's $6.88\times 10^{-4}m^3$ volume without interpenetration (Figure\ \ref{fig:sprinkles}b and c). Again in Figure\ \ref{fig:sprinkles}d we remove the volume rendering to highlight the geometry is simulated solely with codimensional line segments.
\add{For both the noodles and sprinkles we render the fitted cylindrical surface $0.2mm$ thinner than $\xi+\hat{d}$.}

\paragraph{Spherical Granules on cloth}
In the extreme C-IPC can even simulate small spherical granules by treating individual vertices as particles endowed with a strict thickness offset.
We test resolution of non-intersecting particle trajectories by dropping a  column of 50K particles onto a cloth with fixed corners (Figure\ \ref{fig:sandoncloth}). Collisions remain non-intersecting even at high-speed and with the cloth. Particles at rest capture tight spherical packings (see zoom in) confirming offsets are preserved. Once particles pile in the center we release a cloth  corner letting them flow to the ground. To model each particle we compute mass with $\rho=1600kg/m^3$ and $V=4/3\pi r^3 \approx 4.2\times10^{-9}m^3$, and set thickness with a C-IPC offset of $2mm$ and $\hat{d}$ of $1mm$. 
Simulating particles this way then resembles the Discrete Element Method (DEM)\ \cite{cundall1979discrete,yue2018hybrid} where particles are ``coated'' with an interaction zone (analogously $\xi < d < \xi+\hat{d}$ in our setting) for contact forces determination based on the depth, area, or volume of the zones' intersections\ \cite{jiang2020hybrid}.
DEM is a critical tool in geomechanics where it provides a reasonable simplification by ignoring particle deformation while accurately capturing momentum and inter-particle contact forces. 
Here C-IPC relates contact forces directly to distance via our barrier function and so enables seamless coupling with deformable materials as in the cloth here. Unlike DEM,  C-IPC then also additionally guarantees intersection-free trajectories for particle volumes defined by offsets.  

\begin{figure*}[t]
    \centering
    \includegraphics[width=\linewidth]{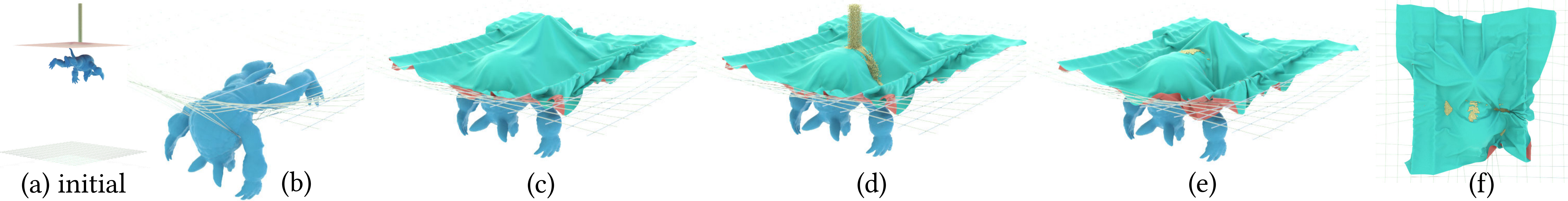}
    \caption{\textbf{All-in.} We simulate objects of all codimensions coupled together via C-IPC contact in a scene with an armadillo volume, a shell cloth, and a \del{sand} column formed by individual particles released in sequence to fall onto a net formed by interleaved rods.}
    \label{fig:allIn}
    \vspace{-0.2cm}
\end{figure*}

\paragraph{All-in}
Finally we test coupling all codimension together.
We drop an FE armadillo volume, a shell cloth and a particle \del{sand} column onto a net formed by interleaved rods (Figure\ \ref{fig:allIn}a), applying a C-IPC offset of $0.5mm$ and $\hat{d}$ of $1mm$ for all domains.
We successively drop these codomains one at a time. 
First the dropped armadillo is tangled and caught between the net's rods (Figure\ \ref{fig:allIn}b). Next the cloth falls upon the armadillo (Figure\ \ref{fig:allIn}c) and finally the \del{sand} column pours onto cloth and flows into the two resulting folds of cloth supported by the armadillo and the rod net (Figure\ \ref{fig:allIn}d, e, and f (top view)).

\subsection{New Stress Tests for Cloth Simulation}
\label{sec:clothStressTestExp}

Finally we propose and evaluate C-IPC on a set of three new, challenging cloth simulation examples designed to require extreme robustness and accuracy in modeling elastodynamics with frictional contact, strain limiting and thickness. To our knowledge C-IPC is the first method to achieve simulations of these scenarios and behaviors.

\paragraph{Pulling cloth over codimensional needles} Cloth contacts against sharp geometries notoriously stress contact simulations with snagging and sedropping and then dragging a cloth panel across a needle bed formed by line segments (Figure\ \ref{fig:needlebedpull}). With a large time step size of $h=0.02s$ and strain limit of $1.0608$, the cloth comes safely to a static rest on the bed of needles without jittering. We pull the left side of the cloth at $1m/s$ across the segment tips and observe rapid sliding along and then off the top of the bed of needles without snagging nor stretching artifacts (strain limits are satisfied throughout).\del{vere stretching artifacts.} Likewise, if these contact are sliding across asperities with large displacement (i.e., large time step and/or large speed) these challenges only grow.
Here we test C-IPC's handling of extreme cases with both challenges, by In cases like these, were we have extreme stretch from sharp contacts, C-IPC's adaptive strain-limiting stiffness is especially important to ensure that time step solves remain numerically feasible and efficient.

\paragraph{Tablecloth trick}
The classic tablecloth trick attempts to pull a cloth out from under a table setting with the aim of keeping the dinnerware upright and ideally staying mostly in place.  
Key to making this trick work is a pull with sufficient speed so that sliding acceleration of the cloth can overcome friction forces with little or no pull on the settings\footnote{\url{www.juggle.org/table-cloth-pull-trick}}. 
Simulation of this effect then requires tight coupling of elasticity and strain limiting with accurate resolution of friction and contact as well as robust handling of sliding contact with sharp edges. This is needed to correctly model the balances between friction, the large stresses and forces from pulling, and the normal forces from  heavy dinnerware.  
In Figure\ \ref{fig:teaser} we set a table for C-IPC with both heavy 
and stiff dinnerware (volumetric FE models). We simulate pulling the strain-limited ($1.0608$ upper bound) tablecloth, time-stepped at $h=0.01s$, with increasing speeds. As we start with lower pulling speeds, as expected, objects do not follow the cloth. As we increase speed we then get less messy results and finally, with a pull at $4m/s$, the full setting remains upright and on the table with just a bit of movement (we also note detailed folds begin to be formed by the strain-limited cloth pulled at high speed). Finally, with a high-speed pull of $8m/s$ the tablecloth pulls out smoothly with almost no change to the setting at all. Here, for the latter two successful pulls we observe extreme forces from the moving boundary conditions pulling the cloth at a high speed and the large friction forces exerted by contact with the heavier dinnerware.

\paragraph{Twisted cylinder.}
Finally, we test thickness modeling, contact resolution and buckling under extreme and increasing stress.
We start with a $1m$-wide cloth cylinder ($0.25m$ radius) modeled by a shell with 88K nodes, with a thickness offset of $1.5mm$ and $\hat{d}$ of $1mm$, time-stepped at $h=0.04s$.
We then simultaneously twist the cylinder and slowly move the two sides together at $72^\circ/s$ and $5mm/s$ respectively to introduce wrinkling, folding, and eventual tight buckling. In order to allow this scripted torture to continue for the entire 38 seconds length of this sequence we do not apply strain limiting -- so C-IPC must resolve exceedingly high stretching forces here. Likewise, to clarify buckling behavior we do not apply gravity (avoiding droop).
In the first second of simulation we immediately obtain interesting global folds (\del{Figure\ \ref{fig:twistCylinder}a}\add{see our supplemental document and video}). Soon after a thick central cylinder of wound cloth forms. The volume of this cylinder's structure is supported by C-IPC's finite thickness offset (Figure\ \ref{fig:twistCylinder}\del{b} \add{upper left}). For contrast consider in 
Figure\ \ref{fig:twistCylinder}\del{c} \add{upper right} a frame captured at the same time step but now simulated \emph{without} C-IPC's offset. Here we clearly see that correct geometry \emph{can not} be formed without C-IPC's thickness offset (nor can it capture the material's later buckling behavior -- see below) further clarifying the importance of consistent thickness modeling for codimensional models.
Next, \del{in Figure\ \ref{fig:twistCylinder}d} as we continue our C-IPC simulation \emph{with offset} for $32.96s$ of further twisting, our offset-thickened cloth continues to support complex behaviors including the final buckled geometry in Figure\ \ref{fig:twistCylinder}\del{e} \add{bottom}.

\begin{figure*}[t]
    \centering
    \includegraphics[width=0.9\linewidth]{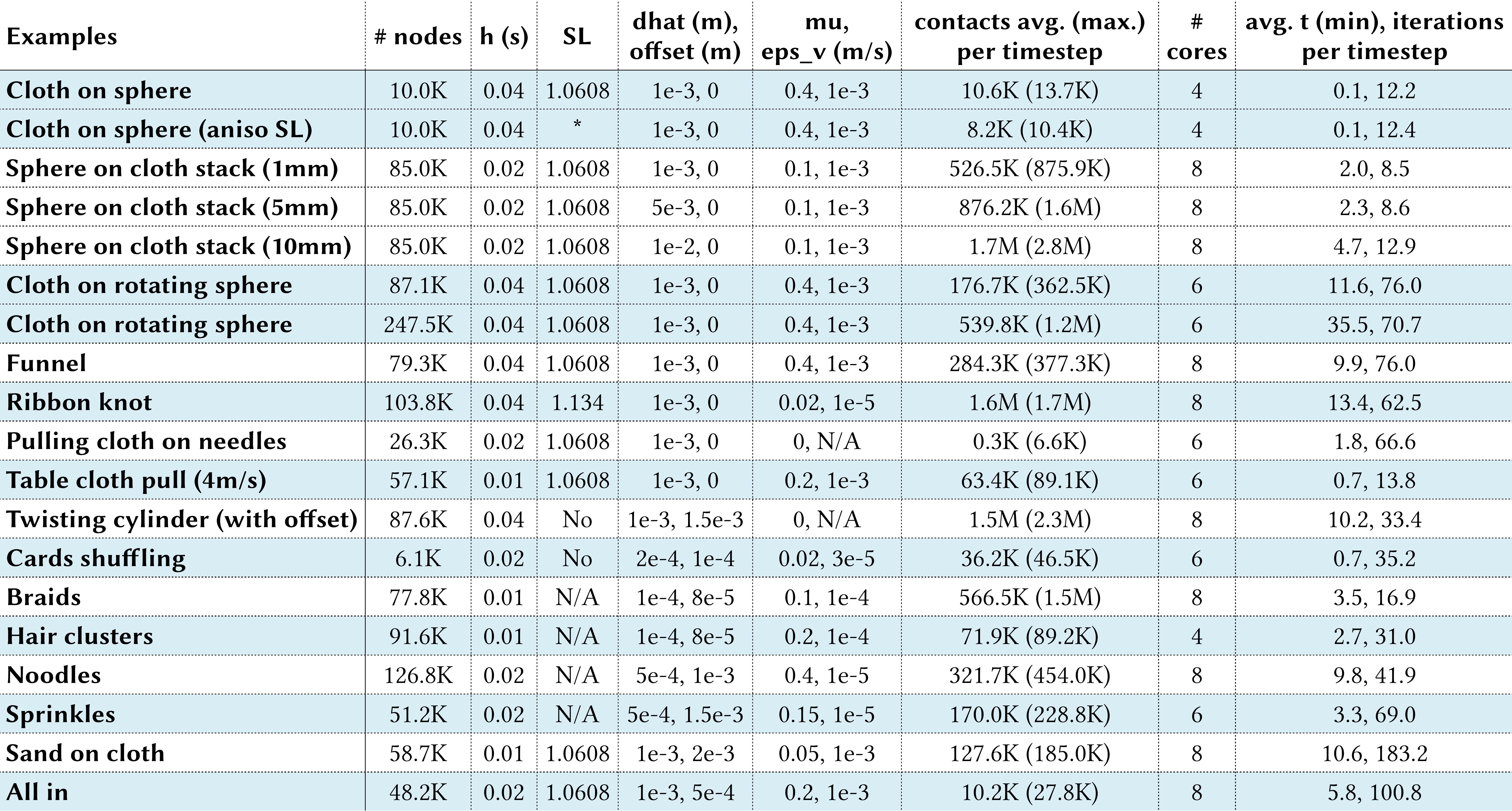}
    \caption{\textbf{Simulation Statistics} for IPC on a subset of our benchmark examples. Remaining garment examples are summarized in Figure\ \ref{fig:garmentTimingTB}. For each simulation we report geometry, time step, strain-limit enforced (SL), thickness settings ($\hat{d}$ and offset), friction parameters ($\mu$ and accuracy $\epsilon_v$), number of contacts processed per time step and machine, as well as average timing and number of Newton iterations per time step solve.  (*The tested anisotropic model has strain limits at $0.14$ for both diagonal entries and $0.063$ for the off-diagonal entry of the Green-Lagrangian strain\ \cite{clyde2017numerical}.)}
    \label{fig:timingTB}
    \vspace{-0.2cm}
\end{figure*}
\section{Conclusion}
C-IPC extends the incremental potential contact framework from volumetric deformables~\cite{Li2020IPC} to codimensional and mixed-dimensional structures with controllable thickness, fully coupled strain limiting, accurate frictional behavior and non-intersection maintained throughout all time steps. Of course as in the original IPC method, in order to provide these non-intersection guarantees IPC requires an initial non-interpenetrating geometry at start of simulation. Currently, e.g. for garments, this can often be easily achieved by staging. However, loosening these constraints to achieve guarantees from a starting tangled state is an important and  exciting direction of future exploration.

\add{As demonstrated in our stress tests, C-IPC consistently converges to optimal solutions even when subject to challenging boundary conditions that push against strain limits and contact barriers. However, (C-)IPC can not progress with boundary conditions that impose infeasible (e.g., intersecting or strain-limit violating) trajectories. Here detecting and/or finding means for IPC to progress through infeasible conditions would be a challenging future extension. 
For friction, C-IPC directly utilizes the smoothed, semi-implicit discretization of maximal dissipation from IPC with the option of lagged iterations. As validated in the original IPC work, we confirm that C-IPC accurately captures correct stick-to-slip thresholds with convergent lagging (see our cloth sliding benchmark). Otherwise, for C-IPC simulation we apply a single lagged iteration and find it sufficient to model complex frictional behaviors (see e.g. the tablecloth pull and rotating sphere) and detailed garment drapes. However, as with the original IPC work (and prior nonsmooth methods) we also note that there is no guarantee of convergence to accurate satisfaction of fully implicit friction relations. An algorithm that achieves this remains an open and important challenge for simulation.}

Along with enabling new and improved applications in cloth, hair and many other codimensional simulation tasks we are also excited to extend C-IPC further for geomechanics applications in the simulation of granular flows -- especially for complex granule shapes which have so far proven challenging to resolve. 

Here we have first focused on providing a reliable simulator for all materials and conditions. As noted in previous works, in many applications the biggest overhead is the many laborious simulation passes required to hand-tune collision parameters to make a scene work at all, and then often many more time-consuming steps to still correct output.
As such, reliability leads to overall increased speed of output. However, although performance for C-IPC, as analyzed above, is more than competitive with state of the art cloth codes when they can offer comparable accuracies, much more can be done to improve C-IPC performance. We look forward to both better leveraging parallel architectures and improving underlying IPC solver methods. Similarly, here we are investigating and extensively testing with \emph{implicit} methods. Another fruitful direction of exploration is \emph{explicit} time stepping. It would be exciting to consider relative behavior of C-IPC and  ACM\ \cite{harmon2009asynchronous} for side-by-side comparison between state of the art implicit and explicit methods providing non-intersection guarantees.\footnote{Towards this end we have been communicating with the ACM authors for over year towards obtaining a representative code for comparison. After hard work on both ends we are hopeful, but licensing concerns required for access to the code have hampered us so far. We hope to have the comparison in future work.} Likewise, we look forward to leveraging the differentiability of our approach. Here we expect that the combined benefits of C-IPC's reliable performance across sweeps of input and its differentiability, should together enable many design and training tasks for garments and complex structures \add{as demonstrated in Macklin et al.\ \shortcite{macklin2020primal} and Geilinger et al.\ \shortcite{geilinger2020add}}. 
Finally, here we have focused on providing strain-limiting solely with upper bounds and while we have not yet seen cases requiring lower bound limits these are easy to apply and so should be interesting for future exploration. 

While proven critical for C-IPC we have also shown that ACCD is an exceedingly simple replacement for pre-existing complex, and often sensitive CCD modules. Here in preliminary testing we have found that in application outside of C-IPC framework ACCD can also provide significant speed-up. We look forward to further applications of ACCD as an extremely easy-to-implement and efficient alternative CCD algorithm for tasks in both simulation and geometry processing.

In summary, independent from elasticity model and time step size, C-IPC guarantees out-of-the-box intersection-free simulation with a strict satisfaction of strain limits (confirmed down to $0.1\%$) and accurately captures geometrically meaningful thickness. This is enabled, in part, by a new, simple, additive CCD (ACCD) method with highly stable and accurate output for challenging, first-time-of-impact tasks. 
\begin{acks}
We thank Jacky (Jiecong) Lu for set-up of ribbon and garment examples; 
Jie Li, Florence Descoubes, and Laurence Boissieux for assistance with ARGUS;
and Rasmus Tamstorf and Etienne Vouga for valuable discussion.
This work was supported in part by \grantsponsor{_NSF}{NSF}{https://nsf.gov/} \grantnum{_NSF}{CAREER IIS-1943199}, 
\grantnum{_NSF}{CCF-1813624}, and \grantnum{_NSF}{ECCS-2023780}.
\end{acks}

\bibliographystyle{ACM-Reference-Format}
\bibliography{main}



\section*{Supplementary material}

# Supplement to Codimensional Incremental Potential Contact

MINCHEN LI, University of California, Los Angeles & University of Pennsylvania & Adobe Research
DANNY M. KAUFMAN, Adobe Research
CHENFANFU JIANG, University of California, Los Angeles & University of Pennsylvania

## Contents

## 1 VOLUMETRIC SHELL V.S. DISCRETE SHELL

Linear volumetric elements could be used as an alternative to simulate thin materials with geometric thickness. However, as their bending is realized by shearing deformation, a shearing penalty serves as an extra source of bending stiffness, making their bending behavior artificially stiffened.

Here we use the cylinderical cloth twisting example to demonstrate this artifact in a side-by-side comparison between the codimensional twist, simulated with C-IPC a volumetric twist simulated by IPC [Li et al. 2020]. Following IPC we generate a volumetric mat mesh by extruding the codimensional cylinder surface (87.6K nodes) in both outward and inward radial directions and then sealing the circular gaps on the two sides with a single layer of triangles to create a closed volumetric cylinder shell mesh. We then apply Tetgen [Si 2015] to tetrahedralize this volume (without additional Steiner

Authors' addresses: Minchen Li, University of California, Los Angeles & University of Pennsylvania & Adobe Research, minchernl@gmail.com; Danny M. Kaufman, Adobe Research, danny.kaufman.cs@gmail.com; Chenfanfu Jiang, University of California, Los Angeles & University of Pennsylvania, chenfanfu.jiang@gmail.com.

points) resulting in a 175.2K-node tetrahedral mesh. Here extrusion distances are both $0.75mm$, making the thickness of the volumetric shell $1.5mm$, to match the contact offset $\xi$ used in our codimensional version. However, as thickness values used to compute volume integrals for mass and elasticity in the codimensional version utilize a cotton material ($0.318mm$), we apply the same Poisson's ratio but set the density and Young's modulus in the volumetric version to $0.318/1.5 = 0.212\times$ that of the codimensional version to unify material properties and so make the two models comparable.

Applying identical boundary conditions, the volumetric simulation clearly obtains artificially stiffened bending, with smoother and less frequent wrinkles, and noticeable gaps between the touching layers. See Figure 1, top for the volumetric results and bottom for the comparable C-IPC version. As there are $2\times$ as many nodal DOF in the volumetric model than in C-IPC, the volumetric simulation is also, likewise, more expensive. For the first 200 time steps, the C-IPC simulation, per time step took 1.7min and 9.5 iterations on average, while the volumetric IPC simulation took 3.4min and 13.1 iterations. In addition, we observe that our C-IPC simulation also generates more contact pairs to process (848K at max) per time step than the volumetric simulation (474K at max) due to the finer wrinkling behavior obtained.

## 2 MEMBRANE LOCKING W.R.T. RESOLUTION AND BENDING STIFFNESS

As discussed in our introduction, increasing resolution can alleviate membrane locking. However, shell models can continue to suffer severe numerical stiffening in bending modes as we vary cloth materials even, as we show here, at high resolution.

To investigate resolution dependence in membrane locking, we test the cloth drape on the static sphere with mesh resolutions varying from 8K up to 246K nodes. We first simulate these drapes with the default $1\times$ (reported cotton) membrane stiffness *without* strain limiting. In Figure 2 we see that, for this material, as resolution increases, membrane-locking artifacts becomes less and less severe. Specifically, at higher resolutions with 85K and 246K nodes, locking artifacts are negligible except for sharp turns on some wrinkle profiles.

However, the severity of membrane locking behaviors vary with both resolution *and* material. Here, membrane locking is essentially an artificial stiffening of bending by an additional bending penalty sourced from the membrane energy. If (as in the above example) the original bending to be modeled is already large, membrane locking artifacts can more easily be reduced with increased resolution. On the other hand, if we simulate the same example set up as above, except now with a $0.01\times$ bending stiffness (as applied in Figure 6, bottom of our primary text), increasing resolution is not nearly as helpful. For example, now even at a 246K node resolution, we

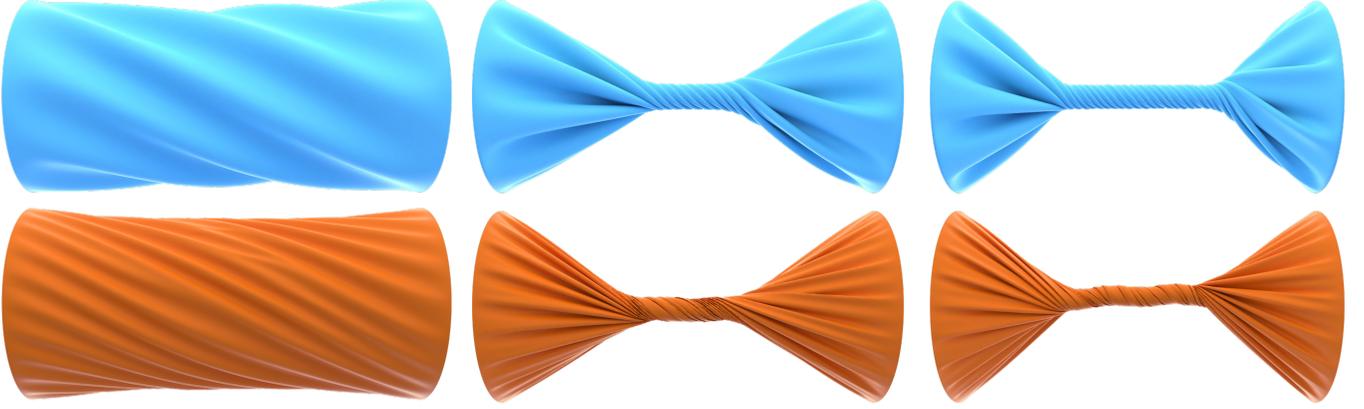

Fig. 1. **Volumetric Shell v.s. Discrete Shell.** From left to right, we compare the 5th, 100th, and 200th time step result between a volumetric version (top) and our codimensional version (bottom) of the cylinder cloth twisting example with the same material properties and boundary conditions. The smoother and lower frequency wrinkles, and the larger gaps between the touching layers in the volumetric results clearly demonstrate the artificially stiffened bending behavior.

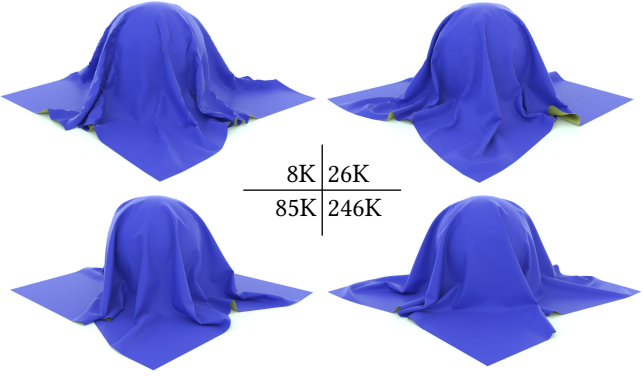

Fig. 2. **Membrane Locking w.r.t. Resolution.** As resolution increases, here membrane locking-artifacts in the cloth on sphere example, simulated using 1× the smallest membrane stiffness *without strain limiting*, becomes less severe and eventually nearly negligible at 85K nodes and 246K nodes, except for sharp turns on some of the wrinkle profiles.

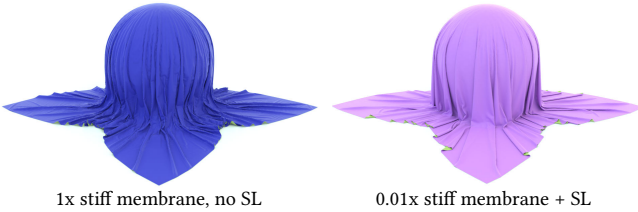

Fig. 3. **Membrane Locking under Small Bending.** With 0.01× bending stiffness, membrane-locking artifacts remain severe even at a resolution of 246K nodes when simulated with 1× the smallest membrane stiffness *without* strain limiting (left). Compare to our C-IPC result with strain limiting (right).

still see obvious membrane-locking artifacts; see Figure 3, left and compare with C-IPC's strain-limited result in Figure 3, right.

## 3 ISOTROPIC STRAIN LIMITING DERIVATIVE DERIVATIONS

We require derivatives of the singular value decomposition (SVD):

$$\frac{\partial S_{ii}}{\partial F} = u_i v_i^T, \quad \frac{\partial^2 S_{ii}}{\partial F^2} = \frac{\partial (u_i v_i^T)}{\partial F} = \frac{\partial u_i}{\partial F} v_i^T + u_i \frac{\partial v_i}{\partial F}^T \quad (1)$$

where $\frac{\partial u_i}{\partial F}$ and $\frac{\partial v_i}{\partial F}$ can be found in Xu et al. [2015].

Then if we denote the barrier $b_{t,i}$ for $i$-th singular value of triangle $t$, we have

$$\begin{aligned}\frac{\partial b_{t,i}}{\partial x} &= \frac{\partial b_{t,i}}{\partial S_{t,ii}} \frac{\partial S_{t,ii}}{\partial F_t} \frac{\partial F_t}{\partial x} \\ \frac{\partial^2 b_{t,i}}{\partial x^2} &= \frac{\partial S_{t,ii}}{\partial x}^T \frac{\partial^2 b_{t,i}}{\partial S_{t,ii}^2} \frac{\partial S_{t,ii}}{\partial x} + \frac{\partial b_{t,i}}{\partial S_{t,ii}} \frac{\partial^2 S_{t,ii}}{\partial x^2}\end{aligned} \quad (2)$$

where $\frac{\partial F_t}{\partial x}$ can be found in anisotropic strain limiting implementation detail section.

For the barrier derivatives, we have

$$\begin{aligned}\frac{\partial b_{t,i}}{\partial S_{t,ii}} &= \frac{\kappa_s}{(s-\hat{s})^2}(2(\hat{s}-S_{t,ii})\ln(\frac{s-S_{t,ii}}{s-\hat{s}}) + \frac{(\hat{s}-S_{t,ii})^2}{s-S_{t,ii}}) \\ \frac{\partial^2 b_{t,i}}{\partial S_{t,ii}^2} &= \frac{\kappa_s}{(s-\hat{s})^2}(-2\ln(\frac{s-S_{t,ii}}{s-\hat{s}}) - 4\frac{\hat{s}-S_{t,ii}}{s-S_{t,ii}} + (\frac{\hat{s}-S_{t,ii}}{s-S_{t,ii}})^2)\end{aligned}, \quad (3)$$

then together with

$$\frac{\partial S_{t,ii}}{\partial x} = \frac{\partial S_{t,ii}}{\partial F_t}\frac{\partial F_t}{\partial x} \quad \text{and} \quad \frac{\partial^2 S_{t,ii}}{\partial x^2} = \frac{\partial F_t}{\partial x}^T \frac{\partial^2 S_{t,ii}}{\partial F_t^2}\frac{\partial F_t}{\partial x}, \quad (4)$$

we can efficiently compute

$$\begin{aligned}\frac{\partial b_t}{\partial x} &= \sum_i (\frac{\partial b_{t,i}}{\partial S_{t,ii}} \frac{\partial S_{t,ii}}{\partial F_t}) \frac{\partial F_t}{\partial x}, \\ \frac{\partial^2 b_{t,i}}{\partial x^2} &= \frac{\partial F_t}{\partial x}^T \sum_i (\frac{\partial^2 b_{t,i}}{\partial F_t^2}) \frac{\partial F_t}{\partial x}\end{aligned} \quad (5)$$

where $\frac{\partial^2 b_{t,i}}{\partial F_t^2} = \frac{\partial S_{t,ii}}{\partial F_t}^T \frac{\partial^2 b_{t,i}}{\partial S_{t,ii}^2} \frac{\partial S_{t,ii}}{\partial F_t} + \frac{\partial b_{t,i}}{\partial S_{t,ii}} \frac{\partial^2 S_{t,ii}}{\partial F_t^2}.$

For SPD projection we can then process the 6x6 matrix $\sum_i (\frac{\partial^2 b_{t,i}}{\partial F_t^2})$.

## 4 ANISOTROPIC STRAIN LIMITING DERIVATIONS

### 4.1 Quadratically Approximating Data-Driven Model

The 1st-order derivatives of $\psi$ can be computed as

$$\begin{aligned}
\frac{\partial \psi}{\partial \tilde{E}_{11}} &= a_{11}\eta_1'(\tilde{E}_{11}^2)\tilde{E}_{11} + a_{12}\eta_2'(\tilde{E}_{11}\tilde{E}_{22})\tilde{E}_{22}, \\
\frac{\partial \psi}{\partial \tilde{E}_{22}} &= a_{22}\eta_3'(\tilde{E}_{22}^2)\tilde{E}_{22} + a_{12}\eta_2'(\tilde{E}_{11}\tilde{E}_{22})\tilde{E}_{11}, \\
\frac{\partial \psi}{\partial \tilde{E}_{12}} &= 2G_{12}\eta_4'(\tilde{E}_{12}^2)\tilde{E}_{12}.
\end{aligned} \quad (6)$$

When $\tilde{E} = 0$ the element is in rest shape, and all gradients above are equal to 0.

For the 2nd-order derivatives we have

$$\begin{aligned}
\frac{\partial^2 \psi}{\partial \tilde{E}_{11}^2} &= 2a_{11}\eta_1''(\tilde{E}_{11}^2)\tilde{E}_{11}^2 + a_{11}\eta_1'(\tilde{E}_{11}^2) + a_{12}\eta_2''(\tilde{E}_{11}\tilde{E}_{22})\tilde{E}_{22}^2, \\
\frac{\partial^2 \psi}{\partial \tilde{E}_{11}\partial \tilde{E}_{22}} &= \frac{\partial^2 \psi}{\partial \tilde{E}_{22}\partial \tilde{E}_{11}} = a_{12}\eta_2''(\tilde{E}_{11}\tilde{E}_{22})\tilde{E}_{22}\tilde{E}_{11} + a_{12}\eta_2'(\tilde{E}_{11}\tilde{E}_{22}), \\
\frac{\partial^2 \psi}{\partial \tilde{E}_{22}^2} &= 2a_{22}\eta_3''(\tilde{E}_{22}^2)\tilde{E}_{22}^2 + a_{22}\eta_3'(\tilde{E}_{22}^2) + a_{12}\eta_2''(\tilde{E}_{11}\tilde{E}_{22})\tilde{E}_{11}^2, \\
\frac{\partial^2 \psi}{\partial \tilde{E}_{12}^2} &= 4G_{12}\eta_4''(\tilde{E}_{12}^2)\tilde{E}_{12}^2 + 2G_{12}\eta_4'(\tilde{E}_{12}^2),
\end{aligned} \quad (7)$$

and all other terms are 0. When $\tilde{E} = 0$, we have

$$\frac{\partial^2 \psi}{\partial \tilde{E}_{11}^2} = a_{11}, \quad \frac{\partial^2 \psi}{\partial \tilde{E}_{22}^2} = a_{22}, \quad \frac{\partial^2 \psi}{\partial \tilde{E}_{12}^2} = 2G_{12}$$

$$\frac{\partial^2 \psi}{\partial \tilde{E}_{11}\partial \tilde{E}_{22}} = \frac{\partial^2 \psi}{\partial \tilde{E}_{22}\partial \tilde{E}_{11}} = a_{12}, \quad (8)$$

all constants. Thus to quadratically approximate the data-driven model at $\tilde{E} = 0$, using our barrier model, we just simply need to use the same stiffnesses.

### 4.2 Implementation Details

For simplicity, if we align the cloth weft and warp directions to the coordinate axes, we can easily compute $\tilde{E}$ from the 3x2 deformation gradient $F = [x_2 - x_1, x_3 - x_1][X_2' - X_1', X_3' - X_1']^{-1}$ as

$$\tilde{E}_{ij} = \frac{1}{2}(F_{mi}F_{mj} - \delta_{ij}),$$

so we have

$$\begin{aligned}
\frac{\partial \tilde{E}_{ij}}{\partial F_{kl}} &= \frac{1}{2}(\delta_{il}F_{kj} + F_{ki}\delta_{jl}), \\
\frac{\partial^2 \tilde{E}_{ij}}{\partial F_{kl}\partial F_{mn}} &= \frac{1}{2}\delta_{mk}(\delta_{il}\delta_{nj} + \delta_{ni}\delta_{jl}).
\end{aligned} \quad (9)$$

Then together with $\partial \psi$ above and $dF/dx = \{\begin{bmatrix}-I \\ -I\end{bmatrix}\begin{bmatrix}I \\ 0\end{bmatrix}\begin{bmatrix}0 \\ I\end{bmatrix}\} : B$, with $B = [X_2' - X_1', X_3' - X_1']^{-1}$ we have

$$\frac{dF}{dx} = \begin{bmatrix}-(B_{11}+B_{21}) & & \\ & * & \\ & & * \\ -(B_{12}+B_{22}) & & \\ & * & \\ & & *\end{bmatrix}, \begin{bmatrix}B_{11} & & \\ & * & \\ & & * \\ B_{12} & & \\ & * & \\ & & *\end{bmatrix}, \begin{bmatrix}B_{21} & & \\ & * & \\ & & * \\ B_{22} & & \\ & * & \\ & & *\end{bmatrix}. \quad (10)$$

with $*$ indicating repetition of the top-left entry. We are then able to implement the constitutive model via

$$\frac{\partial \psi}{\partial x} = \frac{\partial \psi}{\partial \tilde{E}}\frac{\partial \tilde{E}}{\partial F}\frac{\partial F}{\partial x} \quad \text{and} \quad \frac{\partial^2 \psi}{\partial x^2} = (\frac{\partial F}{\partial x})^T\frac{\partial^2 \psi}{\partial F^2}\frac{\partial F}{\partial x} \quad (11)$$

where

$$\frac{\partial^2 \psi}{\partial F^2} = (\frac{\partial \tilde{E}}{\partial F})^T\frac{\partial^2 \psi}{\partial \tilde{E}^2}\frac{\partial \tilde{E}}{\partial F} + \frac{\partial \psi}{\partial \tilde{E}}\frac{\partial^2 \tilde{E}}{\partial F^2}.$$

## 5 ROBUST AND EFFICIENT IMPLEMENTATION OF IPC FRICTION

Computing the IPC friction potential, $f_0$ is well defined without division by 0.

Friction gradient/force computation requires a robust implementation that uses algebraically derived $\frac{f_1(||u_k||)}{||u_k||}$, which does not contain division by 0 in construction of $f_1$.

For the friction Hessian, the inner $2 \times 2$ matrix is

$$\frac{f_1'(||u_k||)||u_k|| - f_1(||u_k||)}{||u_k||^3}u_k u_k^T + \frac{f_1(||u_k||)}{||u_k||}I_2.$$

Here we need to first algebraically derive $\frac{f_1'(||u_k||)||u_k||-f_1(||u_k||)}{||u_k||^2}$ which does not contain division by 0, by construction, and note that $\frac{u_k u_k^T}{||u_k||}$ is always bounded and equals 0 when $||u_k|| = 0$, in which case, since $\frac{f_1(||u_k||)}{||u_k||} > 0$ always holds, there is no need for SPD projection.

Let $u_k = (x, y)$,

$$\begin{aligned}
&\frac{f_1'(||u_k||)||u_k|| - f_1(||u_k||)}{||u_k||^3}u_k u_k^T + \frac{f_1(||u_k||)}{||u_k||}I_2 \\
&= \frac{f_1'(||u_k||)}{||u_k||^2}u_k u_k^T + \frac{-f_1(||u_k||)}{||u_k||^3}\begin{bmatrix}x^2 & xy \\ xy & y^2\end{bmatrix} \\
&\quad + \frac{f_1(||u_k||)}{||u_k||^3}\begin{bmatrix}x^2+y^2 & 0 \\ 0 & x^2+y^2\end{bmatrix} \\
&= \frac{f_1'(||u_k||)}{||u_k||^2}u_k u_k^T + \frac{f_1(||u_k||)}{||u_k||^3}\begin{bmatrix}y^2 & -xy \\ -xy & x^2\end{bmatrix} \\
&= \frac{f_1'(||u_k||)}{||u_k||^2}u_k u_k^T + \frac{f_1(||u_k||)}{||u_k||^3}\bar{u}_k \bar{u}_k^T
\end{aligned} \quad (12)$$

where $\bar{u}_k = (-y, x)$. So for $||u_k|| \geq \epsilon_v h$ when we have $f_1'(\cdot) = 0$ and $f_1(\cdot) = 1$, the matrix is always SPD (we must compute using $\bar{u}_k$ as the original formula can still be not SPD due to numerical error of subtraction); otherwise we project the 2x2 matrix.

## 6 FRICTION VALIDATION ON CLOTH

We confirm that C-IPC inherits the same accurate friction modeling behavior from IPC. We place a $0.5m$-wide piece of square cloth ($8K$ nodes) on a $26.565°$-inclined slope (critical friction coefficient equals 0.5), and simulate it for $4s$ to test stick-slip behavior. For each time step we iteratively update the normal forces $\lambda$ and the tangent operators $T$ until convergence to a fully implicit solution. Here, even with our default parameter settings ($\hat{d} = 10^{-3}m$, $\epsilon_v = 10^{-3}m/s$, $h = 0.04s$), C-IPC obtains stiction with the critical friction coefficient $\mu = 0.5$, and obtains sliding even at $\mu = 0.49$ (99.8% of the critical coefficient), and so well-matches real-world behavior.

## 7 TERMINATION CRITERIA

Following IPC [Li et al. 2020], we set Newton iteration stopping tolerance $\epsilon_d$ on the Newton increment (search direction $p$) divided by time step size $h$. This has unit of velocity ($m/s$). Monitoring the scaled Newton increment helps to conveniently generalize convergence across complex scenes containing objects with drastically differing density, dimension, and stiffness scales. However, we observe that the residual-iteration curve when directly using infinity norm [Li et al. 2020] of the scaled Newton increment is non-smooth and oscillatory (although with an appropriately diminishing trend). This makes it nontrivial to be used for early termination as especially used in graphics applications. Therefore, in C-IPC we apply a normalized $L^2$ norm of the scaled Newton increment as the residual measurement:

$$r = \frac{1}{h}\sqrt{\frac{||p||^2}{(n - n_k)}} \quad (13)$$

where $n$ is the total number of nodes in the scene and $n_k$ is the number of kinematic nodes. Here $r$ still maintains units of velocity as a measurement for convergence but with a much smoother residual-iteration curve. In addition, we require a sequence of $l$ iterations all satisfying $r < \epsilon_d$ to terminate the optimization to further facilitate a robust early termination strategy.

In our experiments we use default $l = 3$ and $\epsilon_d = 10^{-3}m/s$ for all examples, except for card shuffling where we apply $\epsilon_d = 5 \times 10^{-4}m/s$ to capture the intricate dynamics.

## 8 ROD BENDING MODULUS

In discrete rods model [Bergou et al. 2008], the bending energy is directly integrated over the length of the rod as

$$E_{\text{bend}}(x) = \sum_{i=1}^{n} \frac{\alpha(\kappa b_i)^2}{\bar{l}_i} \quad (14)$$

Here $\kappa b_i$ is the curvature binormal at a vertex, $\bar{l}_i = |\bar{e}^{i-1}| + |\bar{e}^i|$ where $|\bar{e}^i|$ is the rest length of edge $i$, and $\alpha$ is an adjustable parameter to control the stiffness of bending. As we know when a rod becomes thicker, it is harder to bend, which means we need to manually increase $\alpha$ accordingly and this makes setting up examples inconvenient.

According to Kirchoff rod theory [Dill 1992], the bending energy of rod at a unit length is defined as

$$\frac{1}{2}Er^4\frac{\pi}{4}\kappa^2$$

where $E$ is the bending modulus in $Pa$ unit, $r$ is the radius of the rod, and $\kappa = \frac{\kappa b}{\bar{l}/2}$ is the point-wise curvature [Bergou et al. 2008]. Integrating this quantity over rod length then gives us

$$\begin{aligned} E_{\text{bend}}(x) &= \int_\Omega \frac{1}{2}Er^4\frac{\pi}{4}\kappa^2 dl \\ &\approx \sum_i \frac{\bar{l}_i}{2}\frac{1}{2}Er^4\frac{\pi}{4}(\frac{\kappa b_i}{\bar{l}_i/2})^2 \\ &= \sum_i Er^4\frac{\pi}{4}\frac{(\kappa b_i)^2}{\bar{l}_i} \end{aligned} \quad (15)$$

Combining with Equation 14 we obtain that the $\alpha$ in Bergou et al. [2008] is associated with the rod's radius as

$$\alpha = Er^4\frac{\pi}{4}$$

so now the change of rod thickness under a certain bending modulus $E$ can be automatically reflected in the bending stiffness. This enables us to easily setup rod examples starting from real material thickness and Young's modulus of e.g. hair, iron, etc, without tuning bending stiffness $\alpha$. A more general form can be found in Section 4 of Bergou et al. [2010].

## 9 SHELL ENERGIES

Finally we elaborate on how our shell energies are integrated so that thickness changes are also directly included in *elasticity* behavior with Young's modulus $E$ and Poisson's ratio $\nu$ directly used to set stiffnesses.

*Membrane Energy.* We apply the membrane energy in Koiter's shell model from Chen et al. [2018]:

$$\Psi_{\text{ShellM}}(x) = \sum_t V_t \frac{1}{4}(\frac{1}{2}\lambda \text{tr}^2(M_t) + \mu \text{tr}(M_t^2)); \quad (16)$$

where $M_t = \bar{F}_t^{-1}F_t - I$, with $F_t$ and $\bar{F}_t$ the first fundamental form of triangle $t$ in world space and material space respectively, $\mu$ and $\lambda$ are the Lamé parameters for shells, and $V_t = |\bar{t}|\xi_t$ is the volume weighting with triangle rest area, $|\bar{t}|$, and thickness, $\xi_t$.

*Bending Energy.* We apply the discrete shell (hinge) bending energy from Grinspun et al. [2003]:

$$\Psi_{\text{ShellB}}(x) = \sum_e k_e(\theta_e - \bar{\theta}_e)^2||\bar{e}||/\bar{h}_e \quad (17)$$

where $k_e$ is the bending stiffness, $\theta_e$ and $\bar{\theta}_e$ are the dihedral angles of edge $e$ in world space and material space respectively, $||\bar{e}||$ is the rest length of edge $e$, and $\bar{h}_e$ is $\frac{1}{3} \times$ of the average of the heights of the two triangles incident to the edge $e$ [Grinspun et al. 2003]. Here we apply the parameterization in Tamstorf and Grinspun [2013]

$$k_e = \frac{E}{24(1-\nu^2)}\xi_e^3, \quad (18)$$

setting the bending stiffness $k_e$ to half of the flextural rigidity so that it is automatically adjusted with the thickness $\xi_e$, and we can instead intuitively set the Young's modulus $E$ and Poisson's ratio $\nu$ according to measured cloth material parameters [Penava et al. 2014].

*Material Settings.* In Penava et al. [2014], we find densities, Young's moduli, and Poisson's ratios for a range of cloth materials. As most cloth are anisotropic, the Young's moduli and Poisson's ratios vary with directions (see [Penava et al. 2014], Table 3). Therefore, we pick the smallest Young's modulus and Poisson's ratio among all directions per material to perform our experiments. For example, from Penava et al. [2014], Table 3, we see cotton has the smallest Young's modulus $0.821 MPa$ in $45°$, and the smallest Poisson's ratio $0.243$ at $90°$. However, as discussed in our main text, despite applying this smallest measured stiffness value we still observe membrane locking at moderate simulation resolutions. Instead C-IPC applies a smaller scaled ($0.1\times$ or even $0.01\times$ the smallest measured value) Young's modulus to avoid membrane locking *combined* with constitutive strain-limiting barrier to ensure no stretching artifacts. Our applied strain limits are set based on Penava et al. [2014], Table 2, where the measured strain range reflects the elasticity range. For example, cotton's largest measured strain range is $6.08\%$ at $30°$, and we use this value to set its strain limit. Applying C-IPC to accurately model the anisotropic properties of cloth to closely match measurements, and likewise to properly handle plastic deformation and cloth fracture all remain interesting future work.



\end{document}